\documentclass[prd,showpacs,preprintnumbers,amsmath,amssymb,floatfix]{revtex4}

\usepackage{graphicx}% Include figure files

\usepackage{bm}% bold math
\usepackage{amsfonts}
\usepackage{array}
\usepackage{float}
\usepackage{epstopdf}
\usepackage{color}

\begin{document}

\title{Anisotropic Compact stars in the Buchdahl model: A comprehensive study}

\author{S. K. Maurya}
\email{sunil@unizwa.edu.om}
\affiliation{Department of Mathematical and Physical Sciences, College of Arts and Science, University of Nizwa, Nizwa, Sultanate of Oman}

\author{Ayan Banerji}
\email{ayan\_7575@yahoo.co.in}
\affiliation{Astrophysics and Cosmology Research Unit, University of KwaZulu Natal, Private Bag X54001, Durban 4000, South Africa}

\author{M. K. Jasim}
\email{mahmoodkhalid@unizwa.edu.om }
\affiliation{Department of Mathematical and Physical Sciences, University of Nizwa, Nizwa, Sultanate of Oman}

\author{J. Kumar \& A. K. Prasad}
\email{jitendark@gmail.com}
\affiliation{Department of Applied Mathematics, Central University of Jharkhand, Ranchi-835205, India.}

\author{Anirudh Pradhan}
\email{pradhan.anirudh@gmail.com}
\affiliation{Department of Mathematics, Institute of Applied Sciences \& Humanities, GLA University, Mathura-281 406, Uttar Pradesh, India}

\date{\today}
\begin{abstract}
In this article we present a class of relativistic solutions describing spherically
symmetric and static anisotropic stars in hydrostatic equilibrium.  For this purpose, we consider a particularized metric potential, namely, Buchdahl ansatz [Phys. Rev. D \textbf{116}, 1027 (1959).] which encompasses almost all the known analytic solution to the spherically symmetric, static Einstein field equations(EFEs) with a perfect fluid source, including in particular the Vaidya-Tikekar and Finch-Skea. We here developed the model by considering anisotropic spherically symmetric static  general relativistic configuration that plays a significant effect on the structure and properties of stellar objects. We have considered eight different cases for generalized Buchdahl dimensionless parameter $K$, and analyzed them in an uniform manner. As a result it turns out that all the considered cases are valid at every point in the interior spacetime. In addition to this, we show that the model satisfies all the energy conditions and maintain hydrostatic equilibrium equation.  In the frame work of anisotropic hypothesis, we consider analogue objects with similar mass and radii such as LMC X-4, SMC X-1, EXO 1785-248 \emph{etc} to restrict the model parameter arbitrariness.
Also, establishing a relation between pressure and density in the form of $P = P (\rho)$,
we demonstrate that EoSs can be approximated to a linear function of density.
 Despite the simplicity of this model, the obtained results are satisfactory.
\end{abstract}


\maketitle

\section{Introduction}
%%%%%%%%%%%%%%%%%%%%%%%%%%%%%%%%%%%%%%
In astrophysics, studying the structural properties and formation of compact objects
such as neutron stars (NSs) and quark stars (QSs), have attracted much attention
to the researchers in the context of General Relativity (GR), as well as widely developing  modified theories of gravity. Crudely, compact stars are the final stages in the
evolution of ordinary stars which become an excellent testbeds for the study of
highly dense matter in an extreme conditions. In recent times a number of compact objects with high densities have been discovered \cite{Lattimer}, which are often observed as pulsars, spinning stars with
strong magnetic fields. Our theoretical understanding about compact star is rooted in the Fermi-Dirac statistics, which is responsible for the high degeneracy pressure that holds up the star against gravitational collapse was proposed by Fowler in 1926 \cite{Fowler}. Shortly afterwards, using Einstein's special theory of relativity
and the principles of quantum physics,  Chandrasekhar showed that \cite{Chandrasekhar1,Chandrasekhar2} white dwarfs are compact stars which is supported solely by a degenerate gas of electrons, to be stable if the maximum size of a stable white dwarf, approximately $3 \times 10^{30} $ kg (about 1.4 times the mass of the Sun).

As of today, there is no comprehensive description of extremely dense matter in a strongly
interacting regime.  A possible theoretical description of such nuclear matter in
extreme densities may consist not only of leptons and nucleons but also several exotic components in their different forms and phases such as hyperons, mesons, baryon resonances as well as strange quark matter (SQM). Therefore, a real composition of matter distribution in the interior of compact objects remains a question for deeper examination. The most general spherically symmetric matter distribution usually thought to be an isotropic fluids, because astrophysical observations support isotropy.
A possible theoretical algorithm was proposed by Fodor \cite{Fodor} that can generate any number of physically realistic pressure and density profiles for isotropic distributions without evaluating integrals.

On one hand, when densities of compact objects are normally above the nuclear matter density, one can expect the appearance of unequal principal stresses, the so-called anisotropic fluid. This usually means that two different kind of pressures inside these compact objects viz., the radial pressure and the tangential pressure \cite{Herrera1}.  This leads to the anisotropic condition that radial pressure component, $p_r$ is not equal to the components in the transverse directions, $p_t$. This effect was first predicted in 1922 by J.H. Jeans \cite{Jeans}
for self-gravitating objects in Newtonian regime. Shortly later,  in the context of GR, Lema$\hat{\textit{i}}$tre \cite{Lemaitre} had also consider the local anisotropy effect and showed that one can relax the upper limits imposed on the maximum value of the surface gravitational potential. Ruderman \cite{Ruderman} gave an interesting picture about more realistic stellar models
and showed that a star with matter density ($\rho > 10^{15} gm/cm^3$), where the nuclear interaction become relativistic in nature, are likely to be anisotropic.

The inclusion of anisotropic effect on compact objects was first considered by Bowers and Liang \cite{Bowers} in 1974. They studied static spherically symmetric configuration and analyzed the hydrostatic equilibrium equation, modified from of it's original form to include the anisotropy effects. Moreover, they provided the results by making
comparison with the stars filled with isotropic fluid. Heintzmann and Hillebrandt \cite{Heintzmann} have investigated neutron star models at high densities with an anisotropic equation of state, and found for arbitrary large anisotropy there is no limiting mass for neutron star. Though the maximum mass of a neutron star still lies beyond 3-4 $M_{\bigodot}$. A lot of works have been carried out 
in deriving new physical solutions with interior anisotropic fluids. Herrera and Santos \cite{Herrera1} reviewed and discussed about possible causes for the appearance of local anisotropy  in self gravitating systems with an examples in both Newtonian and general relativistic context. In \cite{Mak}, a class of exact solutions of Einstein's gravitational field equations have been put forward for the existence of anisotropy in star models. In addition above Harko and his collaborators \cite{Harko1,Harko2,Harko3,Harko4,Harko5} have done some significant work on anisotropic matter distribution. For new exact interior solutions to the Einstein field equations, Chaisi and Maharaj \cite{Maharaj1}
have studied the gravitational behaviour of compact objects under strong gravitational fields. 
Very recently an analysis based on the linear quark EoS for finding the equilibrium
conditions of an anisotropically sustained charged spherical body has been revisited by Sunzu \emph{et al} \cite{Maharaj2}. The studies developed in \cite{Maurya2016,Maurya2017,Maurya1,Maurya2,Deb,Kalam} form part of a quantity of works where the influence of the anisotropic effect on the structure of static  spherically symmetric compact objects are analyzed.
In favour of anisotropy Kalam \emph{et al} \cite{Kalam1} have developed a star model and showed that 
central density depends on anisotropic factor.  For recent investigations, there have been important efforts in
describing relativistic stellar structure in \cite{Bhar,Ratanpal,Takisa,Maharaj3}. The algorithm for solutions of Einstein field equation via. single monotone function have already been discovered by authors \cite{Lake,Herrera2008,Maurya2017a}.

On the other hand, spherical symmetry also allows more general anisotropic fluid configuration
with an EoS. In fact, if the EoS of the material composition of a compact star is
known, one can easily integrate the Tolman-Oppenheimer-Volkoff (TOV) equations to 
extract the geometrical information of a star. For example, linear EOS was used by Ivanov \cite{Ivanov} for charged static spherically symmetric perfect fluid solutions. This situation has been extended by Sharma and Maharaj \cite{Maharaj4} for finding an exact solution to the Einstein field equations with an anisotropic matter distribution. In Ref. \cite{Herrera2}, Herrera and Barreto had considered polytropic stars with anisotropic pressure. Solutions of Einstein's equations for anisotropic fluid distribution with different EoS have been found in \cite{Ivanov1,Rahaman,Nasim,Isayev,Deb,Ivanov2}.
But, in case EoS of the material composition of a compact star is not yet known except 
some phenomenological assumptions, one can introduce a suitable metric \emph{ansatz} for one of the metric 
functions to analyze the physical features of the star. Such a method was initially proposed by Vaidya-Tikekar \cite{Vaidya} and Tikekar \cite{Tikekar}, prescribed an approach of assigning different
geometries with physical 3-spaces (see \cite{komathiraj,Patel,Maharaj5,Gupta} and references therein). Similar type of metric ansatz was considered by Finch and Skea \cite{Finch}  
satisfying all criteria of physical acceptability according to Delgaty and Lake \cite{Delgaty}.
As a consequence, problem of finding the equilibrium configuration of a stellar structure for
anisotropic fluid distribution have been found in \cite{Herrera2129,Paul,Paul1,Maharaj6}.

In the present paper, we consider fairly general Buchdahl ansatz \cite{Buchdahl} for the metric potential. Such an assumption makes Einstein's field equations tractable and cover almost all physically tenable known models of super dense star. Actually, Vaidya and Tikekar \cite{Vaidya} particularized
Buchdahl ansatz by giving a geometric meaning, prescribing specific 3-spheroidal geometries for 4 dimensional hypersurface. This spheroidal condition has been found very useful for finding an exact solution of the Einstein field equations, which is not easy in many other cases. Such 
particular assumption was considered by Kumar \emph{et al} \cite{Kumar1,Kumar}, and comprehensively studied charged compact objects for isotropic matter distribution. Sharma et al.\cite{Sharma}
 have obtained the maximum possible masses and radii for different values of surface density for Vaidya-Tikekar space time.  

Neutron stars, the remnants of the gravitational collapse of $\sim 8$ to $20~ M_{\odot}$ main-sequence stars, in which fundamental physics can be probed in an extreme conditions via astrophysical observations. The structure of such stars depend on the EoS of nuclear matter under extreme conditions. Thus, neutron stars are an excellent probe for the study of dense and strongly-interacting matter. More specifically, the mass-radius of a neutron star is directly related to the EoS of neutron-rich matter \cite{Oertel}, and this could be achieved through the independent measurement of their mass and radius \cite{Steiner2010,Ozel,Ozel1,Guver}. 

From an observational viewpoint, our understanding about neutron stars has changed drastically in the last decade after the discovery of pulsar PSR J1614-2230 \cite{Demorest} as 1.97 $M_{\odot}$. The most significant progress in determining the properties of neutron stars, such as their masses and radii, which is necessary for constraining the equation of state.  However, obtaining an accurate measurements of both the mass and radius of neutron stars are more difficult. Till date, only in a few cases mass and radius of compact stars have been estimated by exploiting a variety of observational techniques, including, in particular radio observations of pulsars and X-ray spectroscopy for example during thermonuclear bursts \cite{Ebisuzaki,Damen,Steiner} or in the quiescent state of low mass X-ray binaries \cite{Brown,Marino}.  It is therefore great important to understand the maximal mass value of such objects which is still an open question but recent observations estimate this limit as $\sim$ 2 $M_{\odot}$, while, for the pulsar J0348+0432, it is 2.01 $M_{\odot}$ \cite{Antoniadis}. Recent studies have reported massive neutron stars to be such as PSR J1614+2230 ($\sim$ 1.97 $M_{\odot}$ \cite{Demorest}), Vela X-1 ($\sim$ 1.8 $M_{\odot}$ \cite{Rawls}) and 4U 1822-371 ($\sim$ 2 $M_{\odot}$ \cite{Darias}). X-ray pulsations with
a period of 13.5 s were first detected in LMC X-4 by Kelley \emph{et al.} \cite{Kelley}. However, the maximal limit of neutron star mass can increases considerably due to strong magnetic field
inside the star.

Thus, neutron stars are very peculiar objects, and observational data about their macroscopic properties (mainly the mass-radius $M-R$ relation) can also be used for studying accurate derivation consistent with the observations. In this paper, we discuss the possibility of extendable range of Buchdahl dimensionless parameter $K$ (a measure of deviation from sphericity) to explore a class of neutron stars in the standard framework of General Relativity. In our model, we do not prescribe the EOS; rather we apply two step method to examine the possibility of using the anisotropy to obtain spherically symmetric 
configurations with Buchdahl metric potential. In order to constrain the
value of model parameters, we consider analogue objects with similar mass and radii such as LMC X-4 \cite{Kelley}, SMC X-1 \cite{Rawls}, EXO 1785-248 \cite{ozel}, SAX J1808.4-3658 (SS2) \cite{Elebert}, Her X-1 \cite{Abubekerov}, 4U 1538-52 \cite{Rawls}, PSR 1937+21 \cite{Freire}, and Cen X-3 \cite{Rawls} to those stars in Buchdahl anisotropic geometry.

The paper begins with the introduction in Sec.I , then
we introduce the relevant Einstein equations for the case of spherical symmetry static spacetime in standard form of Schwarzschild-like coordinates in Sec. \ref{gre}. In Sec. \ref{ESM}, we assume anisotropic pressure in the modeling of realistic compact stellar structures. In the same section we derive the field equations by using coordinate transformation and found eight possible solutions for positive and negative values of Buchdahl parameter $K$. In Sec. \ref{ES}, We discuss the junction conditions and determine the constant coefficient. We also presented the mass-radius relation and surface redshift of the stellar models in same section \ref{ES}. 
The Sec. \ref{PF}, includes detailed analysis of physical features and obtained results are compared with  data from observation along with equation of state (EOS) of the compact star. Concluding remarks have been made in Sec. \ref{FR}.

\section{General relativistic equations}\label{gre}
Let us consider the spacetime being static and spherically symmetric, which describes the interior of the object can be written in the following form

 \begin{equation}\label{eq1}
ds^{2} = -e^{\nu(r) } \, dt^{2}+e^{\lambda(r)} dr^{2} +r^{2}(d\theta^{2} +\sin^{2} \theta \, d\phi^{2}),
\end{equation}

where the coordinates ($t$, $r$, $\theta$, $\phi$) are the Schwarzschild-like coordinates and $\nu(r)$ and $\lambda(r)$ are arbitrary functions of the radial coordinate $r$ alone, which yet to be determined. The Einstein 
tensor is $G_{\mu\nu}=R_{\mu\nu}-\frac{1}{2}g_{\mu\nu}R$, with $R_{\mu\nu}$ and $g_{\mu\nu}$  being respectively the Ricci and the metric tensors, and $R$ being the Ricci scalar ( with the assumption of natural units $G = c = 1$).

Here, we consider the matter contained in the sphere is described by anisotropic fluid. Thus, the structure of such an energy-momentum tensor is then expected to be of the form
\begin{equation}\label{eq2}
T_{\mu\nu} = (\rho + p_{t})u_{\mu}\, u_{\nu} - p_{t}(g_{\mu\nu}) + (p_{r} - p_{t}) \chi_{\mu} \chi_{\nu}, 
\end{equation}
where $u_{\mu}$ is the four-velocity and $\chi_{\mu}$ is the unit spacelike
vector in the radial direction. Thus, the Einstein field equation, $G_{\mu\nu} = 8\pi T_{\mu\nu}$
provides the following set of gravitational field equations 
\begin{eqnarray}
\label{eq3}
\kappa \, \rho(r) &=& \frac{\lambda '}{r} e^{-\lambda } +\frac{(1-e^{-\lambda } )}{r^{2}},\\
\label{eq4}
\kappa \, p_{r}(r) &=& \frac{\nu'}{r} e^{-\lambda } -\frac{(1-e^{-\lambda } )}{r^{2}},\\
\label{eq5}
\kappa \, p_{t}(r) &=& e^{-\lambda }\left[\frac{\nu''}{2} -\frac{\lambda' \nu'}{4} +\frac{\nu'^{2} }{4} +\frac{\nu'-\lambda '}{2r} \right] ,
\end{eqnarray}

where the prime denotes a derivative with respect to the radial coordinate, $r$ and $\kappa=8\,\pi$. Here, $\rho$ is the energy density, while the quantities $p_{r}$ is the pressure in the direction of $\chi^{\nu}$ (radial pressure)
 and $p_{t}$ is the pressure orthogonal to $\chi_{\nu}$ (transversal pressure). Note that pressure isotropy is not required by spherical symmetry, it is an added assumption \cite{Mak,Herrera2001}.
Consequently, $\Delta$ = $p_{t} - p_{r}$ is denoted as the anisotropy factor according to Herrera and Leon \cite{Herrera3}, and it's measure the pressure anisotropy of the fluid. It is to be noted that at the origin of the stellar configuration $\Delta = 0$, i.e.  $p_{t} = p_{r} = p$ is a particular case of an isotropic pressure. Using Eqs. (\ref{eq4}) and (\ref{eq5}), one can obtain the simple form of anisotropic factor, which yield 

\begin{equation}\label{delta1}
\Delta =\kappa \, (p_{t} -\, p_{r} ) = e^{-\lambda } \left[\frac{\nu''}{2} -\frac{\lambda' \nu'}{4} +\frac{\nu'^{2} }{4} -\frac{\nu'+\lambda '}{2r} -\frac{1}{r^{2} } \right]\,  + \frac{1}{r^{2} }. 
\end{equation}
However, a force due to the anisotropic pressure is represented by $\Delta/r$, which is repulsive, if $p_{t} > p_{r}$, and attractive if $p_{t} < p_{r}$ of the stellar model. For the considered matter distribution when $p_{t} > p_{r}$ allows the construction of more compact objects, compared to isotropic fluid sphere \cite{Gokhroo}. Note that this is a system of 3 equations with 5 unknowns. Thus, the system of equations is undetermined, and by assuming suitable conditions we have to reduce the number of unknown functions.

\section{Exact solution of the models for anisotropic stars}\label{ESM}
In this section we establish a procedure for generating a new anisotropic solution of the Einstein field equations from a known metric ansatz due to Buchdahl \cite{Buchdahl} that covers almost all interesting solutions. We use the widely studied metric ansatz given by
  \begin{equation} \label{lambda1}
  e^{\lambda}=\frac{K\,(1+Cr^2)}{K+Cr^2}, ~~~~~ \textrm{when} ~~~ K < 0~~~ \textrm{and}~~ K > 1,
 \end{equation}
 where $K$ and $C$ are two parameters that characterize the geometry of the star. Note that the ansatz for
 the metric function $g_{rr}$ in (\ref{lambda1}) was proposed by Buchdahl \cite{Buchdahl} to develop a viable
 model for a relativistic compact star. The choice of the metric potential is physically well motivated 
 (especially the energy density must be non-singular and decreasing outward) and has been used by many in the past to construct viable stellar models. In addition to above the metric function (\ref{lambda1}) is also positive and free from singularity at $r=0$ and monotonic increasing outward. Here, we will illustrate how an analytic Buchdahl model could be extendable for positive and negative values of spheroidal parameter $K$.
 In the following analysis we pull out the range of $0< K< 1$, where either the energy density or pressure will be negative depending on the two parameters.
 It is interesting to note that one can recover the Schwarzschild interior solution when $K=0$ and for  $K=1$ the hypersurfaces $\{t = \mbox{constant} \}$ are flat.
 In a more generic situation, one could recover the,   Vaidya and Tikekar \cite{Vaidya} solution when $C = -K/R^{2}$, Durgapal and Bannerji \cite{Durgapal} when $K=-2$. The solutions for charged and uncharged perfect fluid were considered by Gupta \textit{et al} \cite{Gupta2005,Gupta2004}, but none of them were well behaved within the proposed range of parameter $K$. However, in the present study we obtain the well behaved solution for some values of $K$ by introducing anisotropy parameter $\Delta$, which provides monotonically decreasing sound speed within the compact stellar model.
 
 As a next step in our analysis we introduce the transformation $e^{\nu}=Y^{2}(r )$ \cite{Ivanov,Kumar1,Kumar}, and substituting the value of $e^{\lambda}$ into the Eq. (\ref{delta1}), one arrives in the following relations 
 \begin{equation}\label{Diff1}
 \frac{d^2Y}{dr^2}-\left[\frac{K+2\,K\,Cr^2+C^2r^4}{r\,(K+Cr^2)\,(1+Cr^2)} \right]\,\frac{dY}{dr}+\left[\frac{C\,(1-K)\,C^2r^4}{r^2\,(K+Cr^2)\,(1+Cr^2)}-\frac{\Delta\,K\,(1+Cr^2)}{(K+Cr^2)}\,\right]\,Y = 0. 
 \end{equation}
The Eq. (\ref{Diff1}) having two unknowns namely $Y(r)$ and $\Delta$. While in order to solve for $Y$, 
we will follow the approach in \cite{Mak}.  Hence, we choose the expressions for anisotropy parameter 
$\Delta=\frac{\Delta_{0}\,C^2r^2}{(1+Cr^2)^2}$.  The constant $\Delta_{0} \geq 0$, with the assumption that
$\Delta_{0} = 0$, corresponding to the isotropic limit. As argued in \cite{Mak}, that $\Delta_{0}$ is the 
measure of anisotropy of the pressure distribution inside the fluid sphere, while at the center the anisotropy vanishes, i.e. $\Delta(0) = 0$. With hindsight, for chosen anisotropy parameter the interior solutions ensure the regularity condition at the centre also.
Therefore, with this choice of $\Delta$ and using an appropriate transformation~~ $Z=\sqrt{\frac{K+Cr^2}{K-1}}$,  the Eq. (\ref{Diff1})  becomes a hypergeometric differential equation of the form 
\begin{equation}\label{Diff2}
(1-Z^2)\,\frac{d^2Y}{dZ^2}+Z\,\frac{dY}{dZ}+(1-K+\Delta_{0}\,K)\,Y=0. 
\end{equation}

Our aim here is to solve the system of the above hypergeometric Equation (\ref{Diff2}) by using Gupta-Jasim \cite{Gupta2004} two step method (See appendix (A)). In this framework we consider two cases for the spheroidal parameter $K$  

\subsubsection*{\textbf{Case I. For $ K < 0 $  i.e  K is negative}}
Now we differentiate the Eq.(\ref{Diff2}) with respect to $Z$ and use another  substitution $Z=\sin{x}$ and $\frac{dY}{dZ}=\psi$, then we have
\begin{equation}\label{eq10}
\frac{d^2\,\psi}{dx^2}+(2-K+\Delta_{0}\,K)\,\psi=0,  
\end{equation}
where $\frac{d\psi}{dx}=\cos{x}~~\frac{d^2Y}{dZ^2}$ and $\frac{d^2\psi}{dx^2}=\cos^2{x}~\frac{d^3Y}{dZ^3}-\sin{x}~\frac{d^2Y}{dZ^2}$, respectively. 
In this approach the above equation turns out to be a second order homogeneous differential equation with constant coefficients, and depends on the two parameters $K$ and $\Delta_{0}$. It is now interesting to classify the each solutions of  Eq. (\ref{eq10}) briefly
\begin{subequations}
\begin{align}
\textrm{Case~ Ia:} ~~~~ \psi=A_{1}\,\cosh (n\,x)+B_{1}\,\sinh (n\,x),~~~~~~~~~\textrm{if} ~~~~~~~~~~ 2-K+\Delta_{0}\,K=-n^2 \label{11}\\
\textrm{Case~ Ib:} ~~~~ \psi=C_{1}\,\cos (n\,x)+D_{1}\,\sin (n\,x),~~~~ ~~~~~~~~  \textrm{if}~~ 2-K+\Delta_{0}\,K=n^2 ~~(\ne 1) \label{12}\\
\textrm{Case~ Ic:} ~~~~ \psi=E_{1}\,\cos (x)+F_{1}\,\sin (x),~~~~~~~~~~~~~~~~~~\textrm{if} ~~~~~~~~~~~2- K+\Delta_{0}\,K=1 \label{13}\\
 \textrm{Case~ Id:} ~~~~\psi=G_{1}\,x+H_{1},~~~~~~~~~~~~~~~~~~~~~~~~~~~~~~~~~~~  \textrm{if} ~~~~~~~~~~2- K+\Delta_{0}\,K=0\label{14}
\end{align}
\end{subequations}
where $A_{1}$, $B_{1}$, $C_{1}$, $D_{1}$, $E_{1}$, $F_{1}$, $G_{1}$  and $H_{1}$ are arbitrary constant of integration, with $x=\sin^{-1}Z= \sin^{-1}\sqrt{\frac{K+Cr^2}{K-1}}$. Now, using (\ref{lambda1}) into the (\ref{eq3}) from which simple manipulations of the Einstein equations lead to the expression of energy density ($K < 0$) as 
\begin{equation}
\frac{\kappa\,\,\rho}{C}= \frac{(3-K + K\,\sin^2x-\sin^2x)}{K\,(K-1)\,\cos^4x}. \label{15}
\end{equation}
Subsequently, other EFEs relating to the metric potential and substituting different values of $Y$ (which is determined by substituting $dY/dZ = \psi$~ and ~$d^2Y/dZ^2 = d\psi/dZ$ in Hypergeometric equation Eq.(\ref{Diff2}), one can obtain\\
\\
\textbf{Case Ia:} ~~ $2-K+\Delta_{0}\,K=-n^2$ 
\begin{eqnarray}
  Y(x)&=&\frac{1}{(n^2+1)} \left[\, \cosh (n x)\,(\,A_{1} \sin x +B_{1}\,n\,\cos x \,) + \sinh(n\,x)\, (\,A_{1}\,n\,\cos x + B_{1} \sin x \,)  \,\right], \label{16}\\
 \frac{\kappa\,p_{r}}{C}&=& \frac{2\,(n^2+1)}{(1-K)\,K\,\cos^2x}\,\left[\,\frac{A_{1}\,\cosh(n\,x)+B_{1}\,\sinh(nx)}{\cosh (n x)\,(\,A_{1} +B_{1}\,n\,\cot x \,) + \sinh(n\,x)\,(\,A_{1}\,n\,\cot x + B_{1} \,) }\,\right]+\frac{1}{K\,\cos^2x}, \label{17}\\
\frac{\kappa\,p_{t}}{C}&=& \frac{2\,(n^2+1)}{(1-K)\,K\,\cos^2x}\,\left[\,\frac{A_{1}\,\cosh(n\,x)+B_{1}\,\sinh(nx)}{\cosh (n x)\,(\,A_{1} +B_{1}\,n\,\cot x \,) + \sinh(n\,x)\,(\,A_{1}\,n\,\cot x + B_{1} \,) }\,\right] +\Upsilon . \label{18}
\end{eqnarray} 
\textbf{Case Ib:}  ~~ $2-K+\Delta_{0}\,K=n^2~(\ne 1)$ 
\begin{eqnarray}
 Y(x)&=&\frac{1}{(1 - n^2)}\,\left[\,\sin x \,[\, C_{1}\,\cos(n x) + D_{1}\,\sin(n x)\,]-\,n\, \cos x\, [C_{1}\,\sin(n x)-D_{1} \cos(n x)\,] ~\right],\label{19}\\
\frac{\kappa\,p_{r}}{C}&=& \frac{2\,(1-n^2)}{(1-K)\,K\,\cos^2x}\,\left[\frac{C_{1}\,\cos (n\,x)+D_{1}\,\sin (n\,x)}{ \,C_{1}\,\cos(n x) + D_{1}\,\sin(n x)-\,n\, \cot x\, [C_{1}\,\sin(n x)-D_{1}\cos(n x)\,]}\,\right]+\frac{1}{K\,\cos^2x}, \label{20}\\
\frac{\kappa\,p_{t}}{C}&=& \frac{2\,(1-n^2)}{(1-K)\,K\,\cos^2x}\,\left[\frac{C_{1}\,\cos (n\,x)+D_{1}\,\sin (n\,x)}{ \,C_{1}\,\cos(n x) + D_{1}\,\sin(n x)-\,n\, \cot x\, [C_{1}\,\sin(n x)-D_{1} \cos(n x)\,]}\,\right]+\Upsilon. \label{21}
\end{eqnarray} 
\textbf{Case Ic:} ~~ $2-K+\Delta_{0}\,K=1$
\begin{eqnarray}
 Y(x)&=&\frac{1}{4}\,\left[\,E_{1}\,(2x+\sin 2x)-F_{1}\,\cos 2x\,\right], \label{22}\\
\frac{\kappa\,p_{r}}{C}&=& \frac{8\,\sin x}{(1-K)\,K\,\cos^2x}\,\left[\frac{E_{1}\,\cos (x)+F_{1}\,\sin (x)}{ \,E_{1}\,(2x+\sin 2x)-F_{1}\,\cos 2x}\right]+\frac{1}{K\,\cos^2x}, \label{23}\\
 \frac{\kappa\,p_{t}}{C}&=& \frac{8\,\sin x}{(1-K)\,K\,\cos^2x}\,\left[\frac{E_{1}\,\cos (x)+F_{1}\,\sin (x)}{ \,E_{1}\,(2x+\sin 2x)-F_{1}\,\cos 2x}\right]+\Upsilon. \label{24}
\end{eqnarray}
\textbf{Case Id:} ~~ $2-K+\Delta_{0}\,K=0$
\begin{eqnarray}
 Y(x)&=& A\,(\cos x + x\,\sin x)+ B \sin x ,\label{25}\\
\frac{\kappa\,p_{r}}{C}&=& \frac{2\,\sin x}{(1-K)\,K\,\cos^2x}\,\left[\frac{G_{1}\,x+ H_{1}}{ G_{1}\,(\cos x + x\,\sin x)+ H_{1} \sin x}\right]+\frac{1}{K\,\cos^2x}, \label{26}\\
 \frac{\kappa\,p_{t}}{C}&=& \frac{2\,\sin x}{(1-K)\,K\,\cos^2x}\,\left[\frac{G_{1}\,x+ H_{1}}{ G_{1}\,(\cos x + x\,\sin x)+H_{1} \sin x}\right]+\Upsilon, \label{27}
\end{eqnarray} 
where $\Upsilon= \frac{\Delta_{0}\,K\,[\,(K-1)\,\sin^2x-K\,]+(1-K)^2\,\cos^2x}{(1-K)^2\,K\,\cos^4x}$.

\subsubsection*{\textbf{Case II. For $K >1 $  i.e  K is Positive}}
Here, we extend our analysis by considering the positive values of $K$ and
 to solve the Eq. (\ref{Diff2}) we adopt a similar approach to differentiate the Eq. (\ref{Diff2}) with respect to $Z$. For this purpose we use another substitution $Z=\cosh{x}$ (hyperboloidal case) and $\frac{dY}{dZ}=\psi$,  equation (\ref{Diff2}) takes the form
\begin{equation}
\frac{d^2\,\psi}{dx^2}-(2-K+\Delta_{0}\,K)\,\psi=0, \label{28}
\end{equation}
where $\frac{d\psi}{dx} = \sinh{x}~\frac{d^2Y}{dZ^2}$, and $\frac{d^2\psi}{dx^2}= -\sinh^2{x}~\frac{d^3Y}{dZ^3}+\cosh{x}~\frac{d^2Y}{dZ^2}$, respectively. To solve the second order homogeneous differential equation (\ref{28}) we consider the following cases 
\begin{subequations}
\begin{align}
\textrm{Case~ IIa:} ~~~~ \psi=A_{2}\,\cos (n\,x)+B_{2}\,\sin (n\,x)~~~~~~~~~~~~~ \textrm{if} ~~~~~~~~~~ 2-K+\Delta_{0}\,K=-n^2, \label{29}\\
\textrm{Case~ IIb:} ~~~~ \psi=C_{2}\,\cosh (n\,x)+D_{2}\,\sinh (n\,x),~~~~~~~\textrm{if}~~~ 2-K+\Delta_{0}\,K=n^2 ~~(\ne 1) \label{30} \\
\textrm{Case~ IIc:} ~~~~ \psi=E_{2}\,\cosh (x)+F_{2}\,\sinh (x),~~~~~~~~~~~~~~~\textrm{if} ~~~~~~~~~~~2- K+\Delta_{0}\,K=1 \label{31}\\
\textrm{Case~ IId:} ~~~~\psi=G_{2}\,x+H_{2}~~~~~~~~~~~~~~~~~~~~~~~~~~~~~~~~~~~  \textrm{if} ~~~~~~~~~~2- K+\Delta_{0}\,K=0 ,\label{32}
\end{align}
\end{subequations}
where $A_{2}$, $B_{2}$, $C_{2}$, $D_{2}$, $E_{2}$, $F_{2}$, $G_{2}$ and $H_{1}$ are arbitrary constants of integration, with $x=\cosh^{-1}Z= \cosh^{-1}\sqrt{\frac{K+Cr^2}{K-1}}$. Recalling the Eq. (\ref{lambda1}) and plugged into the relevant equation we obtain the expression of energy density ($K >1$) as
\begin{equation}
\frac{\kappa\,\rho}{C}= \frac{( 3-K+K\,\cosh^2x-\cosh^2x )}{K\,(K-1)\,\sinh^4x}. \label{33}
\end{equation}
Now proceeding as same for $K<0$, we consider the following cases for $K>1$, and pressure components can be developed as follows:\\
\\
\textbf{Case IIa:} ~~ $2-K+\Delta_0\,K= -n^2$ 
\begin{eqnarray}
 Y(x)&=&\frac{1}{(n^2+1)} \left[\,\cosh x\, [ A\,\cos(n x) + B\, \sin(n x) ] + 
 n\,\sinh x \,[ A\, \sin(n x)-B\, \cos(n x) ]  \,\right], \label{34}\\
 \frac{\kappa\,p_r}{C}&=& \frac{2\,(n^2+1)}{(K-1)\,K\,\sinh^2x}\,\left[\,\frac{A_{2}\,\cos (n\,x)+B_{2}\,\sin (n\,x)}{\, [\, A_{2}\,\cos(n x) + B_{2}\, \sin(n x)\, ] + 
 n\,\tanh x\,[\, A_{2}\, \sin(n x)-B_{2}\, \cos(n x)\, ] }\,\right]-\frac{1}{K\,\sinh^2x}~~~\label{35}\\
 \frac{\kappa\,p_t}{C}&=& \frac{2\,(n^2+1)}{(K-1)\,K\,\sinh^2x}\,\left[\,\frac{A_{2}\,\cos (n\,x)+B_{2}\,\sin (n\,x)}{\, [ A_{2}\,\cos(n x) + B_{2}\, \sin(n x) ] + 
 n\,\tanh x\,[ A_{2}\, \sin(n x)-B_{2}\, \cos(n x) ] }\,\right] +\Upsilon_1 . \label{36}
\end{eqnarray}
\textbf{Case IIb:}  ~~ $2-K+\Delta_0\,K=n^2~(\ne 1)$ 
\begin{eqnarray}
 Y(x)&=& \frac{1}{(1 - n^2)}\,\left[\,\cosh x\, [ C_{2}\,\cosh(n x) + D_{2}\, \sinh(n x) ] - 
 n\,\sinh x\,[ C_{2}\, \sinh(n x)-D_{2}\, \cosh(n x) ] \,\right],\label{37}\\
\frac{\kappa\,p_r}{C}&=&\ \frac{2\,(1-n^2)}{(K-1)\,K\,\sinh^2x}\,\left[\,\frac{C_{2}\,\cosh (n\,x)+D_{2}\,\sinh (n\,x)}{\, [\, C_{2}\,\cosh(n x) + D_{2}\, \sinh(n x)\, ] - 
 n\,\tanh x\,[\, C_{2}\, \sinh(n x)-D_{2}\, \cosh(n x) \,] }\,\right]-\frac{1}{K\,\sinh^2x}~~~\label{38}\\
\frac{\kappa\,p_t}{C}&=&  \frac{2\,(1-n^2)}{(K-1)\,K\,\sinh^2x}\,\left[\,\frac{C_{2}\,\cosh (n\,x)+D_{2}\,\sinh (n\,x)}{\, [\, C_{2}\,\cosh(n x) + D_{2}\, \sinh(n x) ] - 
 n\,\tanh x\,[ C_{2}\, \sinh(n x)-D_{2}\, \cosh(n x) ] }\,\right]+\Upsilon_1, \label{39}
\end{eqnarray} 
 \textbf{Case IIc:} ~~ $2-K+\Delta_0\,K=1$
\begin{eqnarray}
 Y(x)&=&\frac{1}{4}\,\left[\,A\,\cosh 2 x + B\, \sinh (2 x) - 2B\,x \,\right],\label{40}\\
\frac{\kappa\,p_r}{C}&=& \frac{8\,\cosh x}{(K-1)\,K\,\sinh^2x}\,\left[\frac{E_{2}\,\cosh (x)+F_{2}\,\sinh (x)}{ \,E_{2}\,\cosh(2 x) + F_{2}\, \sinh (2 x) - 2\,F_{2}\,x }\,\right]-\frac{1}{K\,\sinh^2x} \label{41}\\
 \frac{\kappa\,p_t}{C}&=& \frac{8\,\cosh x}{(K-1)\,K\,\sinh^2x}\,\left[\frac{E_{2}\,\cosh (x)+F_{2}\,\sinh (x)}{ \,E_{2}\,\cosh(2 x) + F_{2}\, \sinh (2 x) - 2\,F_{2}\,x }\,\right]+\Upsilon_1  \label{42}
\end{eqnarray}
\textbf{Case IId:} ~~ $2-K+\Delta_0\,K=0$
\begin{eqnarray}
 Y(x)&=& G_{2}\,( x\,\cosh x - \sinh x ) + H_{2}\,\cosh x \label{43}\\
\frac{\kappa\,p_r}{C}&=& \frac{2\,\cosh x}{(K-1)\,K\,\sinh^2x}\,\left[\frac{G_{2}\,x+ H_{2}}{ G_{2}\,( x\,\cosh x - \sinh x ) + H_{2}\,\cosh x}\,\right]-\frac{1}{K\,\sinh^2x} \label{44}\\
 \frac{\kappa\,p_t}{C}&=& \frac{2\,\cosh x}{(K-1)\,K\,\sinh^2x}\,\left[\frac{G_{2}\,x+ H_{2}}{ G_{2}\,( x\,\cosh x - \sinh x ) + H_{2}\,\cosh x}\,\right]+\Upsilon_1  \label{45}
\end{eqnarray}
where $\Upsilon_1=\frac{\Delta_0\,K\,[\,(K-1)\,\cosh^2x-K\,]-(1-K)^2\,\sinh^2x}{(1-K)^2\,K\,\sinh^4x}$, and we have four set of solutions corresponding to the positive and negative values of $K$. Following the standard procedure for stellar modelling one usually impose some restrictions. In a realistic scenario, one can expect that following conditions satisfy throughout the stellar interior:
\begin{itemize}
\item The interior solution goes up to a certain radius $R$, where the spacetime is assumed not to possess an event horizon,

\item Positive definiteness of the energy density and pressure at the centre, 

\item The density should be maximum at centre and decresing monotonically within $ 0<r < R $ i.e. The density gradient  $ d\rho/dr $ is negative within $ 0<r < R $,

\item The pressure should be maximum at centre and decresing monotonically within $ 0<r < R $ i.e. The pressure gradient $ dp/dr $ is also negative within $ 0< r < R $.

\item The ratio of pressure and density should be less than unity within $ 0<r < R $ i.e. $p/\rho$ should lies between $0$ to $1$ within the stellar model.
\end{itemize}
\begin{figure}[htp!] 
\begin{center}
\includegraphics[width=6cm]{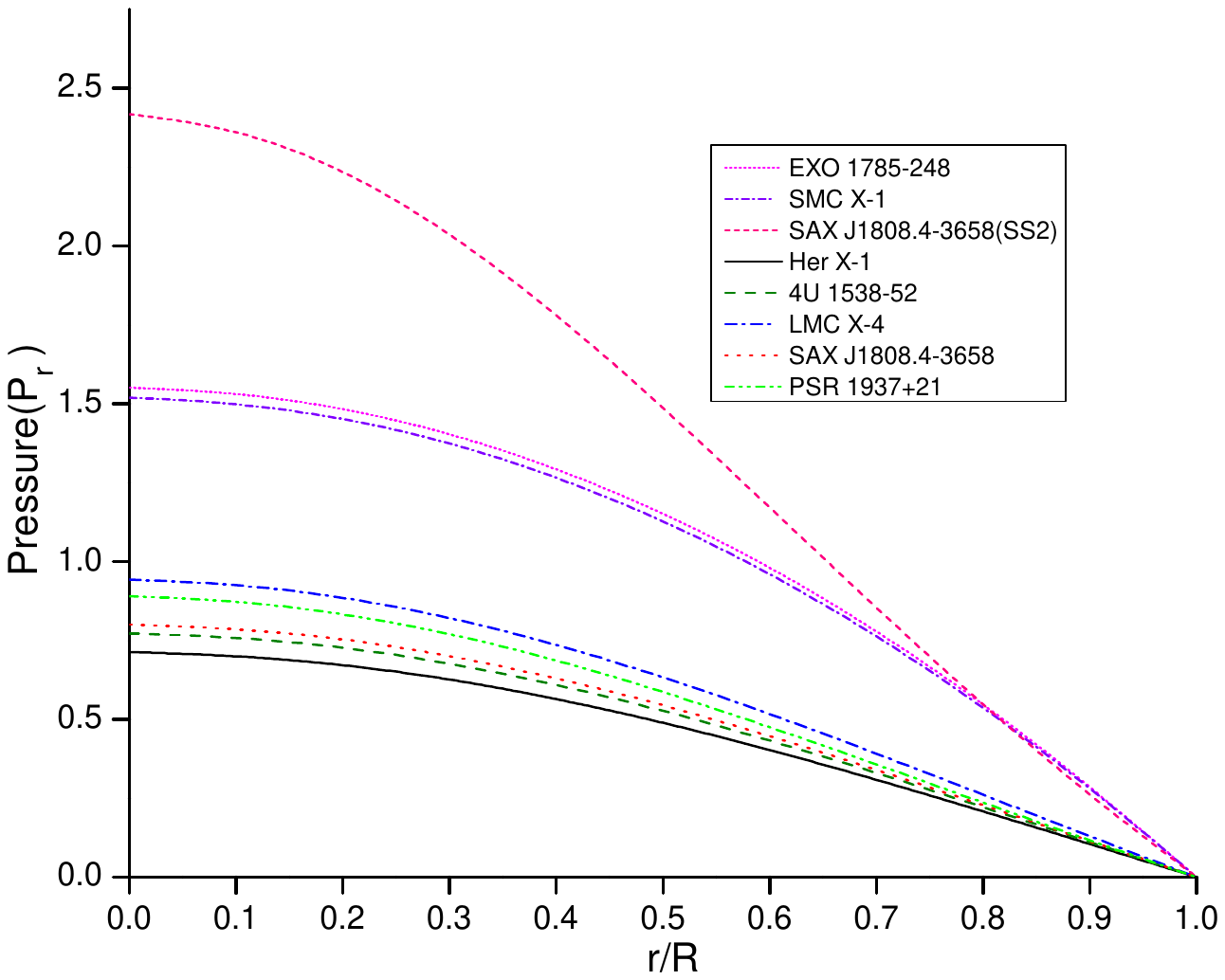}\includegraphics[width=6cm]{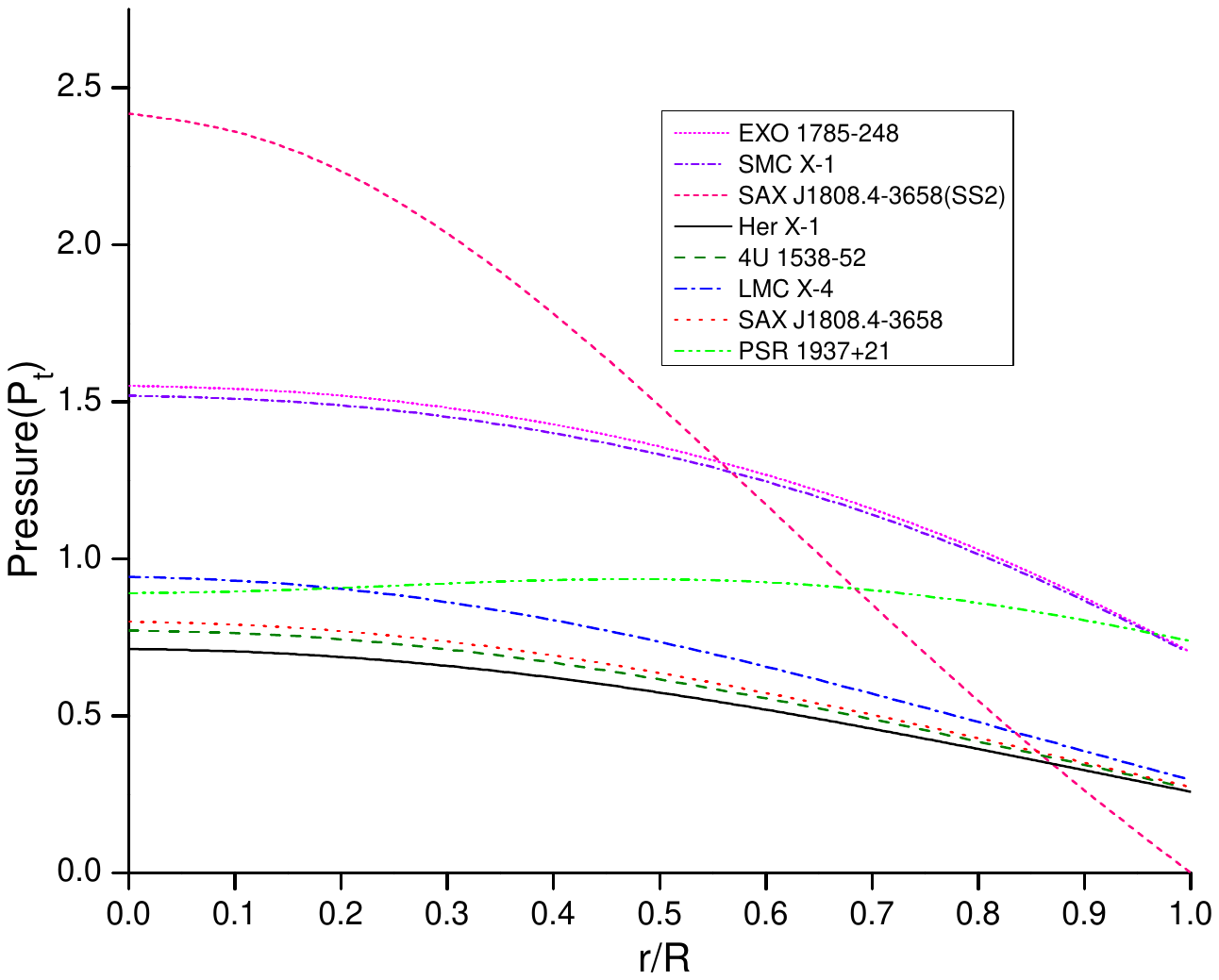}\includegraphics[width=6cm]{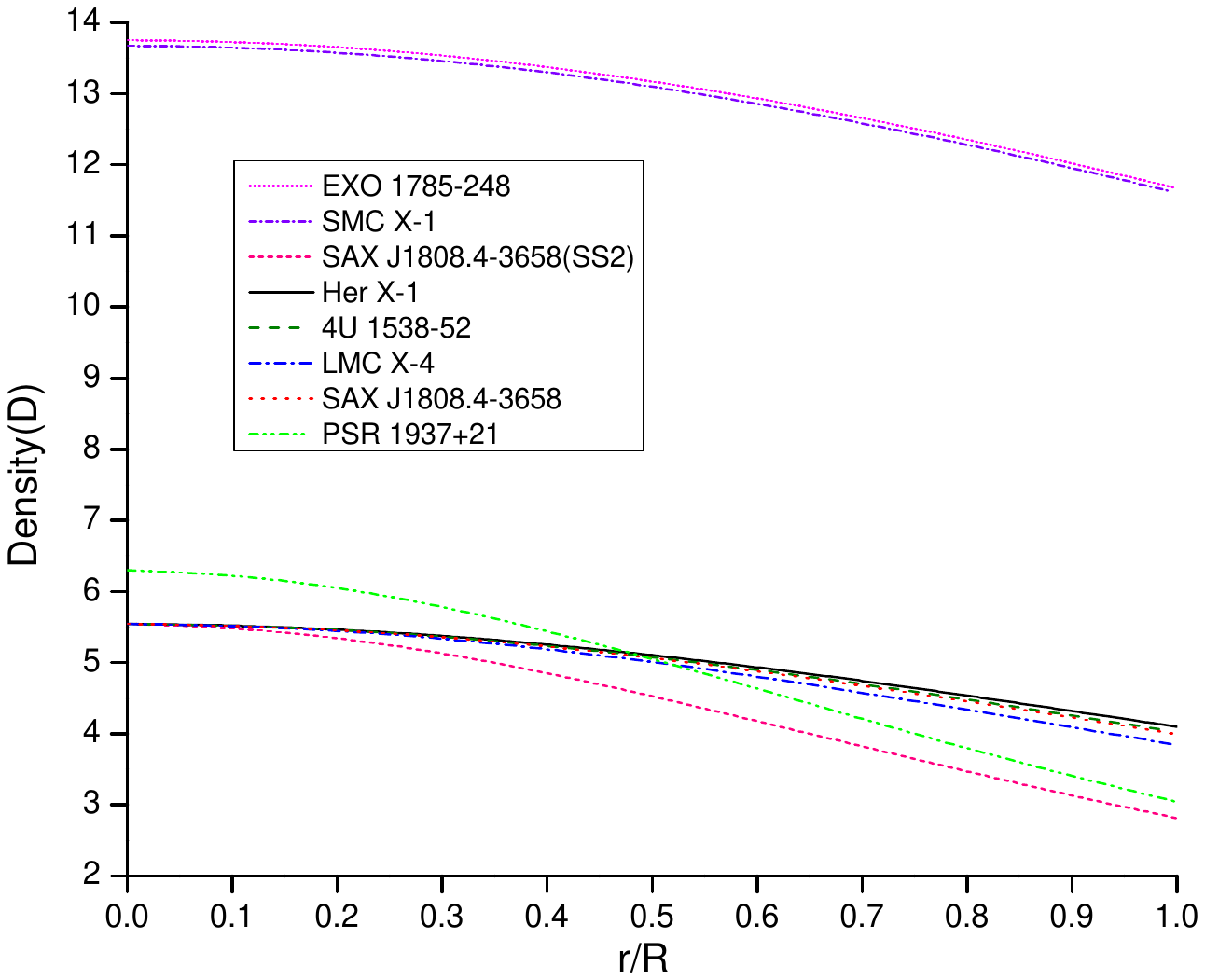}\\
\includegraphics[width=6cm]{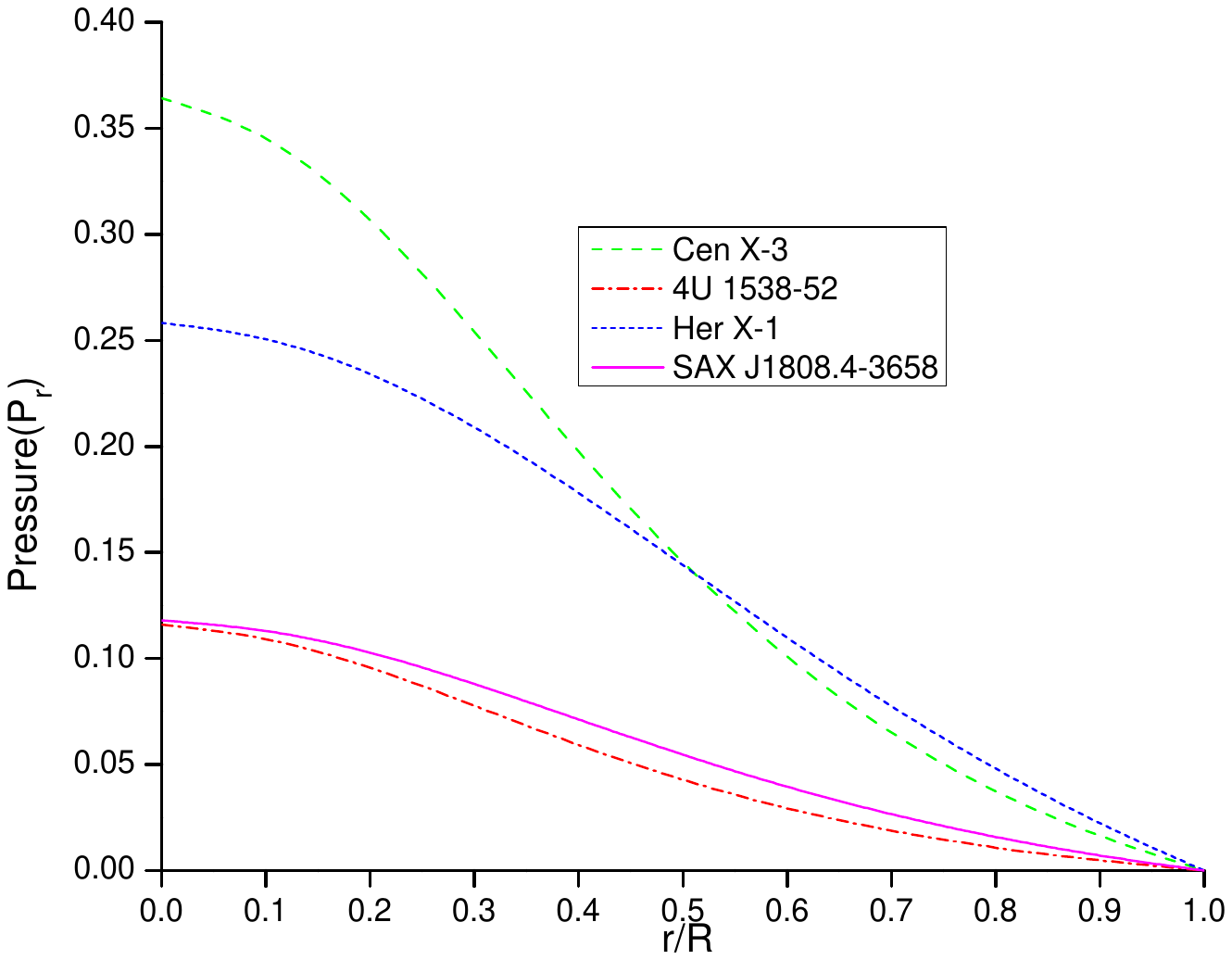}\includegraphics[width=6cm]{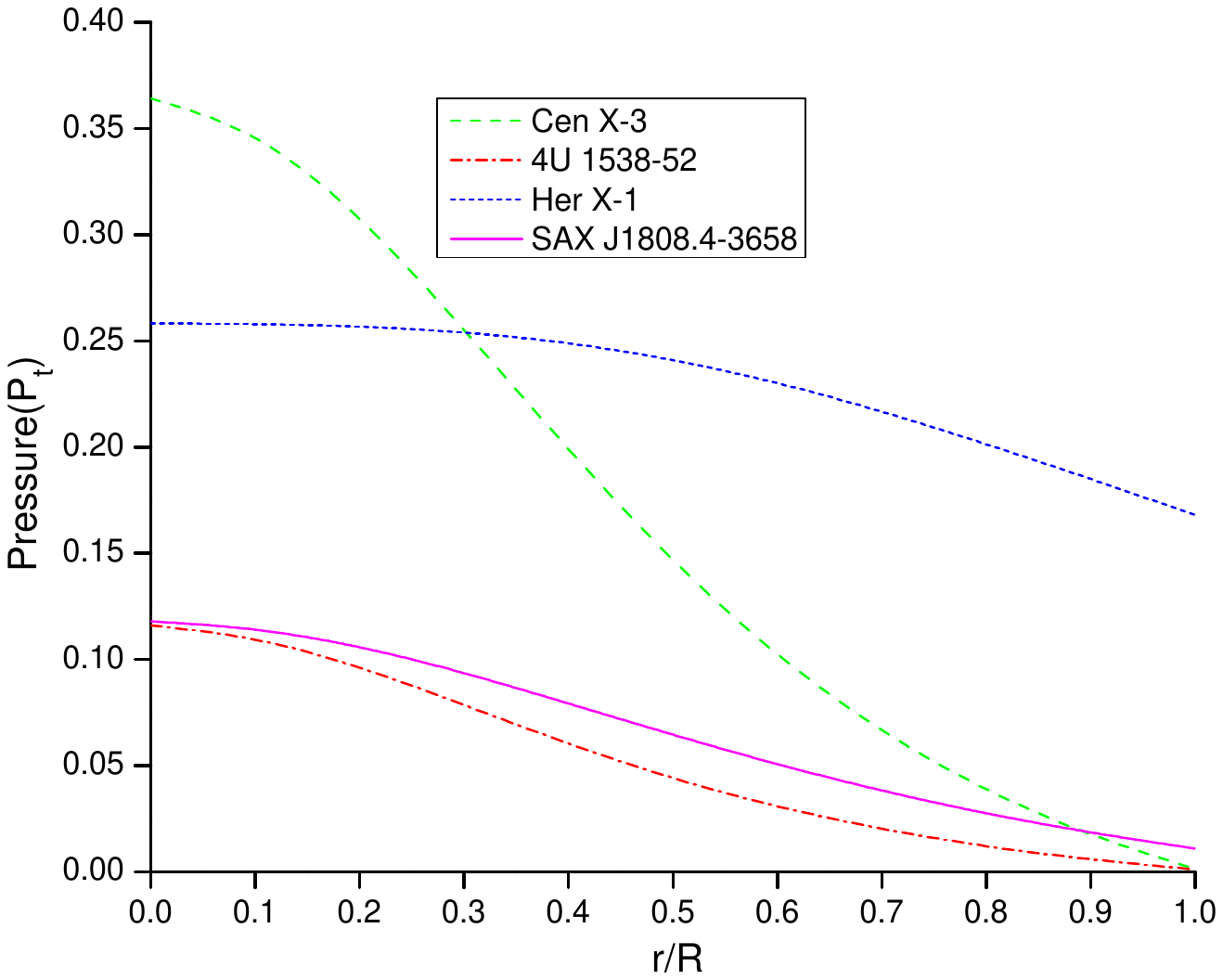}\includegraphics[width=6cm]{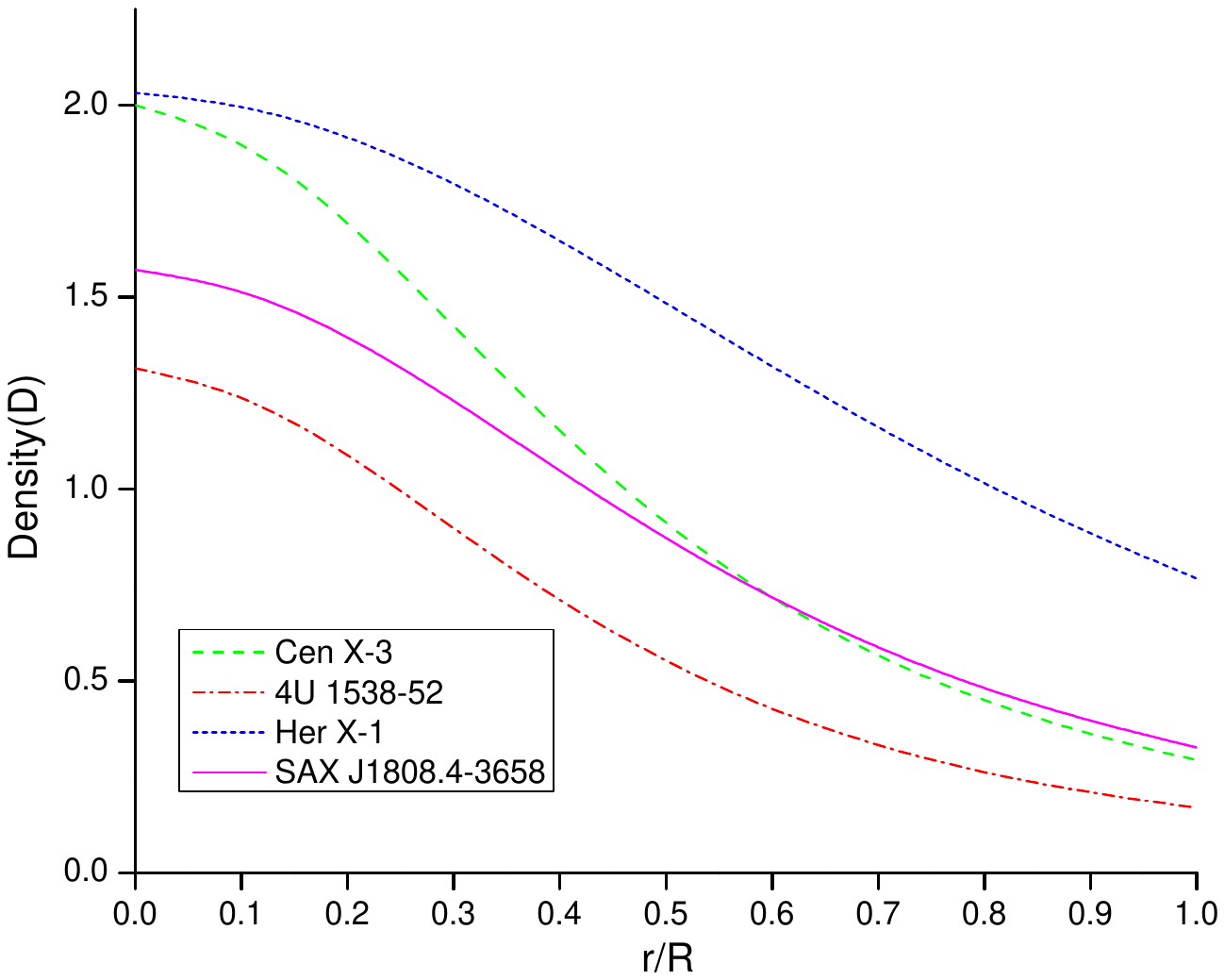}
\caption{From top left to right we have plotted effective radial pressure ($P_r=\kappa\,p_r/C$), effective transverse pressure ($P_t=\kappa\,p_t/C$) and effective energy density ($D=\kappa\,\rho/C$) verses radial coordinate ($r/R$) for Case \textbf{I}, in their normalized forms inside the star. In the lower graphs we repeat the same situation for Case \textbf{II}, where $P_r$ ,~ $P_t$ and $D$ are dimension less. The radial pressure ($p_r$), tangential pressure ($p_t$) and density ($\rho$) can be determined in $CGS$ unit as: $p_r=P_r\times C\,\times\,4.81 \times10^{47}\,dyne$, $p_t= P_t\times C\,\times\,4.81 \times10^{47}\,dyne$, $\rho= D\times C\,\times\,5.35 \times10^{26}\,gm/cm$.   
The values of parameter which we have used for graphical presentation are: (i) $K = -0.27898,~ C = 1.33\,\times10^{-13}\,cm^{-2}, ~ n = 0.1$ for EXO 1785-248 (Ia); (ii) K = -0.28103, $C = 1.52\,\times10^{-13}\,cm^{-2}$, n=0.1 for SMC X-1 (Ia); (iii) K = -1.18, $C = 1.37\,\times10^{-12}\,cm^{-2}$, n=1.783 for SAX J1808.4-3658 (SS2) (Ib); (iv) K = -1.18, $C = 3.07\,\times10^{-13}\,cm^{-2}$ for Her X-1 (Ic); (v) K = -1.18, $C = 3.47\,\times10^{-13}\,cm^{-2}$ for 4U 1538-52 (Ic); (vi) K =-1.18, $C = 3.21\,\times10^{-13}\,cm^{-2}$  for LMC X-4 (Ic); (vii) K = -1.18, $C = 3.49\,\times10^{-13}\,cm^{-2}$  for SAX J1808.4-3658 (Ic); (viii) K = -0.91, $C = 8.82\,\times10^{-13}\,cm^{-2}$ for PSR 1937+21 (Id); (ix) K=3, $C = 3.03\,\times10^{-12}\,cm^{-2}$, n= 0.99 for Cen X-3 (IIa); (x) K = 1.78, $C = 4.71\,\times10^{-12}\,cm^{-2}$, n=0.4796 for 4U 1538-52 (IIb); (xi) K = 3.1, $C = 1.28\,\times10^{-12}\,cm^{-2}$ for Her X-1 (IIc); (xii) K=2.1, $C = 2.78\,\times10^{-12}\,cm^{-2}$ for SAX J1808.4-3658 (IId). See Table 1 for more details.}
\label{f1}
\end{center}
\end{figure}
 These features, positive density, positive pressure, and the absence of
horizons, are the most important features characterizing a star. The task is now to check the well-behaved geometry and capability of describing realistic stars, we plot Figs. \ref{f1} (due to complexity of expression). For our stellar model, depending on the different values of $K$, the behavior of $\rho$, $p_r$ and $p_t$ have been studied. Such analytical representations have been performed by using recent measurements of mass and radius of neutron stars, LMC X-4, SMC X-1, EXO 1785-248, SAX J1808.4-3658 (SS2), Her X-1, 4U 1538-52, PSR 1937+21, Cen X-3 and SAX J1808.4-3658. Detailed expressions and value of constants
have been used in this work is given in Figs. \ref{f1}, and will not be repeated here. It is evident from these plots that energy density is maximum as $r \rightarrow 0$ and decreases towards the boundary. Finally, we move on to describe the results obtained from our calculations, which are illustrated in Fig. \ref{pd}, that anisotropy is zero at centre and positive in the stellar interior, which implies that the tangential pressure ($p_t$) is always greater than the radial pressure ($p_r$).
Finally, using the anisotropic fluid will simplify the comparison with isotropic solutions
and most often used for studying massive compact objects \cite{Gokhroo}. 

 In addition to this central density, central and surface pressure of compact stars 
are presented in table \textbf{II}. It is intriguing to note that maximum density at the
centre $ \sim 10^{15} gm/cm^3$, which is constraint with the argument by Ruderman \cite{Ruderman} for
anisotropic stellar configurations that can describe realistic neutron stars. For example, 
the millisecond pulsar SAX J1808.4-3658 (SS2) \cite{Elebert} with 1.3237 $M_{\odot}$ has the central
density $4.06 \times 10^{15}  gm/cm^3$ ( the other results are given in Table \textbf{II}). 
Moreover,  inside the star, $p_r$ and  $p_t$  $> 0$, and the pressure decreases monotonically as we
move away from the center as evident in Figs. (\ref{f1}). Furthermore, it has been shown
that upper bound on the total compactness of a static spherically symmetric fluid in the form of
$2M/R \leq 8/9$ \cite{Buchdahl}. As one can see, we have explicitly derived Buchdahl's 
inequality for anisotropic fluid star, which matches exactly with the limit derived for 
uniform density star (see Table \textbf{II}).\\ 
\begin{figure} 
\begin{center}
\includegraphics[width=6cm]{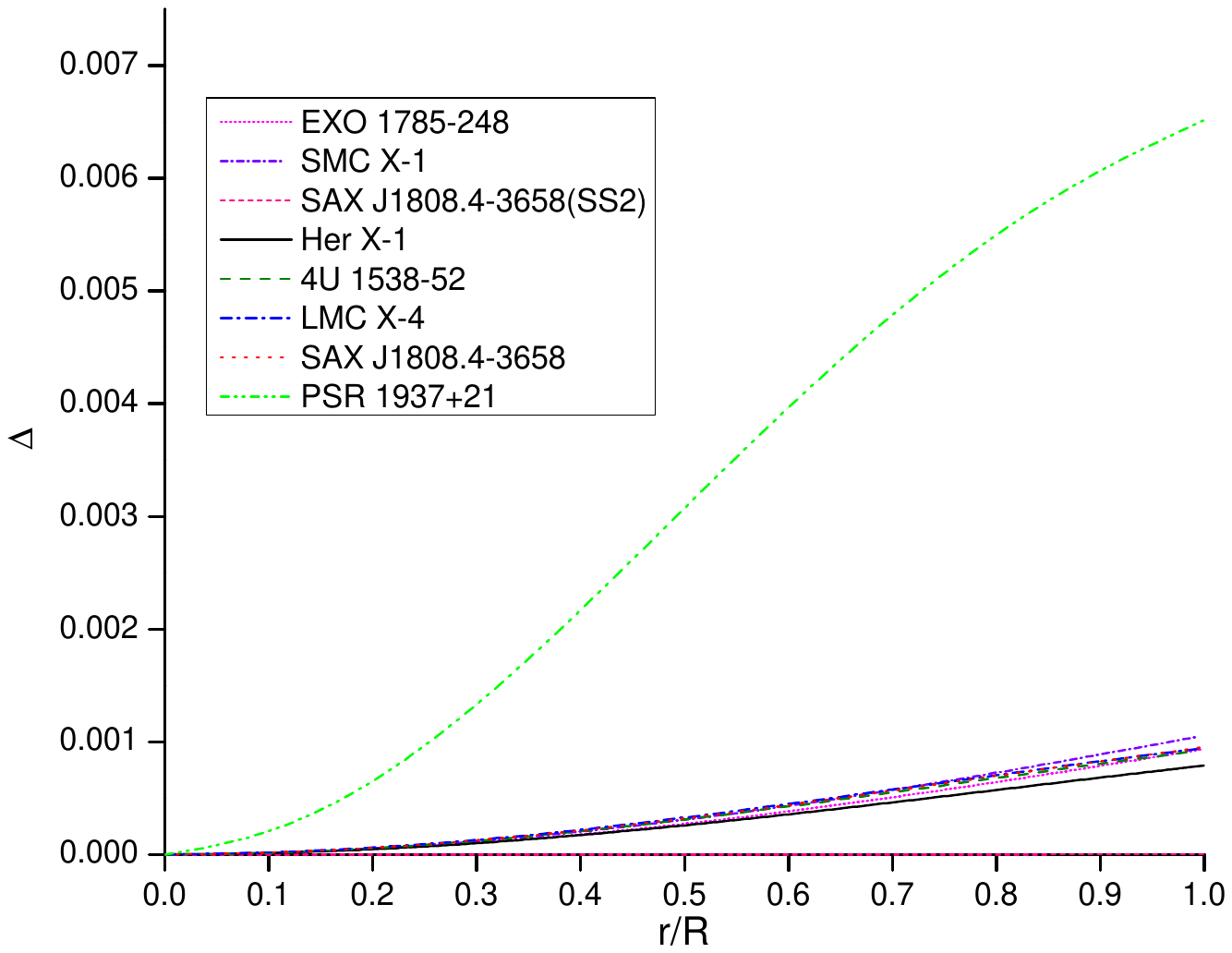}\includegraphics[width=6cm]{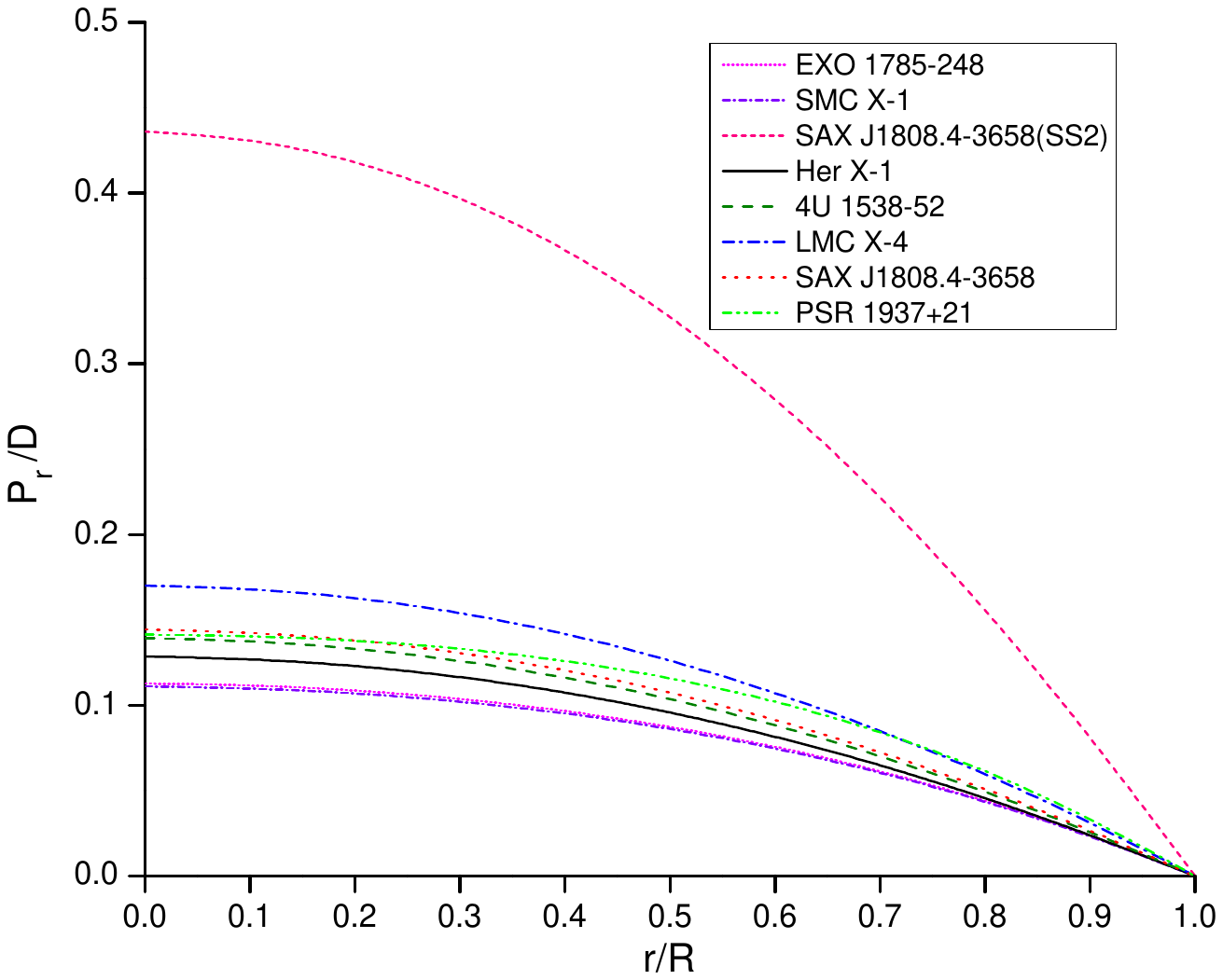}\includegraphics[width=6cm]{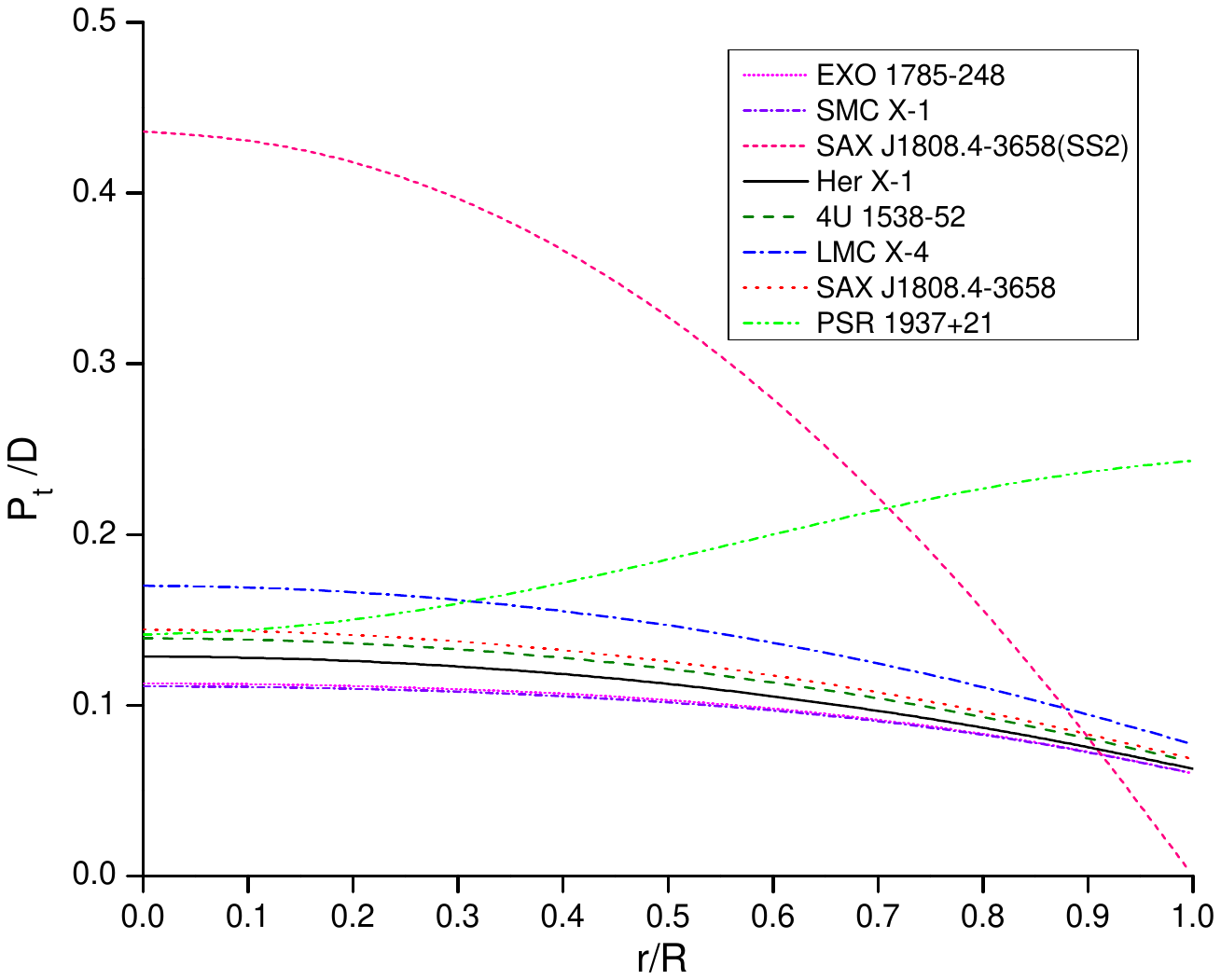}\\\includegraphics[width=6cm]{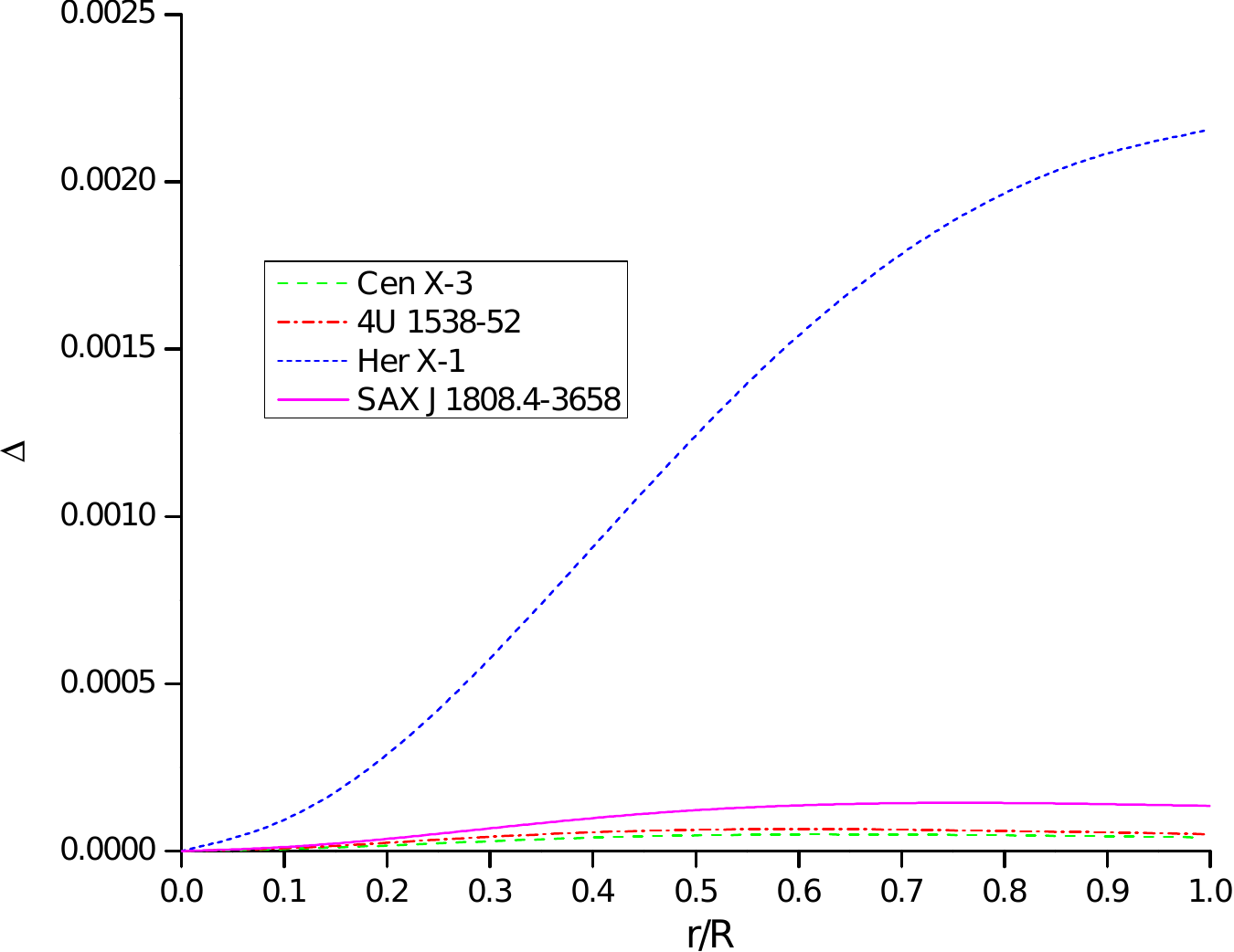}\includegraphics[width=6cm]{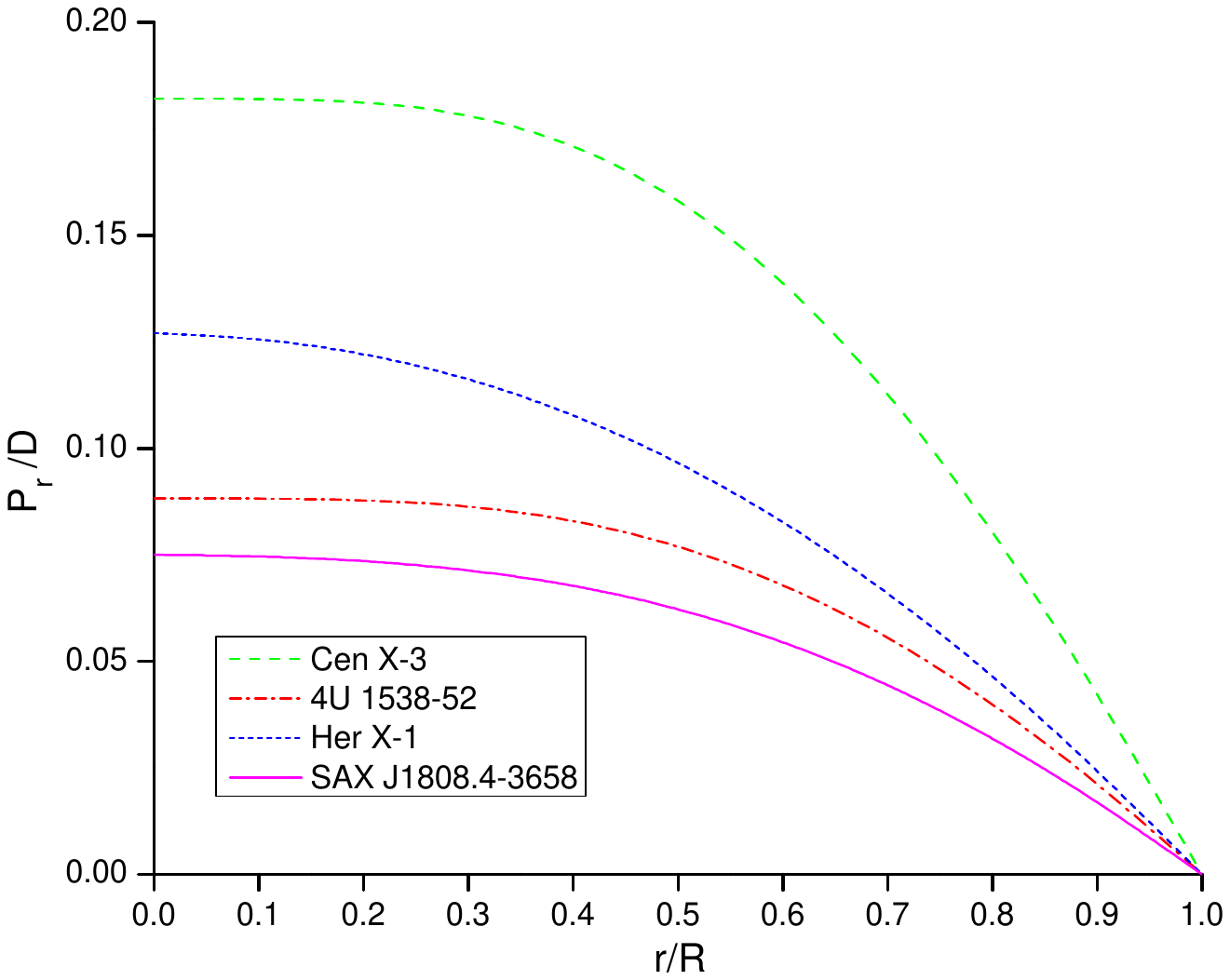}\includegraphics[width=6cm]{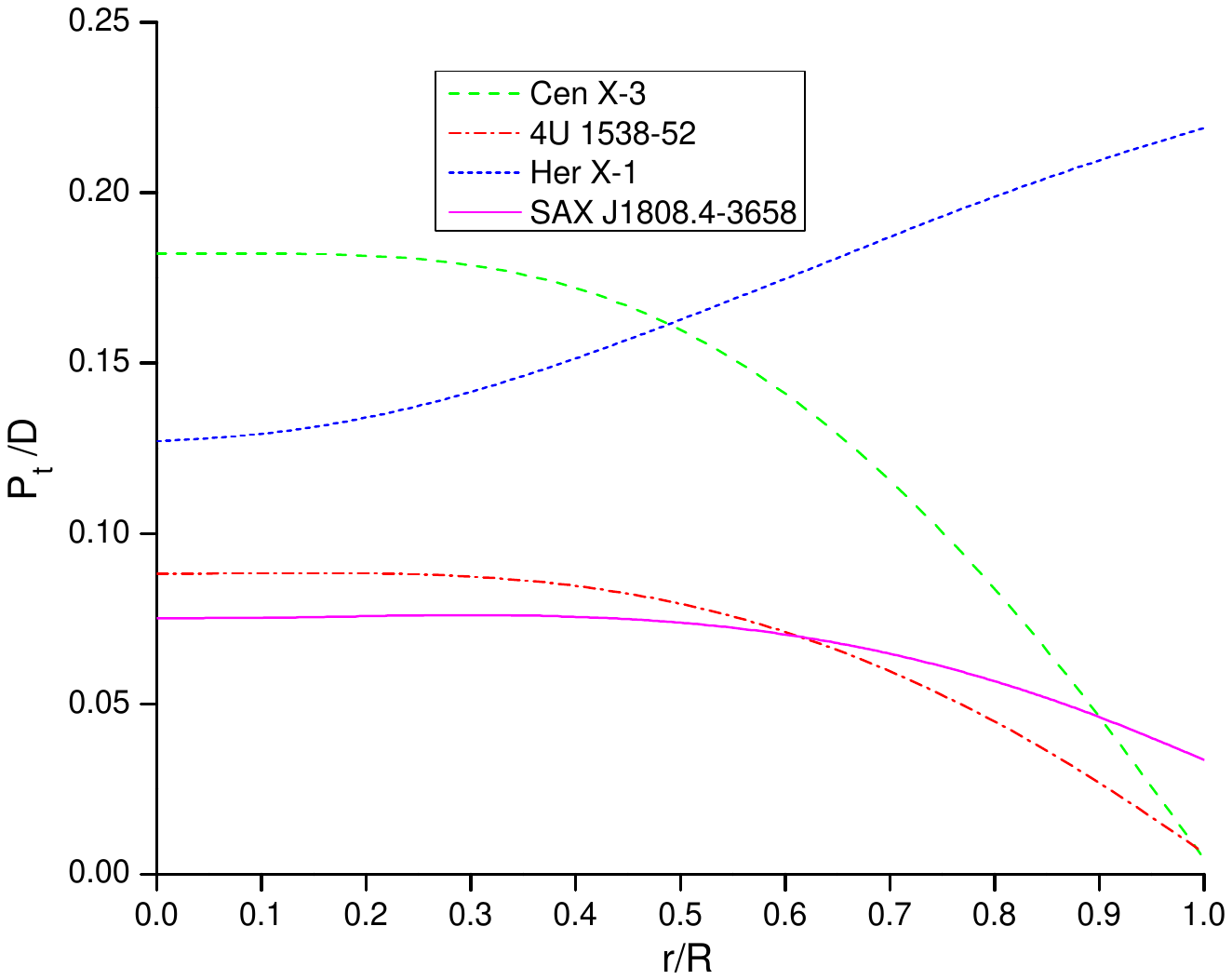}
\caption{Variation of anisotropy factor $\Delta$ (in $km^{-2}$) effective pressure-density ratio $ P_i/D $ vs. radial coordinate r/R for Case \textbf{I } (upper panel) \& \textbf{II} (lower panel). For plotting this graphs, we have employed the same data set as used in Fig. \ref{f1} }.\label{pd}
\end{center} 
\end{figure}
\section{Exterior solutions}
\label{ES}

\begin{figure}[h!]
\begin{center}
\includegraphics[width=8.6cm]{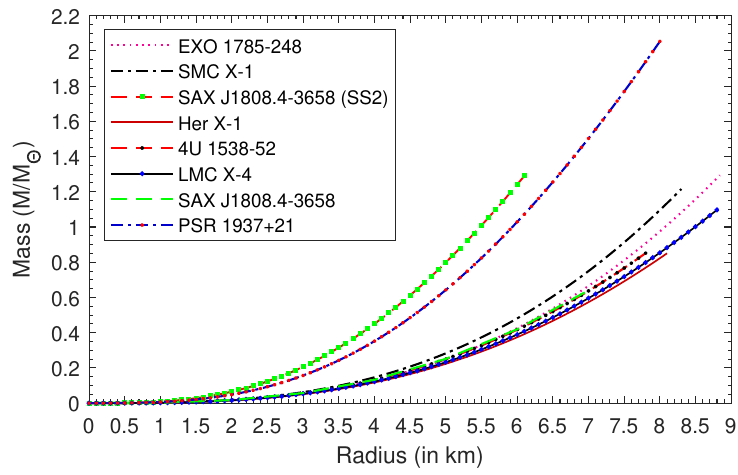} \includegraphics[width=8cm]{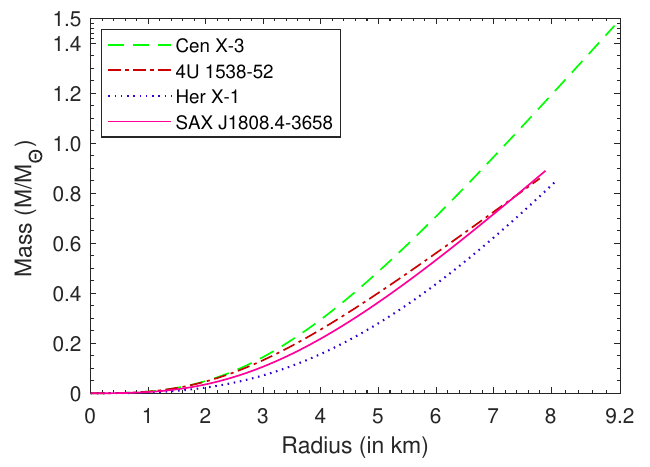}
\caption{Variation of the the total mass normalized in units of Solar mass ($M/M_{\odot}$) with the total radius R for the \textrm{case I} (left panel) and \textrm{case II} (right panel), respectively.}\label{f7}
\end{center} 
\end{figure}

To proceed further, the interior spacetime metric (\ref{eq1}) should be matched with
the Schwarzschild exterior solution at the boundary of the star ($r = R$).
In principle the radius $R$ is a natural parameter, where the radial pressure vanishes i.e. $p_r(R) = 0$. The exterior vacuum solution is then given by the Schwarzschild metric
 \begin{eqnarray}
 ds^{2}= -\bigg( 1-\dfrac{2M}{r}\bigg)dt^{2}+\bigg(1-\dfrac{2M}{r}\bigg)^{-1}dr^{2}+r^{2}(d\theta^{2}+\sin^{2}\theta d\phi^{2}),\label{46}
\end{eqnarray}
where $M$ is the total mass of the gravitational system and it's given by
\begin{eqnarray}
M_{tot}(r) =\int_{0}^{r} 4\pi r^2 \rho~dr. \label{47}
\end{eqnarray}
At this stage the interior solution must be matched to the vacuum exterior Schwarzschild metric. We match two spacetimes across the boundary surface using the Darmois-Israel formalism \cite{Israel}, which are tantamount by the following two conditions across the boundary surface $r = R$
  \begin{eqnarray} 
  e^{-\lambda}=1-\dfrac{2M}{R}, ~~~\textrm{and}~~~ 
  e^{\nu}= y^{2}=1-\dfrac{2M}{R},\label{50}
  \\p_r(r=R)= 0.\label{51}
 \end{eqnarray}
Now, using the conditions (\ref{50}) and (\ref{51}),  we can fix the values of arbitrary constants. Thus, boundary condition provides a full set of expressions for arbitrary constant $ A_{1}$ to $H_{1}$ (when $K< 0$) and $ A_{2} $ to $ H_{2} $ (when $K>1$) as follows: 
\begin{center}
\begin{tabular}{ l | c  }
  $\text{Case Ia}$ & $\frac{A_1}{B_1}= \frac{n\,(1 - K)\,\cosh(n x_1)\, \csc(x_1) + (3 - K + 2\,n^2)\,\sec(x_1)\,\sinh(n\,x_1)}{(-3 + K- 2 n^2)\,\cosh(n x_1)\,\sec x_1 +  n\,(K-1)\,\csc x_1\, \sinh(n x_1)}$ \\  \hline
  $\text{Case Ib}$ & $\frac{C_1}{D_1}= \frac{n\,(K-1)\, \cos(n x_1)\,\csc x_1 + (-3 + K + 2\, n^2)\,\sec x_1\,\sin(n x_1)}{(3 -K -2 \,n^2)\,\cos(n x_1)\,\sec x_1 + (K-1)\,n\,\csc x_1\, sin(n x_1)}$ \\ \hline
  $\text{Case Ic}$ & $\frac{E_1}{F_1}= \frac{(1-K)\,\cos(2 x_1) - 8\,\sin^2 x_1}{4\,\sin 2 x_1 - (K-1)\,(2\,x_1 + \sin(2\,x_1))}$ \\ \hline
 $\text{Case Id}$ & $ \frac{G_1}{H_1}=\frac{2\,\sin x_1 + (1- K)\, \sin x_1}{(K-1)\,\cos x_1 + x_1\,[-2\,\sin x_1 + (K-1)\,\sin x_1]}$ \\ \hline
 $\text{Case IIa}$ & $\frac{A_2}{B_2}=\frac{n\,(K-1)\,\cos(n x_2)\,\textrm{sech($x_2 $)}   + (3 - K + 2\,n^2)\,\textrm{csch($x_2$) } \, \sin (n x_2)}{(-3 + K - 2\,n^2)\,\cos(n x_2)\, \textrm{csch ($x_2$)} + n\,(K-1)\,\textrm{sech($x_2$)}\,\sin(n x_2)}$ \\ \hline
 $\text{Case IIb}$ & $\frac{C_2}{D_2}=\frac{n \, (K-1)\, \cosh(n x_2)\, \textrm{sech($x_2$)} + (-3 + K + 2\,n^2)\, \textrm{csch($x_2$)}\, \sinh(n x_2)}{(3-K - 2\,n^2)\,\cosh(n x_2)\, \textrm{csch($x_2$)} + n\,(K-1)\,\textrm{sech($x_2$)}\, \sinh(n x_2)}$ \\ \hline
 $\text{Case IIc}$ & $\frac{E_2}{F_2}=\frac{-8\,\sinh x_2\,\cosh x_2 + (K-1) (-2\,x_2 + \sinh(2 x_2))}{8\,\cosh^2 x_2 - (K-1)\,\cosh (2 x_2)}$ \\ \hline
 $\text{Case IId}$ & $\frac{G_2}{H_2}=\frac{2\,\cosh x_2 + (1- K) \, \cosh x_2}{-2\,x_2\,\cosh x_2 + (K-1)\,x_2\,\cosh x_2 + \sinh x_2 - K\,\sinh x_2}$
\end{tabular}.
\end{center}
where~~~~$x_1=\sin^{-1}\sqrt{\frac{K+CR^2}{K-1}}$~~~ and ~~~$x_2=\cosh^{-1}\sqrt{\frac{K+CR^2}{K-1}}$. Here, we want to investigate the gravitational mass and radius of neutron stars. 
With the following condition $e^{-\lambda}=1-\frac{2M}{R}$, it is useful to write the total mass in the following form 
\begin{eqnarray}
M=\frac{(K-1)\,CR^3}{2\,K\,(1+CR^2)} ,
\end{eqnarray}
We now present our results for the static neutron star models, showing the total mass M (in solar masses $M_{\odot}$) versus the physical radius R (in km) in Fig. \ref{f7}. In these two figures all values are considered in the same succession as mentioned in Fig. \ref{f1}.
%%%%%%%%%%%%%%%%%%%%%%%%%%%%%%%%%%

%%%%%%%%%%%%%%%%%%%%%%%%%%%%%%%%%%%%%%% 
We shall now use the general relativistic effect of gravitational redshift by the relation
$z_S$ = $\Delta \lambda/\lambda_{e}$ = $\frac{\lambda_{0}-\lambda_{e}}{\lambda_{e}}$, where $\lambda_{e}$ is the emitted wavelength at the surface of a nonrotating star and $\lambda_{0}$ is the observed wavelength received at radial coordinate $r$. In the weak-field limit,  gravitational redshift from the surface of the star as measured by a distant observer $(g_{tt}\rightarrow -1)$, is given by
\begin{eqnarray}
\label{eq24}
1+z_S = \arrowvert g_{tt}(R) \arrowvert ^{-1/2} = \left(1-\frac{2M}{R}\right)^{-1/2},
\end{eqnarray}
\begin{figure}[h]
\begin{center}
\includegraphics[width=6cm]{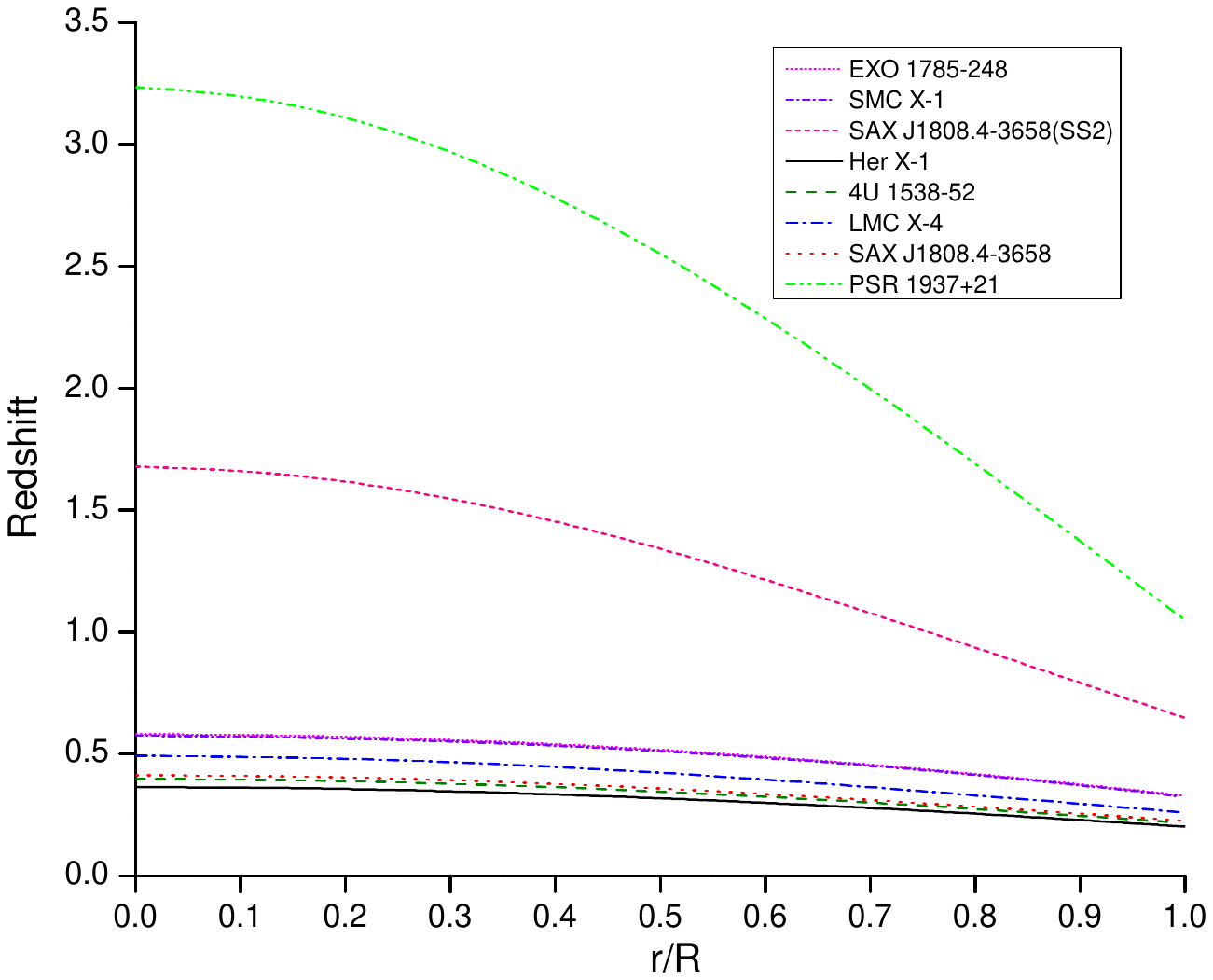} \includegraphics[width=6cm]{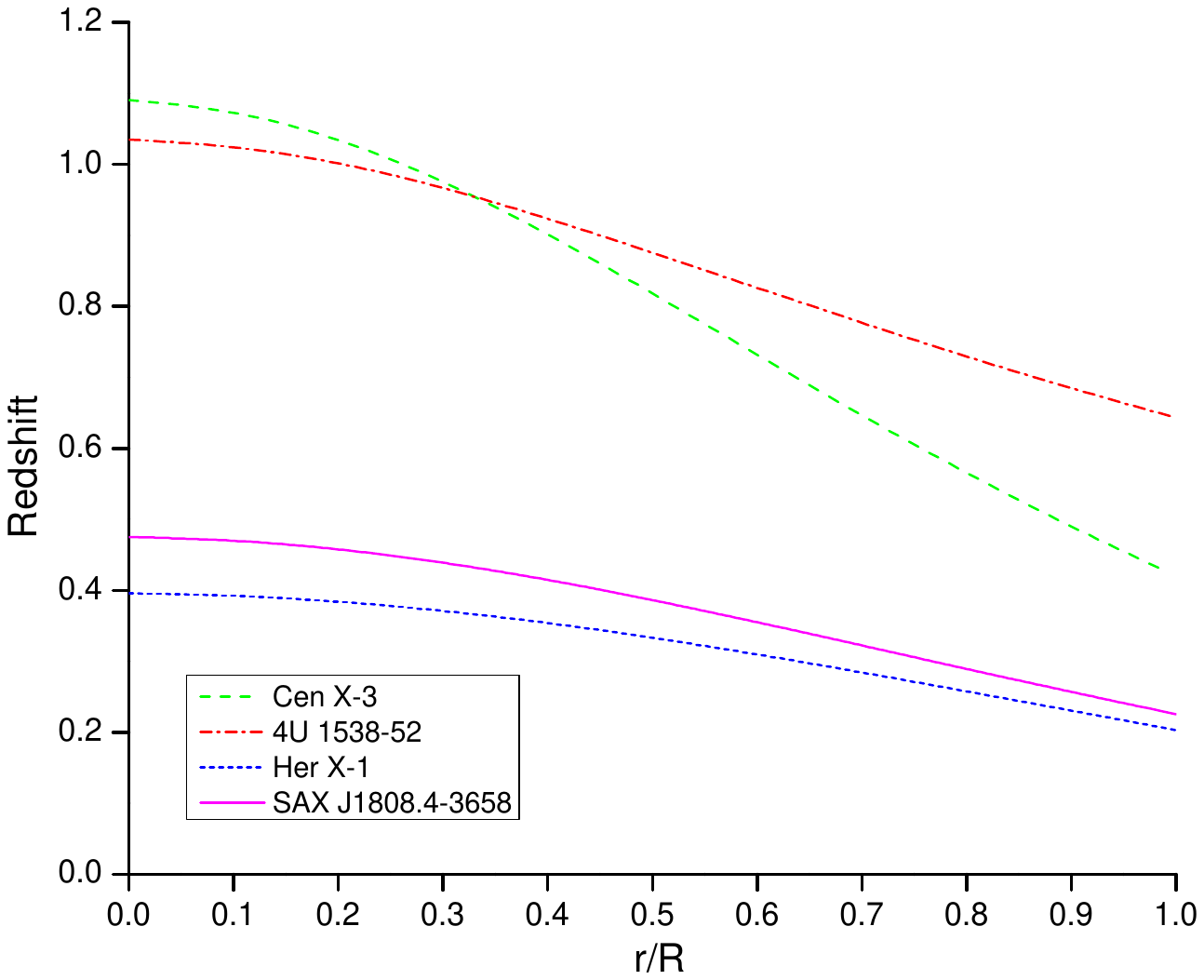}
\caption{Behaviour of redshift (left figure for Case \textbf{II}. and right figure for Case \textbf{II}.) vs. radial coordinate r/R which have been plotted for different compact star candidates.For the purpose of plotting this graph, we have employed the data set of values as same as FIG.\ref{f1}}\label{f5}
\end{center}
\end{figure}
where $g_{tt}(R)$ = $e^{\nu(R)}$ =$\left(1-\frac{2M}{R}\right)$ is the metric function. It was shown earlier by Buchdahl \cite{Buchdahl}, that for spherically symmetric distribution of a prefect fluid
the gravitational redshift is $z_{s} <$ 2. However, different arguments have been put forward 
for the existence of anisotropy star models which turns out to be 3.84, as suggested by \cite{Karmakar,Barraco}.
On the other hand, in studying general restrictions for the redshift for anisotropic stars,
Bohmer and Harko \cite{Boehmer} showed that this value could be increased up to $z_{s} \leq$ 5, which is consistent with the bound  $z_{s} \leq$ 5.211 obtained by Ivanov \cite{Ivanov}. We perform the whole calculations for redshift of the enlisted compact objects by taking the same values, 
which we have used for graphical presentation Fig.\,\ref{f5}.
We are mostly interested bounds on surface redshift for spherically symmetric stellar structures 
and our results are quite satisfactory.

\section{Physical features of anisotropic models}  
\label{PF}
We now study physical properties of the stellar configuration made up of anisotropic
fluids by performing some analytical calculations. We analyzed the stability problem by considering 
modified Tolman-Oppenheimer-Volkoff (TOV) equation and checking the causality conditions
within the fluid. With these one can determinate the value of the speed
of sound across a given star. Finally,  we investigate the type of compact objects 
that might arise from these solutions and to restrict the model arbitrariness.
\subsection{Causality condition}
In addition to the positivity of density and pressure profiles,
we shall pay special and particular attention to the condition of
bounding sound speeds (radial and tangential direction) within the matter distribution.  In essence of this we fix $c = 1$, and investigate the sound speed for anisotropic fluid distribution. It is obvious that the velocity of sound is less than the velocity of light i.e. $0< v_{r}^{2}=dp_{r}/d\rho <1$ $\text{and}$ $0< v_{r}^{2}=dp_{t}/d\rho <1$. The stability of fluid sphere with internal pressure anisotropy was also probed by Herrera \cite{Herrera(2016)} and his collaborators. Here, we consider the Case \textbf{I} \& \textbf{II} separately, and the expression for velocity of sound as follows:\\ 

$\textbf{Case Ia:}$
\begin{eqnarray}
\dfrac{dp_r}{d\rho}&=& \dfrac{N_1}{S_1}, \label{62}\\
\dfrac{dp_t}{d\rho}&=& \dfrac{N_1}{S_1}+\dfrac{\Delta_0\big[2\,\cos^2 x\,\sin x\,(K-1)+4\,\sin x\big((K-1)\,\sin^2 x\,-K\big)\big]}{(K-1)^2\,\cos^5 x}, \label{63}
\end{eqnarray}

$\textbf{Case Ib:}$
\begin{eqnarray}
\dfrac{dp_r}{d\rho}&=& \dfrac{N_2}{S_1}, \label{64}\\
\dfrac{dp_t}{d\rho}&=& \dfrac{N_2}{S_1}+\dfrac{\Delta_0\big[2\,\cos^2 x\,\sin x\,(K-1)+4\,\sin x\big((K-1)\,\sin^2 x\,-K\big)\big]}{(K-1)^2\,\cos^5 x}, \label{65}
\end{eqnarray}

$\textbf{Case Ic}$:
\begin{eqnarray}
\dfrac{dp_r}{d\rho}&=& \dfrac{N_3}{S_1}, \label{66}\\
\dfrac{dp_t}{d\rho} &=& \dfrac{N_3}{S_1}+\dfrac{\Delta_0\big[2\,\cos^2 x\,\sin x\,(K-1)+4\,\sin x\big((K-1)\,\sin^2 x\,-K\big)\big]}{(K-1)^2\,\cos^5 x} ,\label{67}
\end{eqnarray}

$\textbf{Case Id}$:
\begin{eqnarray}
\dfrac{dp_r}{d\rho}&=&\dfrac{N_4}{S_1}, \label{68}\\
\dfrac{dp_t}{d\rho}&=&\dfrac{N_4}{S_1}+\dfrac{\Delta_0\big[2\,\cos^2 x\,\sin x\,(K-1)+4\,\sin x\big((K-1)\,\sin^2 x\,-K\big)\big]}{(K-1)^2\,\cos^5 x} ,\label{69}
\end{eqnarray}

$\textbf{ Case IIa:}$
\begin{eqnarray}
\dfrac{dp_r}{d\rho}&=& \dfrac{N_5}{S_2}, \label{70}\\
\dfrac{dp_t}{d\rho}&=& \dfrac{N_5}{S_5}+\,\dfrac{\Delta_0\big[2\,\cosh x\,\sinh^2 x\,(K-1)-4\,\cosh x\,\big((K-1)\cosh^2 x -K\big)\big]}{(K-1)^2\,\sinh^5 x}, \label{71}
\end{eqnarray}

$\textbf{ Case IIb:}$
\begin{eqnarray}
\dfrac{dp_r}{d\rho} &=& \dfrac{N_6}{S_2}, \label{72}\\
\dfrac{dp_t}{d\rho} &=& \dfrac{N_6}{S_2}+\dfrac{\Delta_0\big[2\,\cosh x\,\sinh^2 x\,(K-1)-4\,\cosh x\,\big((K-1)\cosh^2 x -K\big)\big]}{(K-1)^2\,\sinh^5 x}, \label{73}
\end{eqnarray}

$\textbf{ Case IIc:}$
\begin{eqnarray}
\dfrac{dp_r}{d\rho}&=&\dfrac{N_7}{S_2}, \label{74}\\
\dfrac{dp_t}{d\rho}&=&\dfrac{N_7}{S_2}+\dfrac{\Delta_0\big[2\,\cosh x\,\sinh^2 x\,(K-1)-4\,\cosh x\,\big((K-1)\cosh^2 x -K\big)\big]}{(K-1)^2\,\sinh^5 x} ,\label{75}
\end{eqnarray}

$\textbf{ Case IId:}$
\begin{eqnarray}
\dfrac{dp_r}{d\rho}&=&\dfrac{N_8}{S_2}, \label{76}\\
\dfrac{dp_t}{d\rho}&=& \dfrac{N_8}{S_2}+\dfrac{\Delta_0\big[2\,\cosh x\,\sinh^2 x\,(K-1)-4\,\cosh x\,\big((K-1)\cosh^2 x -K\big)\big]}{(K-1)^2\,\sinh^5 x}, \label{77}
\end{eqnarray}  
where the expressions of used coefficients $N_1,~N_2,~N_3,~N_4,~N_5,~N_6,~N_7,~N_8,~S_1$ and $S_2$ in Eqs.~(\ref{62})-(\ref{77}) are given in the Appendix (B).
\begin{figure}[htp!]
\begin{center}
\includegraphics[width=7cm]{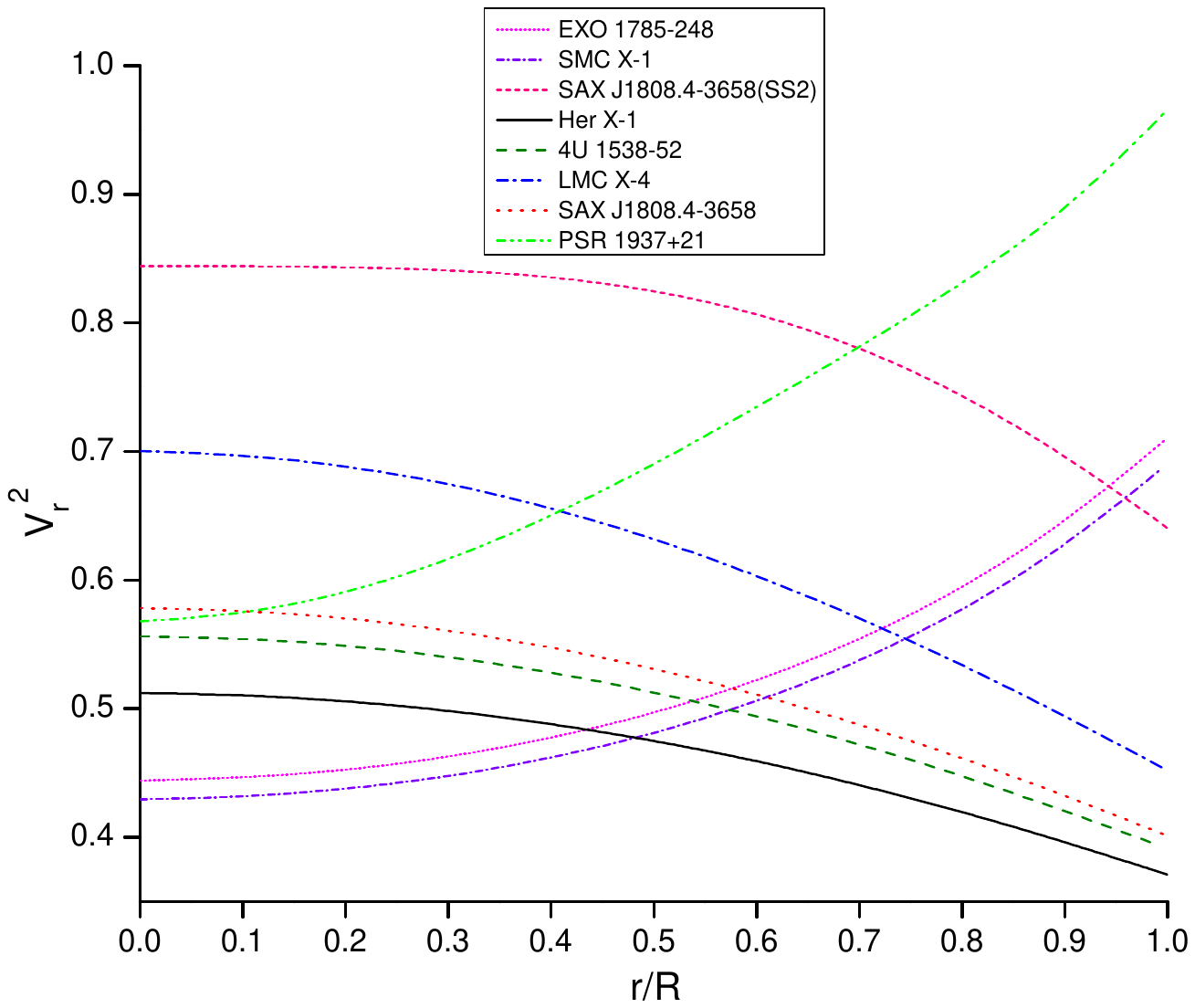}\includegraphics[width=7cm]{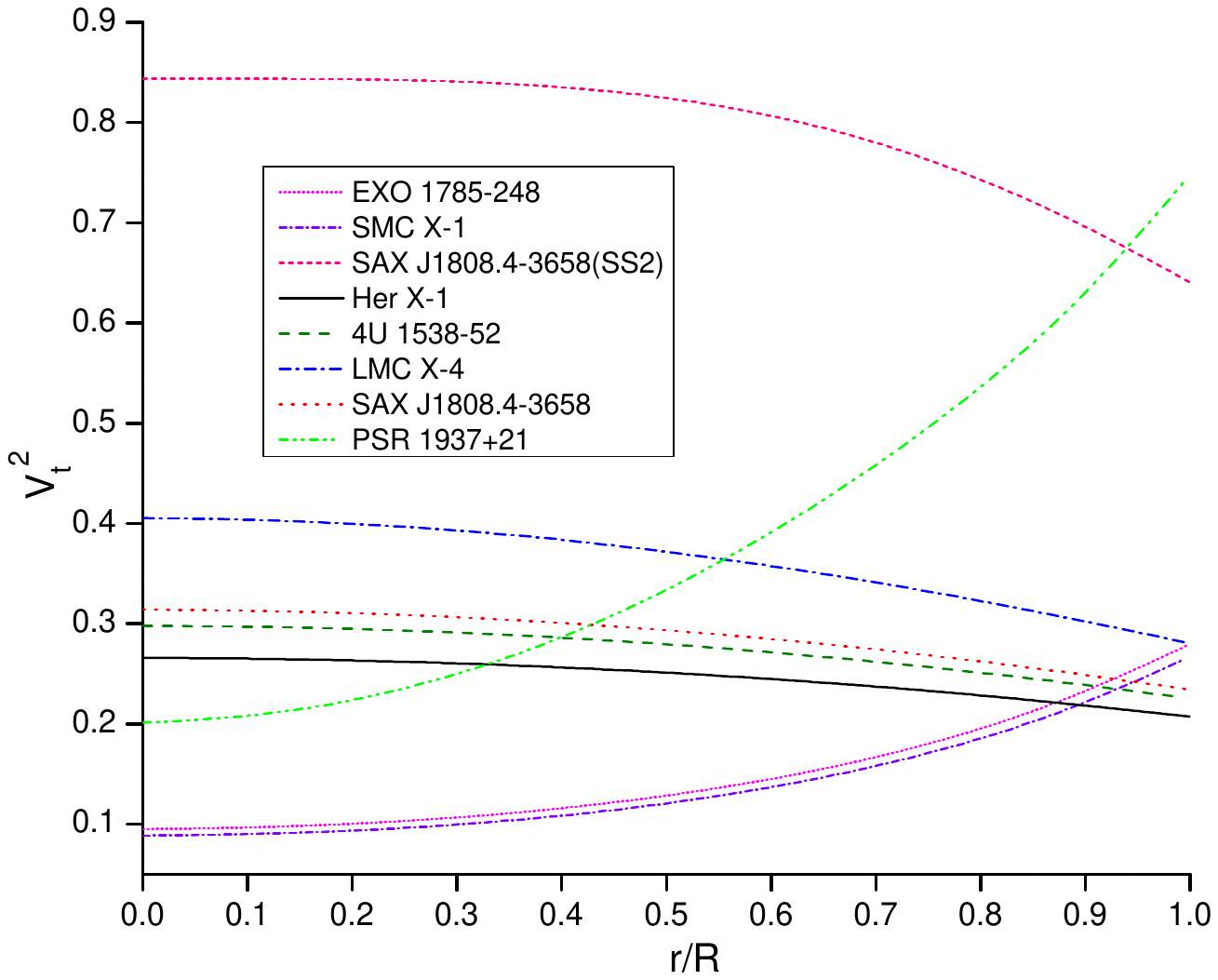}\\\includegraphics[width=7cm]{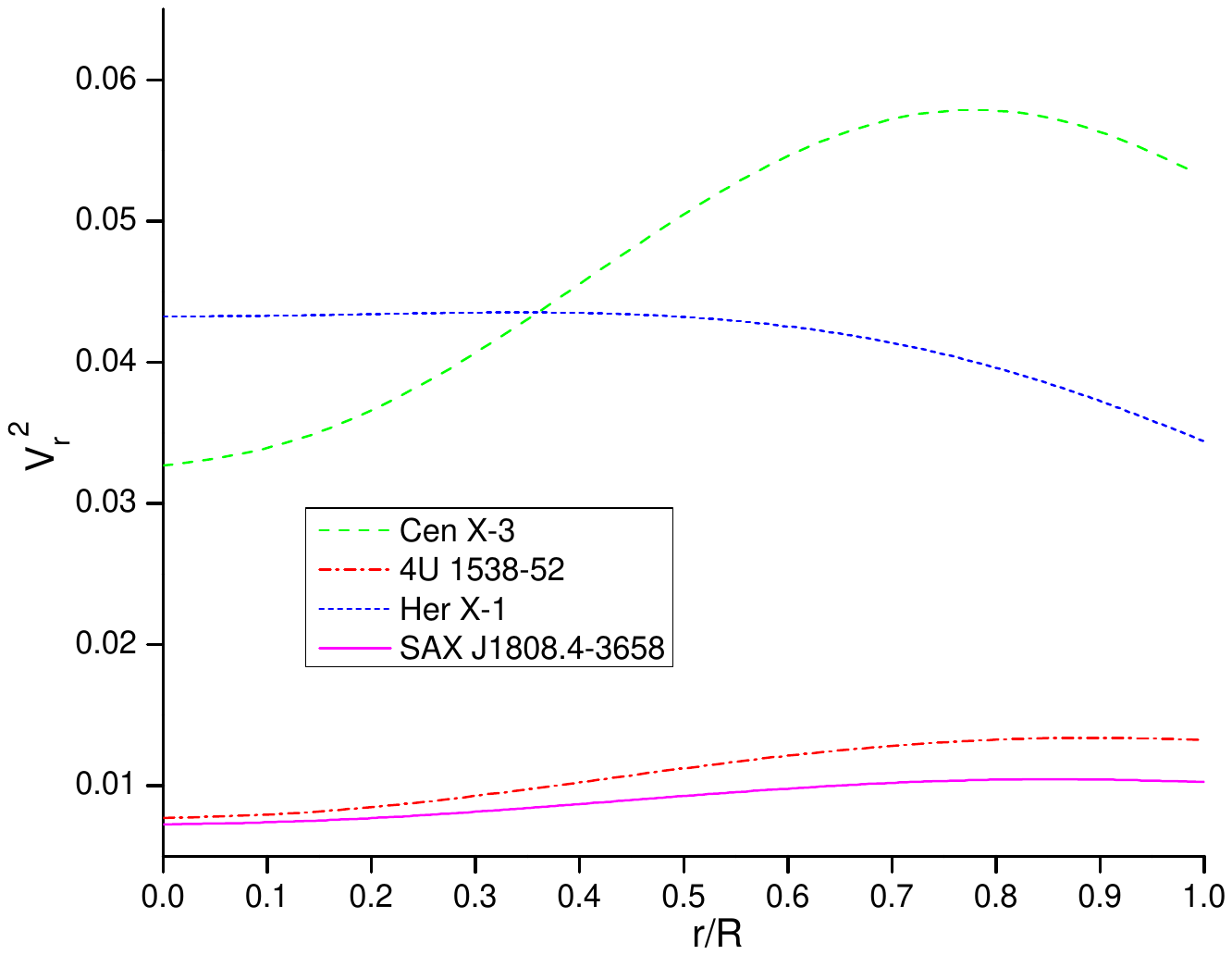}\includegraphics[width=7cm]{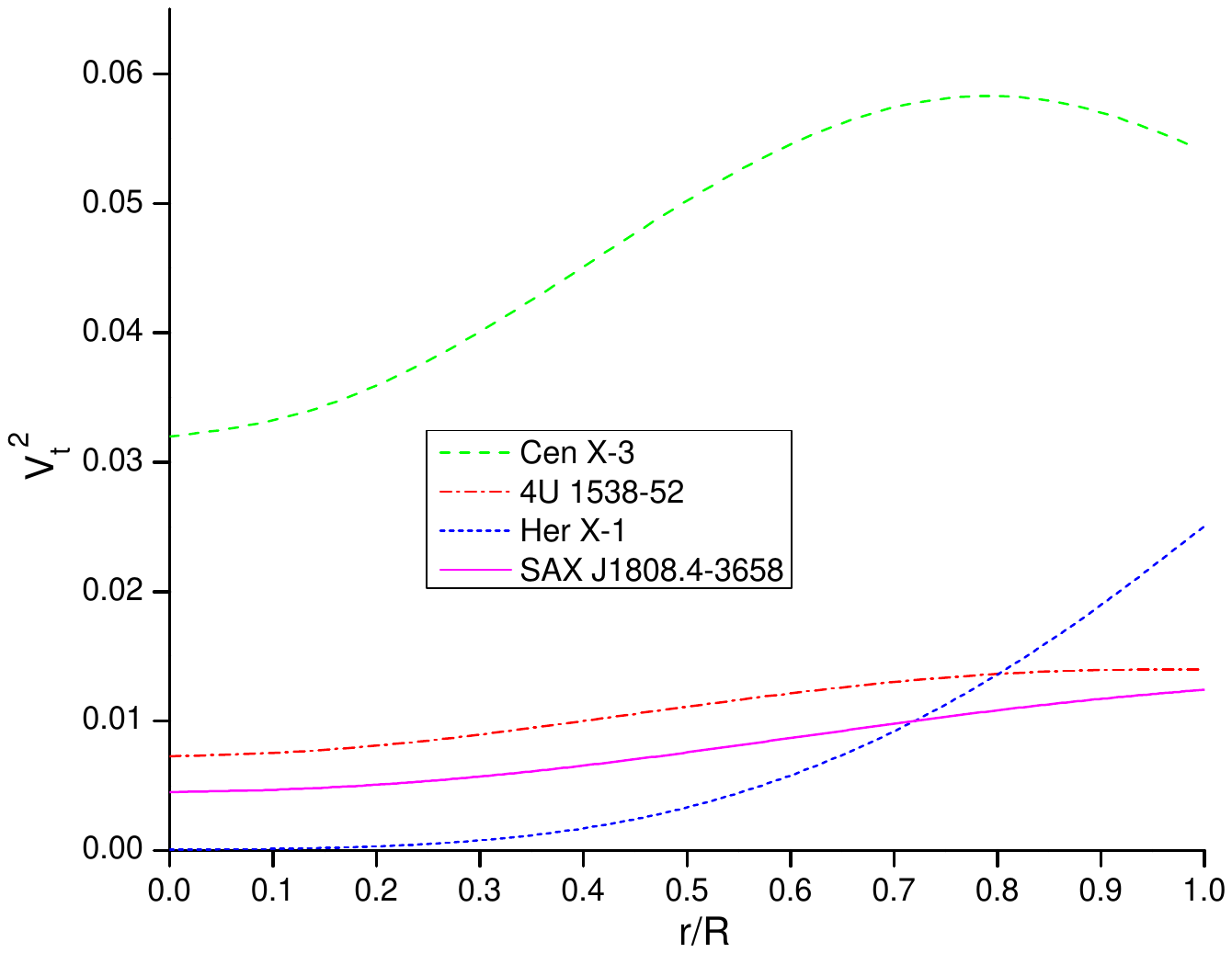}
\caption{Variation of radial and transverse speed of sound have been plotted for respective stellar model for Case \textbf{I} (top figures) \& \textbf{II} (Bottom figures). We use
same data as of Fig. \ref{f1} }. \label{f2}
\end{center}
\end{figure}    

In this analytical approach, we use the graphical representation to represent the velocity of sound due to  complexity of the expression. Considering all expressions for both cases \textbf{1} \& \textbf{2}, we have plotted 
Fig. \ref{f2}. In Fig. \ref{f2} we plot for radial and transverse velocity of sound when $K<0$(top figures) and $K>1$(bottom figures) for compact star candidates LMC X-4, SMC X-1, EXO 1785-248, SAX J1808.4-3658 (SS2), Her X-1, 4U 1538-52, PSR 1937+21, Cen X-3 and SAX J1808.4-3658. Our investigation shows that
our equation of state for anisotropic matter satisfies the causality condition. Form Fig. \ref{f2}, it is interesting to note that the velocity of sound is decreasing for the stars: SAX J1808.4-3658 (SS2) (case Ib), LMC X-4 (case Ic), 4U 1538-52 (case Ic), Her X-1 (case Ic), SAX J1808.4-3658 (case Ic), Her X-1 (case IIc) and increasing for SMC X-1 (case Ia) EXO 1785-248 X-1 (case Ia), PSR 1937+21 (case Id), Cen X-3 (case IIa), 4U 1538-52 (case IIb), SAX J1808.4-3658 (case IId) towards the boundary which  implies that our solution is well behaved for above cases. The decreasing features of the velocities are appearing in the present compact star model due to presence of anisotropy only because velocity of sound is not decreasing for Buchdahal metric in charged as well as  uncharged perfect fluid solution \cite{Gupta2005,Gupta2004}. Now, we focus on investigation of adiabatic index, energy conditions and hydrostatic equilibrium for compact stars in accordance to their mass and radius ratio.
%%%%%%%%%%%%%%%%%%%%%%%%%%%%%%%%%%
\subsection{Adiabatic index}
For a specific energy density, the rigidity of the EOS can be described by the adiabatic index. On the other hand, adiabatic index also characterize the stability of relativistic as well as non relativistic compact star models. Following the work of Chandrasekhar\cite{Chandrasekhar1964}, Many authors~\citep{Hillebrandt1976,Horvat2010,Doneva2012,Silva2015} have discussed the dynamical stability of the stellar system against an infinitesimal adiabatic perturbation corresponding to radial pressure. For any dynamically stable stellar system, Heintzmann and Hillebrandt \cite{Heintzmann}  have suggested that the radial adiabatic index must be more than $\frac{4}{3}$ at all interior points of the compact star. The radial adiabatic index $\gamma_r$ in our system is given as \begin{equation}
\gamma_{r}=\frac{p_r+\rho}{p_r}\,\frac{dp_r}{d \rho}=\frac{p_r+\rho}{p_r}\,v _{r}^{2}, \label{gamma_r} 
\end{equation}
\begin{figure}[h!]
\begin{center}
\includegraphics[width=7cm]{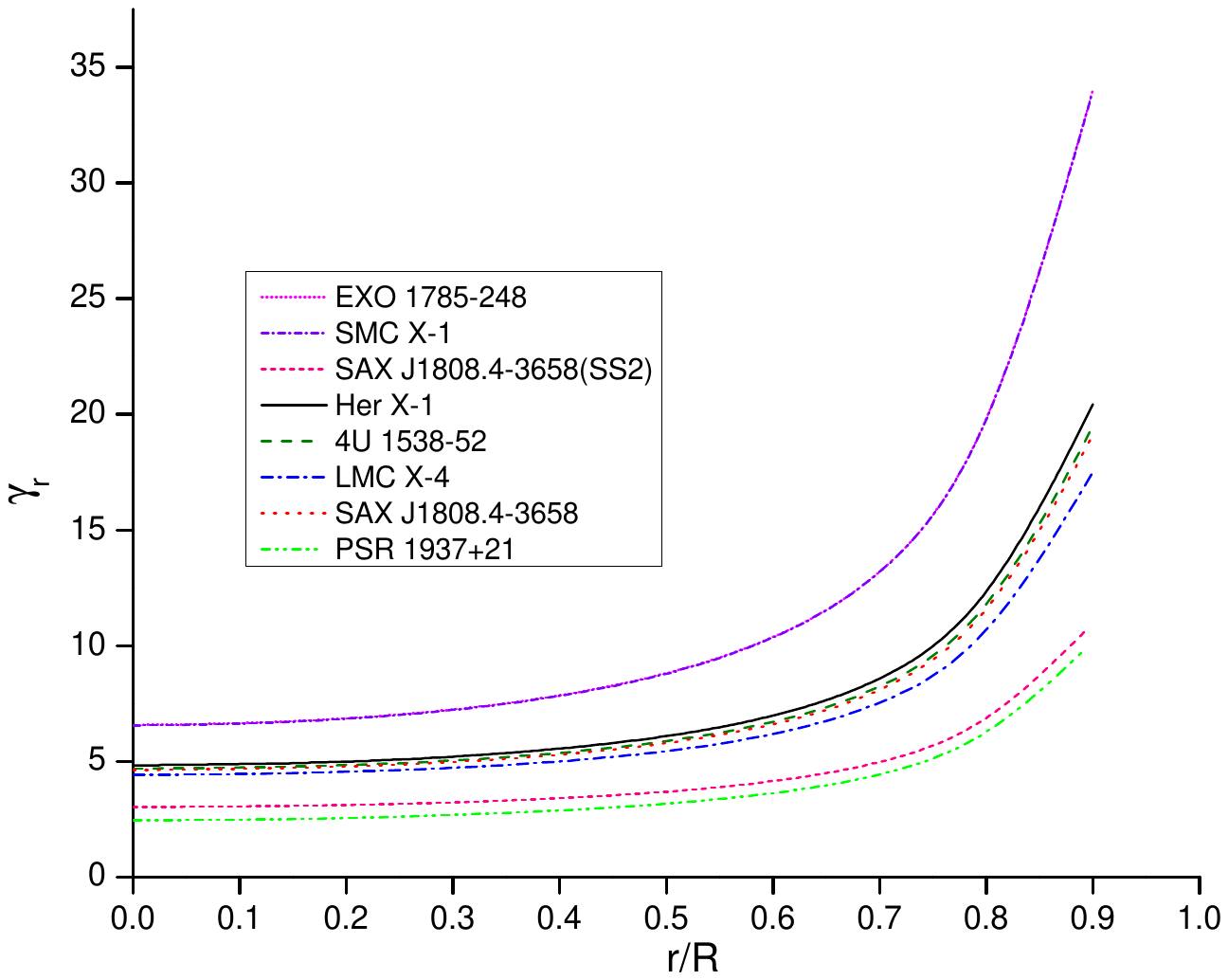}\includegraphics[width=7cm]{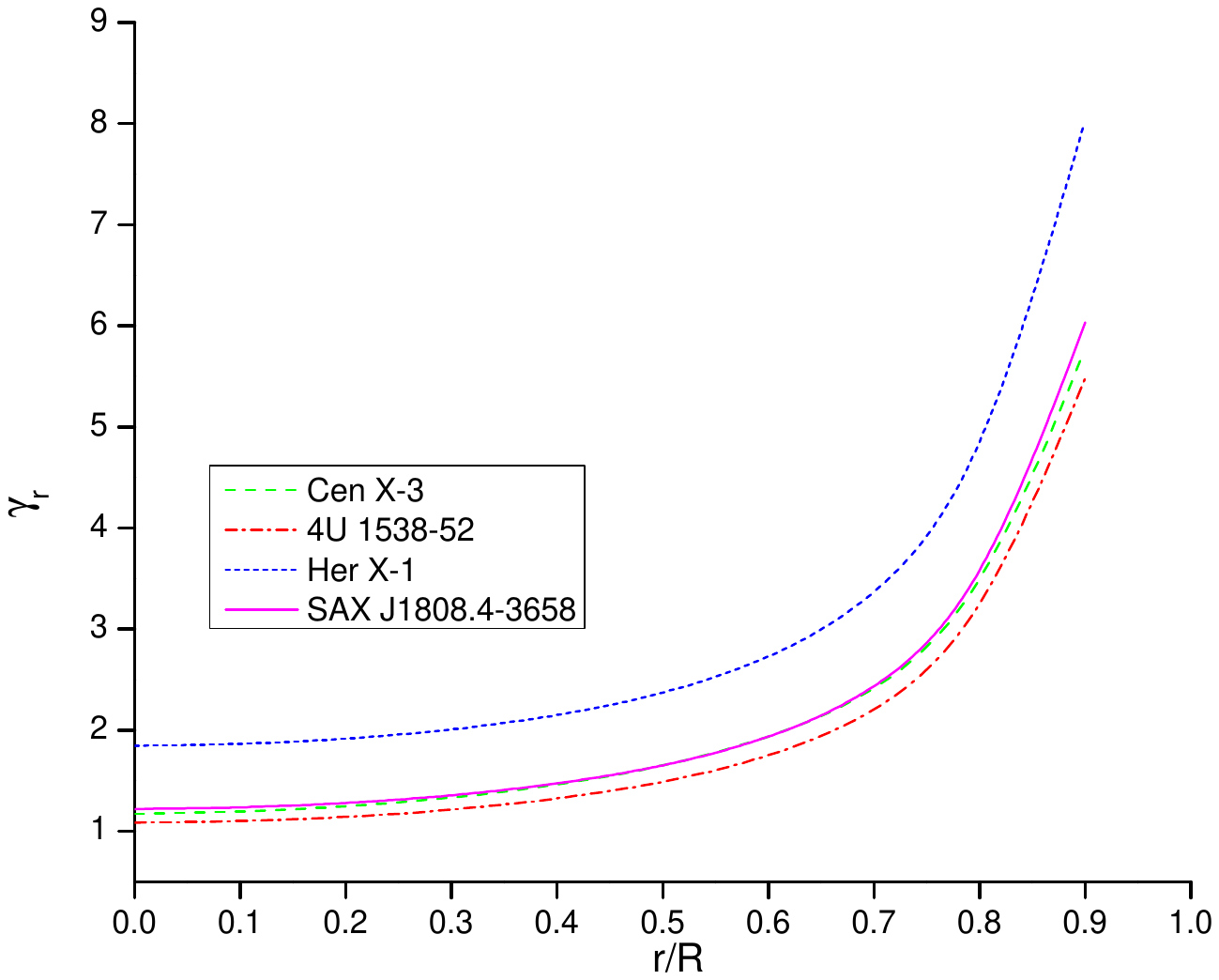}
\caption{Adiabatic index ($\gamma_{r}$) (left figure for Case I and right for Case II) vs. radial coordinate r/R which have been plotted for different compact star candidates.For the purpose of plotting this graph, we have employed the data set of values as same as FIG.\ref{f1}} \label{adia}
\end{center}
\end{figure}
The graphical representation of radial adiabatic index is given by the Fig.(\ref{adia}). For this figure it the clear that value of adiabatic index corresponding to radial pressure is more than $4/3$ at all interior point for each different compact star model.  
%%%%%%%%%%%%%%%%%%%%%%%%%%%%%%%%%%%%%%%
\subsection{ Energy Conditions}
Here we analyze the energy conditions according to relativistic classical
field theories of gravitation. In the context of GR the energy conditions are local inequalities
that process a relation between matter density and pressure obeying certain restrictions.
Many plausible physical constraints have been proposed, such as positive mass theorem \cite{Schoen},
censorship theorem \cite{Olum,Bassett}, singularity theorems \cite{Hawking}, and various constraints
on black hole surface gravity \cite{Visser}, but perhaps the most important and far-reaching applications
are the energy conditions. There are several different ways to formulate the energy conditions, but we will focus here only on (i) the Null energy condition (NEC), (ii) Weak energy condition (WEC) and (iii) Strong energy condition (SEC). In summary -
 \begin{subequations}
\label{781}
\begin{align}
\textbf{NEC}: \rho (r)+  p_r \geq 0, \\
\textbf{WECr} : \rho + p_r \geq 0, ~~ \text{and}~~~\rho(r)\geq 0, \\
\textbf{WECt}: \rho+p_t \geq 0, ~~ \text{and} ~~~\rho(r)\geq 0 , \\
\textbf{SEC}: \rho+ p_r+ 2 p_t \geq 0.
\end{align}
\end{subequations}
%%%%%%%%%%%%%%%%%%%%%%%%%%%%%%%%%%
\begin{figure}[h!]  
\begin{center}
\includegraphics[width=6cm]{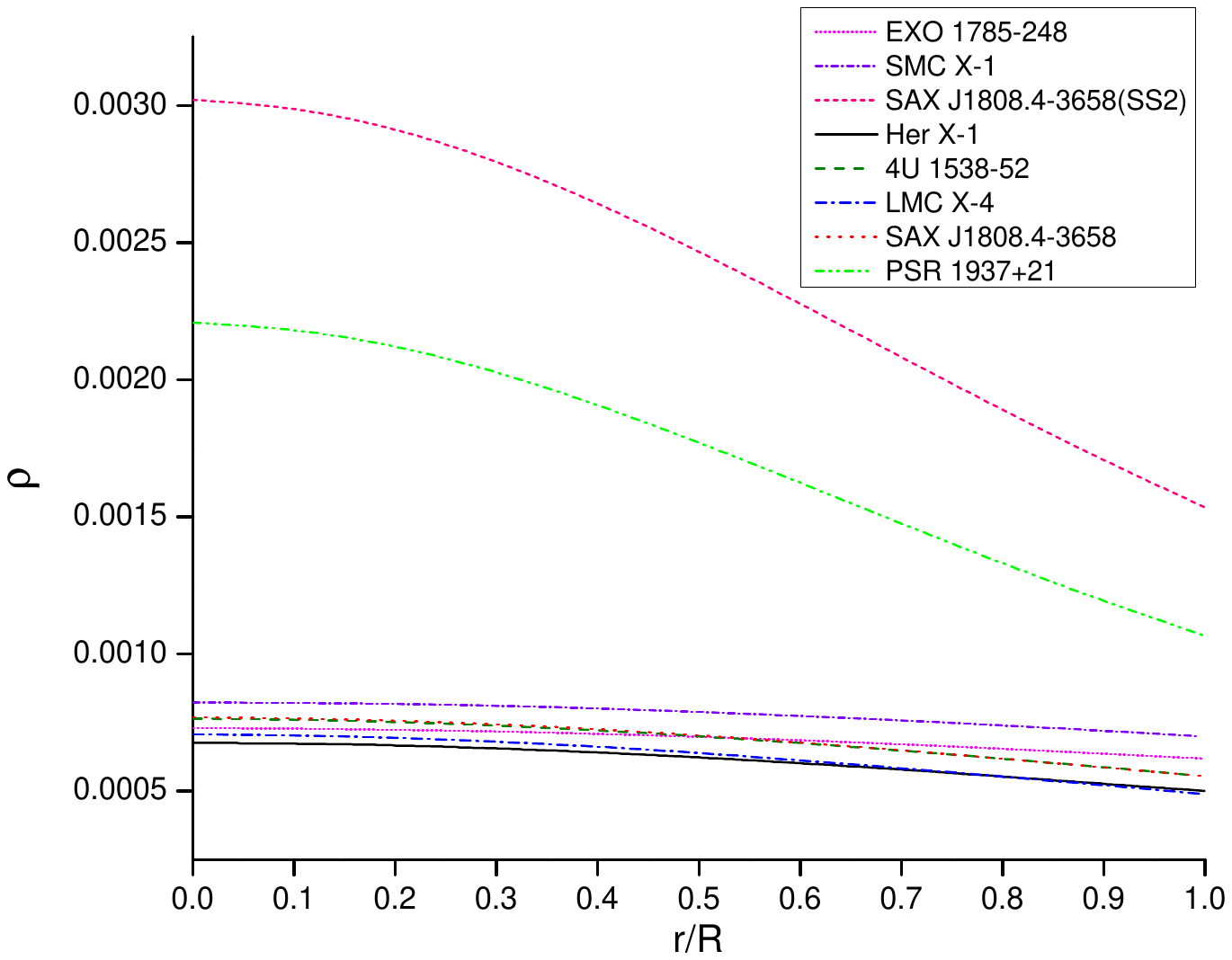}\includegraphics[width=6cm]{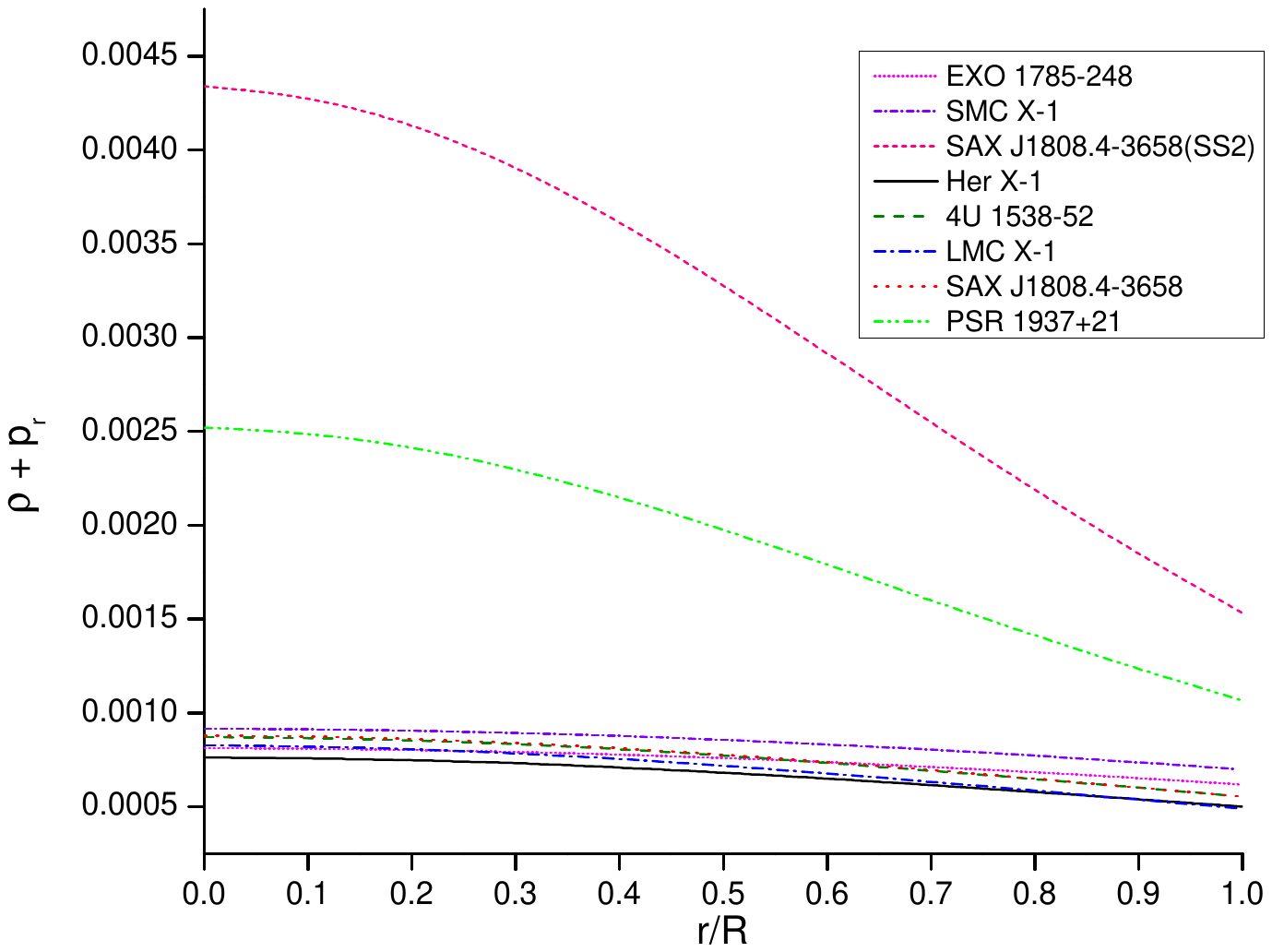}\includegraphics[width=6cm]{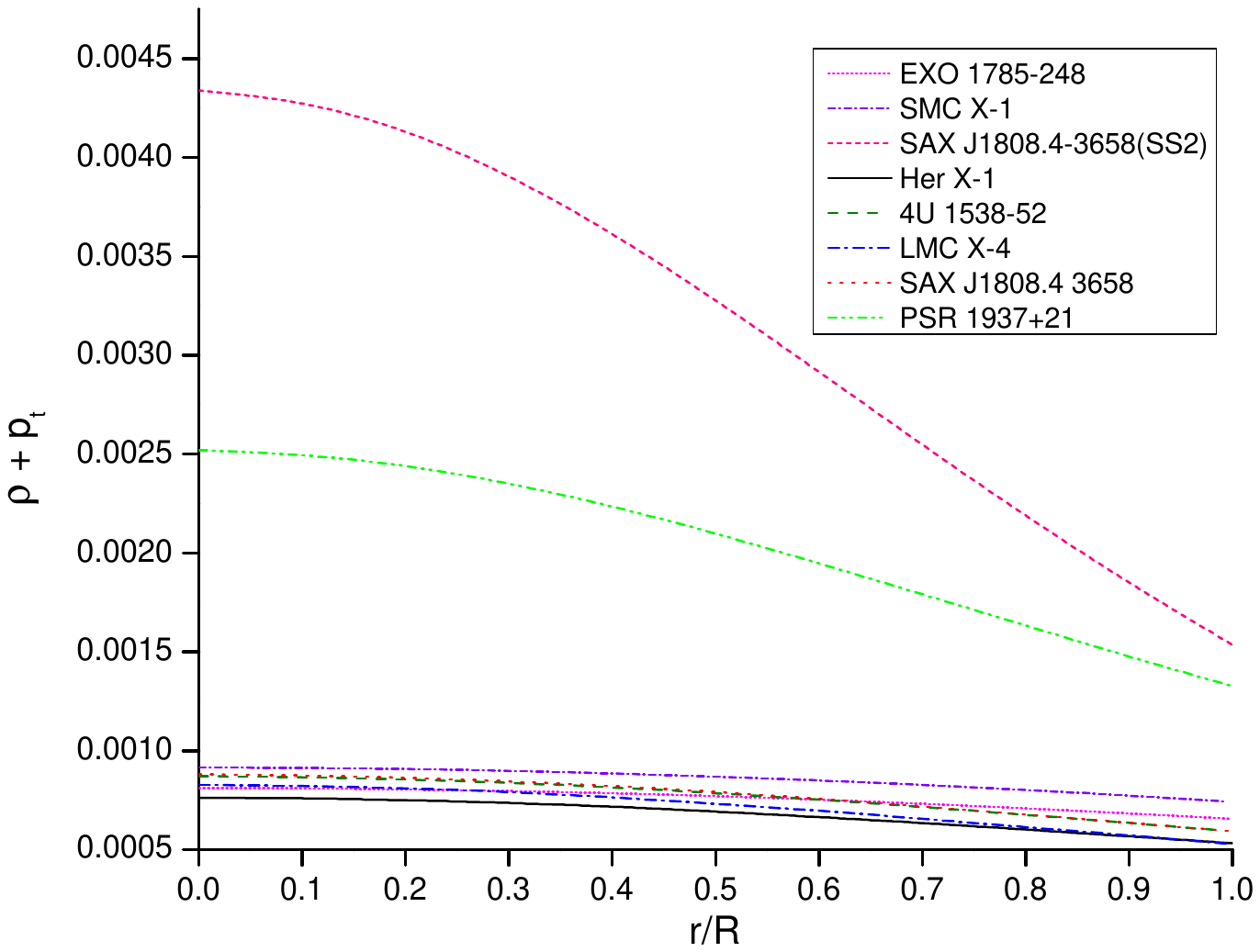}\\\includegraphics[width=6cm]{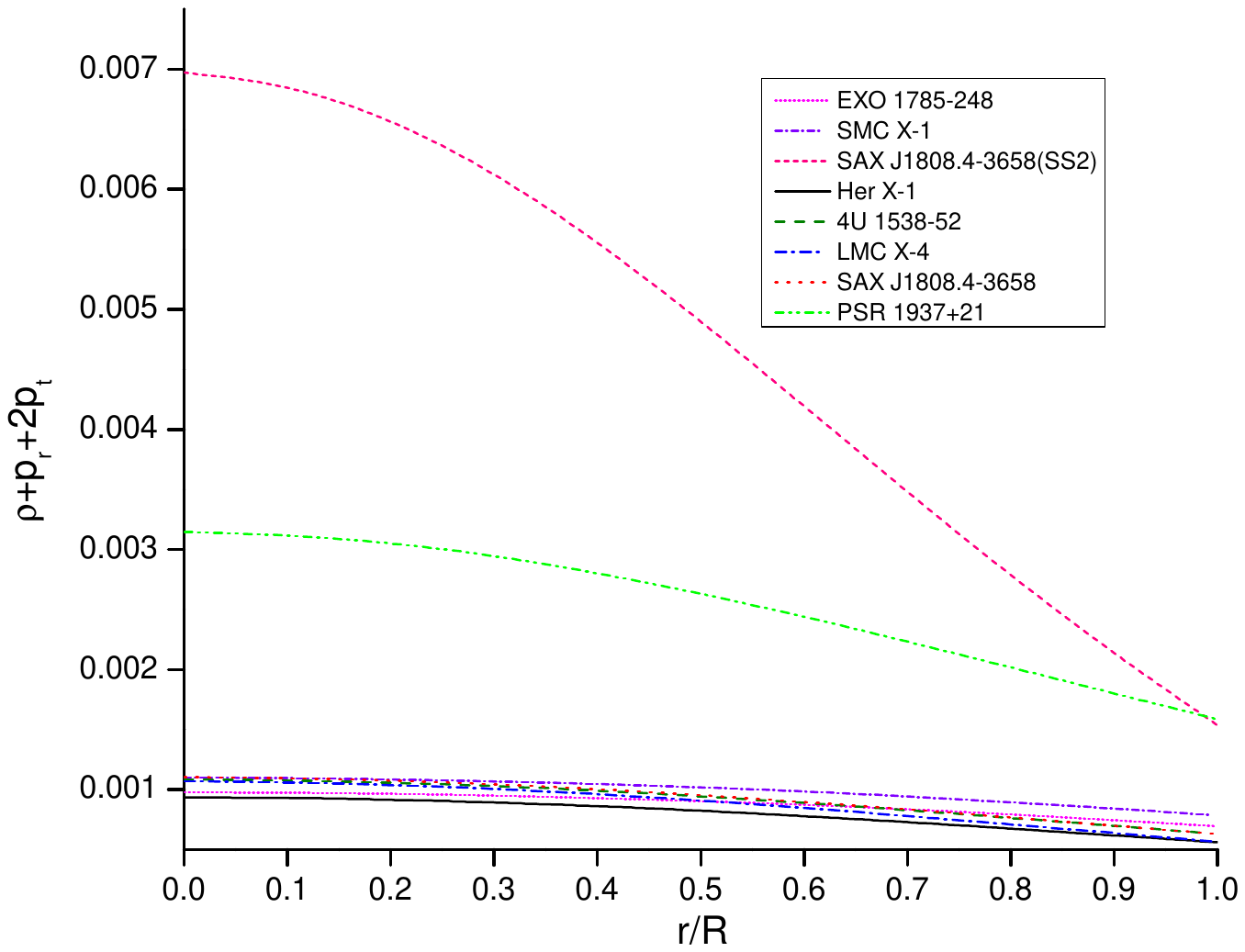}\includegraphics[width=6cm]{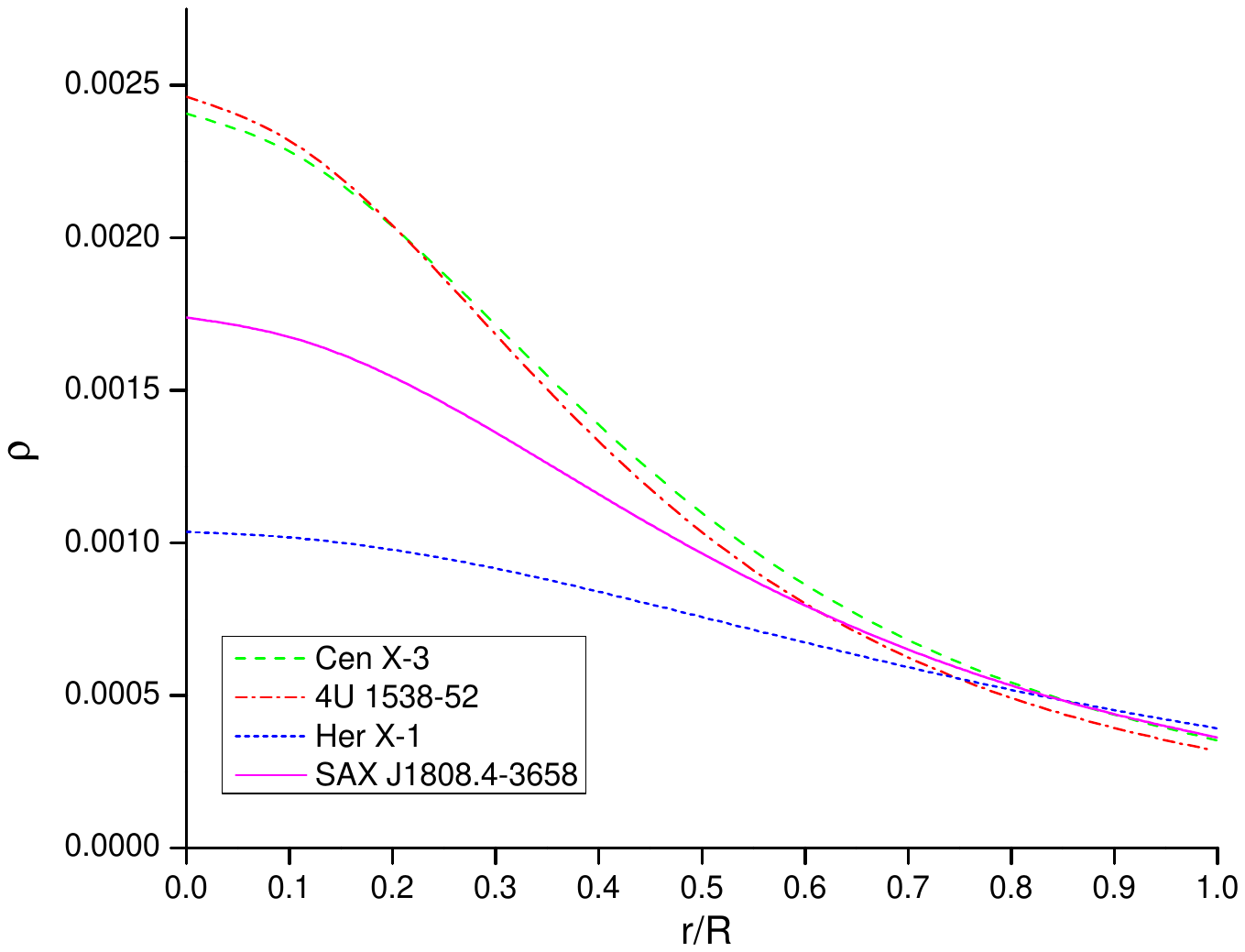}\includegraphics[width=6cm]{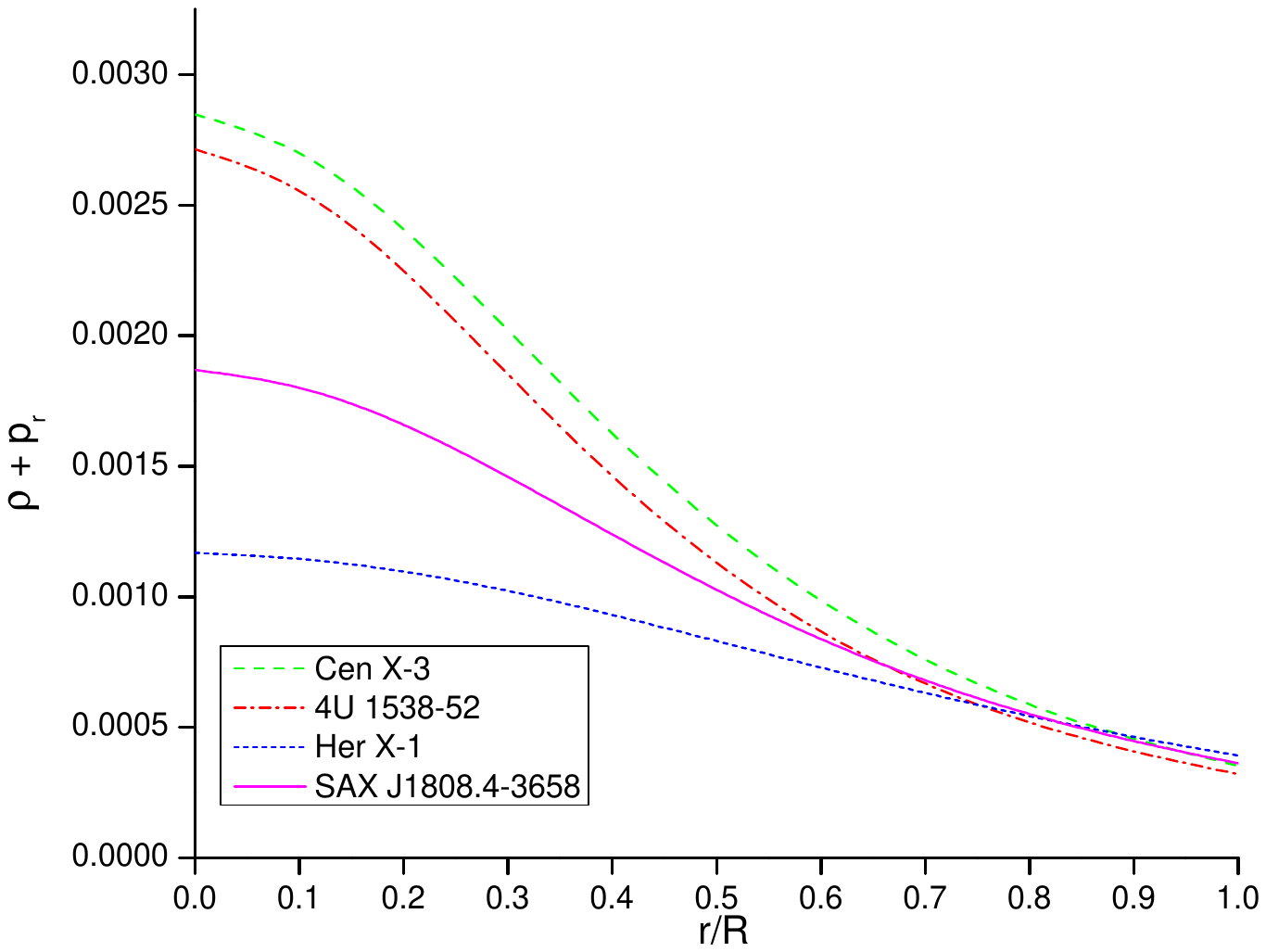}\\\includegraphics[width=6cm]{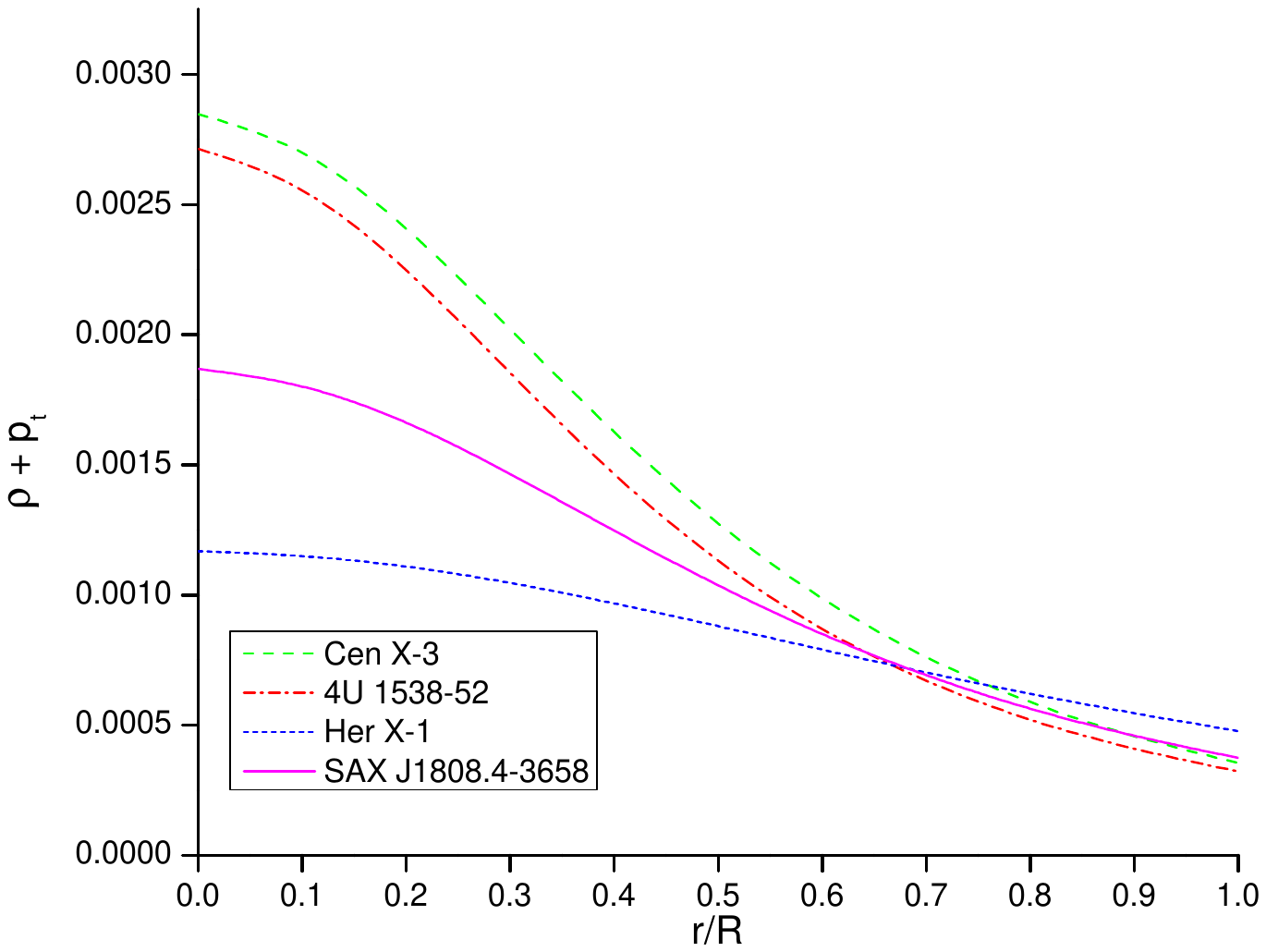}\includegraphics[width=6cm]{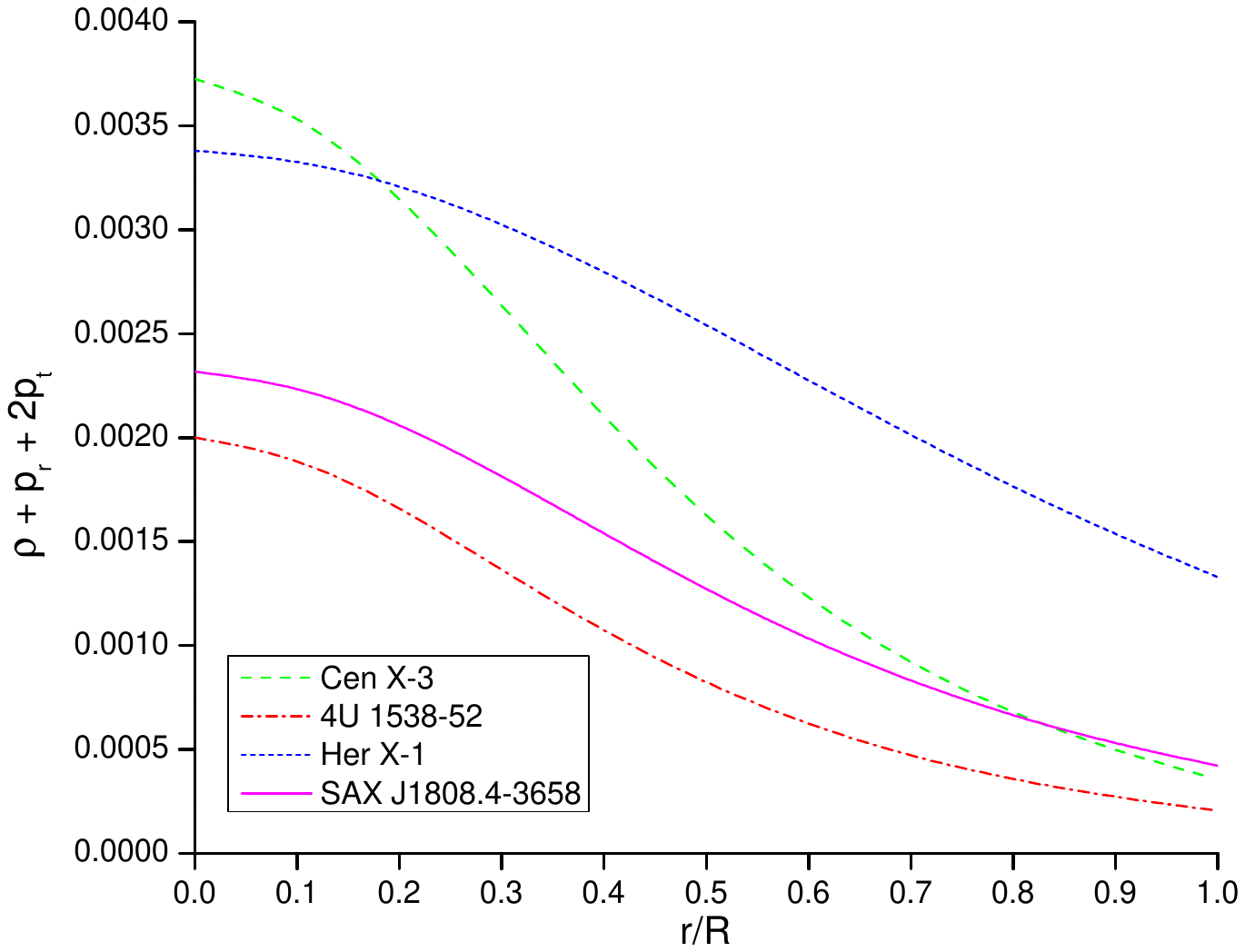}
\caption{The different energy conditions (in $km^{-2}$) diagram in model for Case \textbf{I} \& \textbf{II} have been plotted with respect to radial coordinate $r/R$. where, first four graphs describe energy conditions corresponding Case I while next four graphs for case II.}\label{f3}
\end{center}
\end{figure}
%%%%%%%%%%%%%%%%%%%%%%%%%%%%%%%%%%%%%%%
We derive precisely all the forms of energy conditions for both cases by plugging the 
values of energy density  and respective pressure equations. The resulting graph
Fig. \ref{f3} shows that all the inequalities hold simultaneously for the sources considered here.

%%%%%%%%%%%%%%%%%%%%%%%%%%%%%%%%%%
\subsection{Generalized TOV equation}
The Tolman-Oppenheimer-Volkoff (TOV) equation is used to constrains the structure of a spherically symmetric body (both isotropic and anisotropic fluid models) that is covered regarding stars in hydrostatic equilibrium. Here, we start by explaining different forces, namely, gravitational, hydrostatic and anisotropic forces, respectively. The governing generalized-TOV equation for anisotropic fluid distribution, given by  \cite{Tolman,Volkoff}
\begin{eqnarray}
 -\frac{M_G(\rho+p_r)}{r^2}e^{\frac{\lambda-\nu}{2}}-\frac{dp}{dr}+\frac{2(p_t-p_r)}{r} =0,\label{78}
 \end{eqnarray}
where $M_G$ is the effective gravitational mass inside the fluid sphere of radius `$r$', and defined by
\begin{eqnarray}
M_G(r)=\frac{1}{2}r^2 \nu^{\prime}e^{(\nu - \lambda)/2}.\label{79}
\end{eqnarray}
Now, plugging the value of $ M_G(r) $ in Eq. (\ref{78}), we get
\begin{eqnarray}
-\frac{\nu'}{2}(\rho+p_r)-\frac{dp_r}{dr}+\frac{2(p_t-p_r)}{r} =0,  \label{80}
\end{eqnarray}
The above TOV equation describes the equilibrium condition for 
anisotropic fluid spheres subject to gravitational, hydrostatic and anisotropic 
force due to the anisotropic pressure. Combining all forces we have the following form
%%%%%%%%%%%%%%%%%%%%%%%%%%%%%%%%%%%%%%%
\begin{eqnarray}
F_g+F_h+F_a=0,\label{311}
\end{eqnarray}
Now, we start by explaining the Eq. \ref{311} from an equilibrium point of view, where three different components are gravitational $(F_g)$, hydrostatic $(F_h)$ and anisotropic $(F_a)$ forces, respectively with the following expression 
\begin{eqnarray}
F_g = -\dfrac{\nu'}{2}(\rho+p_r),~~~  F_h=-\dfrac{dp_r}{dr}  ~~~\text{and}~~F_a=\dfrac{2(p_t-p_r)}{r}=\frac{2\Delta}{C}. 
\end{eqnarray}
Here, the anisotropy force $(F_a)$ takes the following form for both cases \textbf{I} and  \textbf{II}, which turns out to be \\  
\begin{eqnarray} 
 \text{Case~~} \textbf{I}:  F_a=\dfrac{\Delta_0\big[ (K-1)\,\sin^2 x-K \big]}{(K-1)^2\,\cos^4 x}~~~  \text{and}~~~  \text{Case~~}\textbf{II}:  F_a=\dfrac{\Delta_0\big[ (K-1)\,\cosh^2 x-K \big]}{(K-1)^2\,\sinh^4 x},
\end{eqnarray}
and the other components are written in an explicitly form \\
 $\textbf{ Case Ia:}$ 
\begin{eqnarray}  
F_h &=& N_1\Big[\dfrac{Cr}{\sin x\,\cos x (K-1)} \Big], \label{82}\\
F_g&=&-\dfrac{Cr\,(1+n^2)}{\sin^2 x\, (K-1)}\left[\,\frac{A_{1}\,\cosh(n\,x)+B_{1}\,\sinh(n\,x)}{\cosh(n x)\,(\,A_{1} +B_{1}\,n\,\cot x \,) + \sinh(n\,x)\,(\,A_{1}\,n\,\cot x + B_{1} \,) }\,\right](\rho+p_r), \label{81}
\end{eqnarray}

$\textbf{ Case Ib:}$
\begin{eqnarray} 
F_h &=& N_2\Big[\dfrac{Cr}{\sin x\,\cos x (K-1)} \Big], \label{84}\\
F_g&=&-\dfrac{Cr\,(1-n^2)}{\sin^2 x\, (K-1)}\left[\,\frac{C_{1}\,\cos(n\,x)+D_{1}\,\sin(n\,x)}{C_1\,\cos(n x)+D_{1}\,\sin(n\,x)-\,n\,\cot x\,(\,C_{1}\,\sin(n\,x) - D_{1}\,\cos(n\,x)) }\,\right](\rho+p_r),  \label{83}
\end{eqnarray}

$\textbf{ Case Ic:}$ 
\begin{eqnarray} 
F_h &=& N_3\Big[\dfrac{Cr}{\sin x\,\cos x (K-1)} \Big], \label{86}\\
F_g &=&-\dfrac{8\,Cr}{\sin x\, (K-1)}\left[\frac{E_{1}\,\cos (x)+F_{1}\,\sin (x)}{ \,E_{1}\,(2x+\sin 2x)-F_{1}\,\cos 2x}\right](\rho+p_r), \label{85}
\end{eqnarray}

 $\textbf{ Case Id:}$
 \begin{eqnarray} 
 F_h &=& N_4\Big[\dfrac{Cr}{\sin x\,\cos x (K-1)} \Big], \label{88}\\
 F_g &=&-\dfrac{Cr}{\sin x\, (K-1)}\left[\frac{G_{1}\,(x)+H_{1}}{G_{1}\,(\cos x + x\,\sin x)+H_{1}\,\sin x}\right](\rho+p_r), \label{87}
\end{eqnarray}

$\textbf{ Case IIa:}$
\begin{eqnarray}  
F_h &=& N_5 \Big[\dfrac{Cr}{\sinh x\,\cosh x (K-1)} \Big], \label{90}\\
F_g &=& -\dfrac{Cr\,(1+n^2)}{\cosh^2 x\, (K-1)}\left[\,\frac{A_{2}\,\cos(n\,x)+B_{2}\,\sin(n\,x)}{A_{2}\,\cos (n\,x) +B_{2}\,\sin(n\,x) + n\,\tanh x\,\big(\,A_{2}\,\sin(n\,x) - B_{2}\,\cos(n\,x)\big) }\right](\rho+p_r),  \label{89}
\end{eqnarray}

 $\textbf{ Case IIb:}$
\begin{eqnarray}  
F_h &=& N_6 \Big[\dfrac{Cr}{\sinh x\,\cosh x (K-1)} \Big], \label{92}\\
F_g &=& -\dfrac{Cr\,(1-n^2)}{\cosh^2 x\, (K-1)}\left[\,\frac{C_{2}\,\cosh (n\,x)+D_{2}\,\sinh (n\,x)}{\, [\, C_{2}\,\cosh(n x) + D_{2}\, \sinh(n x)\, ] - 
 n\,\tanh x\,[\, C_{2}\, \sinh(n x)+D_{2}\, \cosh(n x) \,] }\,\right](\rho+p_r), \label{91}
\end{eqnarray}
 $\textbf{ Case IIC:}$
\begin{eqnarray}  
F_h &=& N_7 \Big[\dfrac{Cr}{\sinh x\,\cosh x (K-1)} \Big], \label{94}\\
F_g &=& -\dfrac{8\,Cr}{\cosh x\, (K-1)}\left[\frac{E_{2}\,\cosh x + F_{2}\,\sinh x}{ \,E_{2}\,\cosh 2x+F_{2}\,(\sinh 2x-2x)}\right](\rho+p_r), \label{93}
\end{eqnarray}

\begin{figure} [h!]
\begin{center}
\includegraphics[width=6cm]{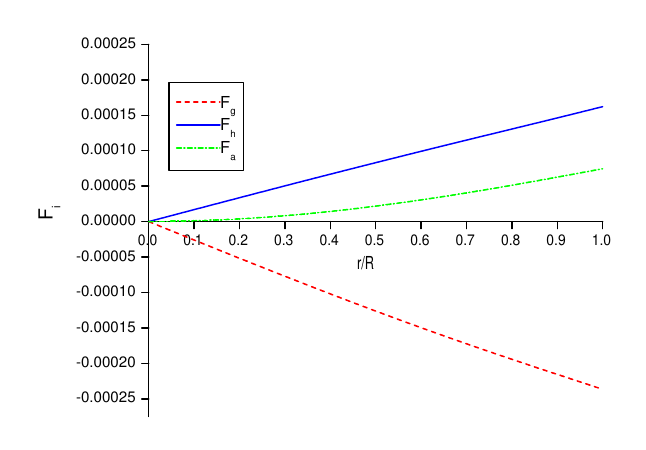}\includegraphics[width=6cm]{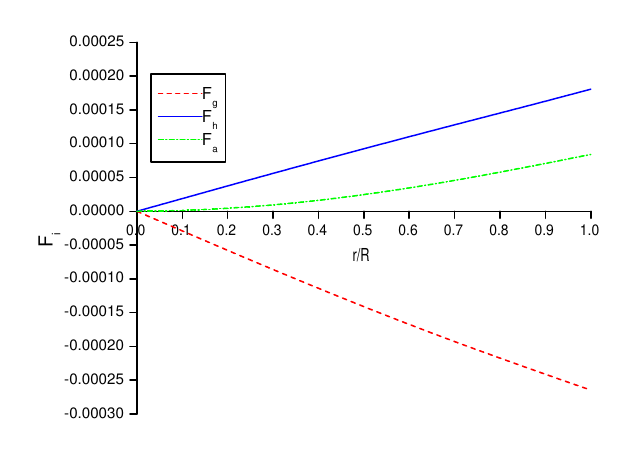}\includegraphics[width=6cm]{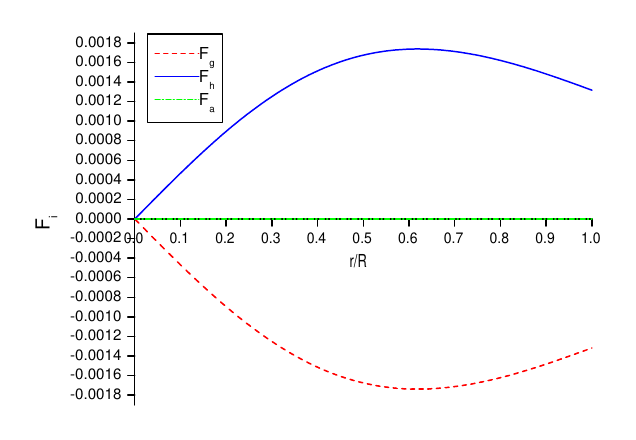}\\\includegraphics[width=6cm]{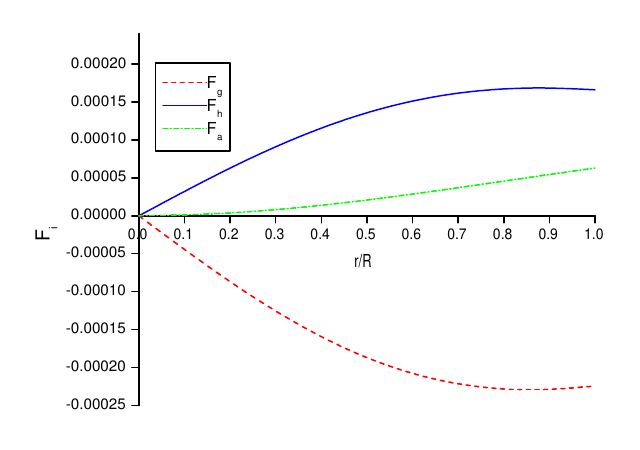}\includegraphics[width=6cm]{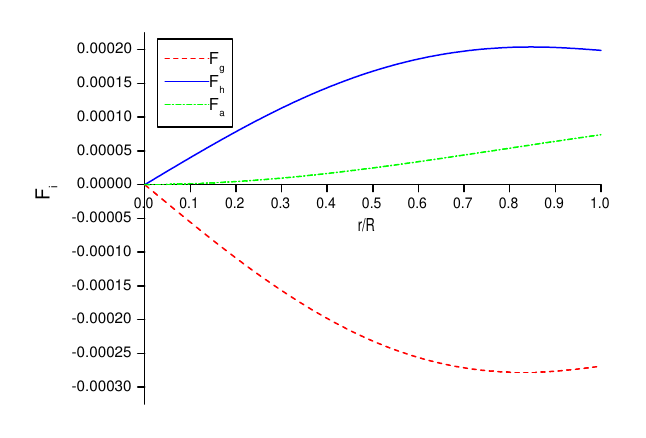}\includegraphics[width=6cm]{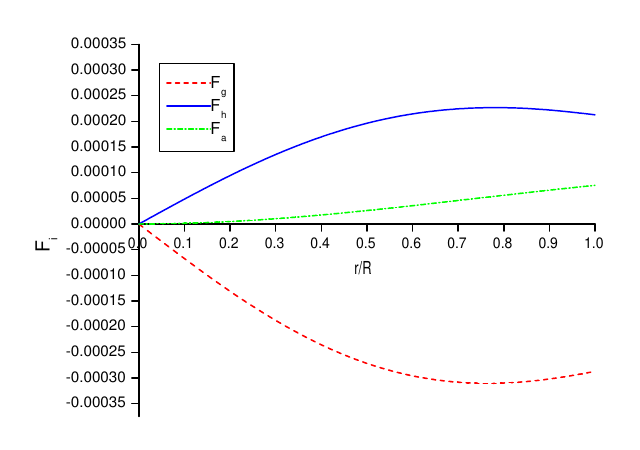}\\\includegraphics[width=6cm]{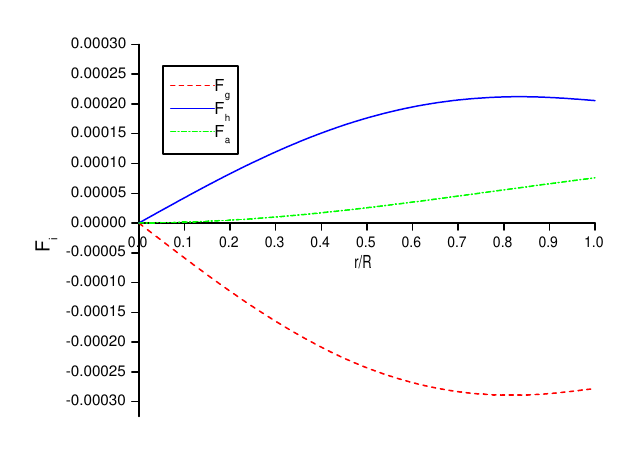}\includegraphics[width=6cm]{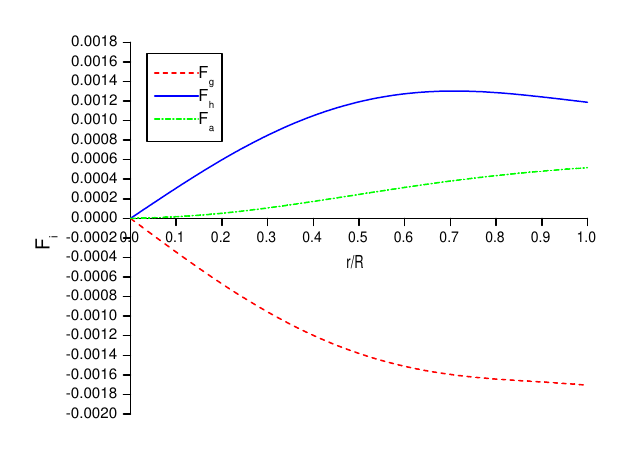}\includegraphics[width=6cm]{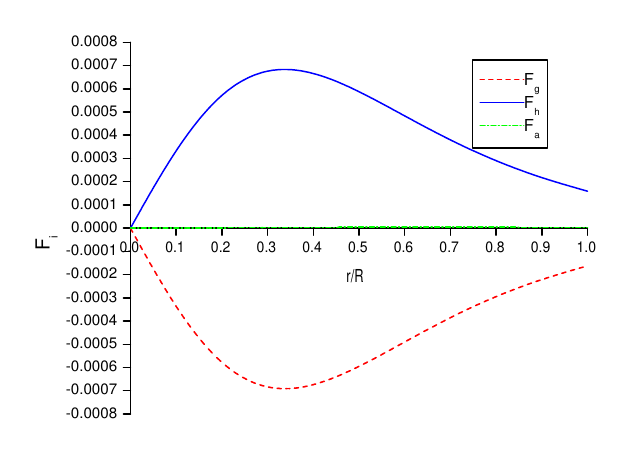}\\\includegraphics[width=6cm]{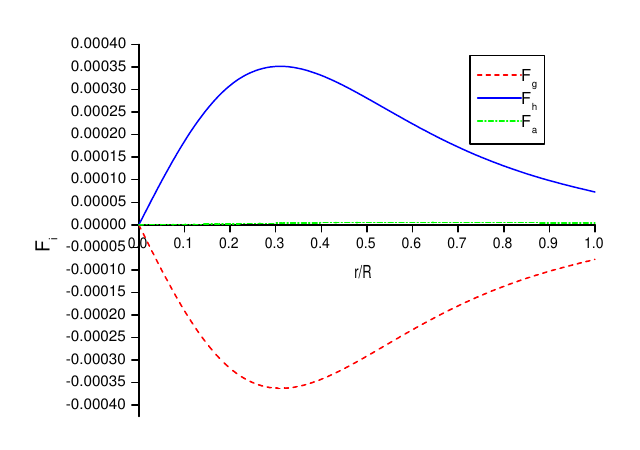}\includegraphics[width=6cm]{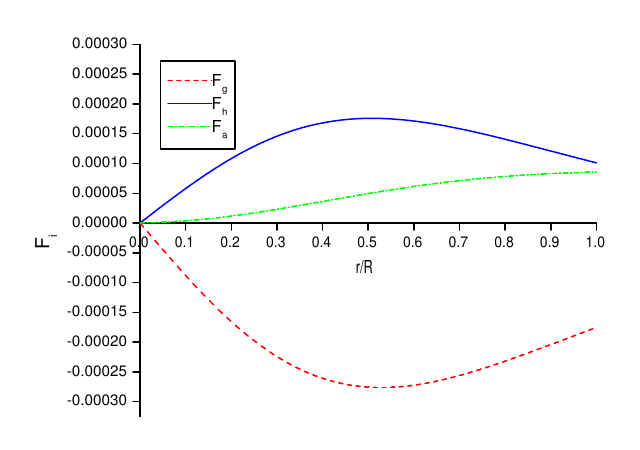}\includegraphics[width=6cm]{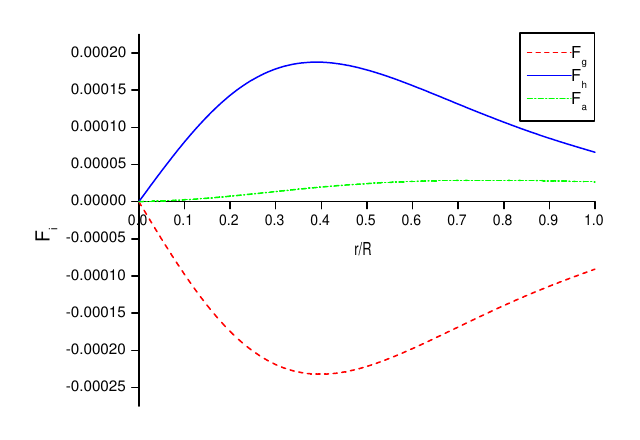} 
\caption{Variation of different forces (in $km^{-3}$ with $G=c=1$) with respect to radial coordinate r/R. For plots we have drawn (i) EXO 1785-248, (ii) SMC X-1, and (iii) SAX J1808.4-3658(SS2)-1, from felt to right in the first row. In the second row (iv) Her X-1,(v) 4U 1538-52, (vi) LMC X-4, have been plotted. In the third and fourth rows (vii) SAX J1808.4-3658, (viii) PSR 1937+21 (ix) Cen X-3, (x) 4U 1538-52,(xi) Her X-1, (xii) SAX J1808.4-3658 have been plotted for Case \textbf{I} \& \textbf{II}, respectively.  }\label{f4} 
\end{center}
\end{figure}

 $\textbf{ Case IId:}$
\begin{eqnarray} 
F_h &=& N_8 \Big[\dfrac{Cr}{\sinh x\,\cosh x (K-1)} \Big], \label{96}\\
F_g&=&  -\dfrac{Cr}{\cosh x\, (K-1)}\left[\frac{G_{2}\,(x)+H_{2}}{G_{2}\,(x\,\cosh x - \sinh x)+H_{2} \cosh x}\right](\rho+p_r) \label{95}
\end{eqnarray} 
In order to evaluate equilibrium conditions, the hydrostatic equilibrium diagrams obtained for the eight different compact stars which are shown in Fig. (\ref{f4}). From a mathematical point of view, one can see form Fig. (\ref{f4}) that the gravitational force ($F_g$) is dominating over the hydrostatic ($F_h$) and anisotropic ($F_a$) forces which is counter balance by joint action of hydrostatic ($F_h$) and anisotropic ($F_a$) forces.  From Fig. \ref{f4}, we see that the force components $F_g$, $F_h$ and $F_a$ of TOV equation are regular finite at the origin as well as on the surface of the star. Moreover, we also observe some other interesting features of force components corresponding to each star which are as follows: The hydrostatic force ($F_h$) and gravitational force ($F_g$) are increasing monotonically throughout within the stellar models and attain its maximum value on the boundary corresponding to the stars (i) EXO 1785-248, (ii) SMC X-1 while for other stars, namely (iii) SAX J1808.4-3658 (SS2)-1,~ (iv) Her X-1,~(v) 4U 1538-52,~ (vi) LMC X-4, (vii) SAX J1808.4-3658,~ (viii) PSR 1937+21 (for case I) and (ix) Cen X-3,~ (x) 4U 1538-52,~(xi) Her X-1,~ (xii) SAX J1808.4-3658 (for case II), the force $F_h$ and $F_g$ increases first and reach its maximum value at the some point $r/R$ within the stellar model and there after start decreases towards the respective boundary. On the other hand, the anisotropic force $F_a$ is increasing monotonically towards the surface boundary corresponding to each obtained star. From the Fig.\ref{f4}, we also note that the anisotropic force $F_a$ has very less effect compare to hydrostatic force $F_h$ and gravitational force $F_g$ for the stars namely (iii) SAX J1808.4-3658(SS2) (for case I) and (ix) Cen X-3, (x) 4U 1538-52 (for case II).   

\begin {table}
\caption{ Values of the model parameters of Case I and Case II for different values of $ K,C,n $ and $ \Delta_0 $ }
\begin{center}
\label{T1}
\begin{tabular}{|c|c|c|c|c|c|c|c|c|}
\hline  
Compact star candidates& $ M(M_{\odot}) $ &  $R$~(km)  & $K$ & $CR^2$ & $n$ & $ \Delta_0 $ & $2-K+\Delta_0\,K$ & Cases \\
\hline
EXO 1785-248 ( $\ddot{\text{O}}$zel \emph{et al.} \cite{ozel}) &  1.3 & 8.849 & -0.27898 & 0.1044 & 0.1 & 8.205 &-0.01 &Case Ia. \\
\hline
SMC X-1(Rawls \emph{et al.} \cite{Rawls}) & 1.04 & 8.301 & -0.28103 &  0.1044 & 0.1 & 8.152 &-0.01 & Case Ia.\\
\hline
SAX J1808.4-3658 (SS2)(Li \textit{et al.} \cite{Li})& 1.3237 & 6.16 & -1.18 & 0.52 & 1.783 & 0.000772 & 3.18 &Case Ib.\\
\hline
Her X-1 ~(Abubekerov \emph{et al.})  \cite{Abubekerov} & 0.85 & 8.1 & -1.18 & 0.2013 &  & 1.85 & 1 & Case Ic.\\
\hline
4U 1538-52 ~(Rawls \emph{et al.} \cite{Rawls}) & 0.87 & 7.866 & -1.18  & 0.2145 &  & 1.85 & 1 & Case Ic.\\
\hline
LMC X-4(Rawls \emph{et al.} \cite{Rawls})& 1.29 & 8.831 & -1.18 &  0.25006 &  & 1.85 & 1 & Case Ic.\\
\hline
SAX J1808.4-3658 ~ (Elebert \textit{et al.} \cite{Elebert}) & 0.9 & 7.951 & -1.18 & 0.2206 &  & 1.85 & 1 & Case Ic.\\
\hline
PSR 1937+21 ~ (Kapoor \textit{et al.} \cite{Kapoor2001}, Xin \textit{et al.}\cite{Xin2001} ) &  2.0833 & 8.04 & -0.91 &  0.5698 &  & 3.198 & 0 & Case Id.\\ 
\hline
Cen X-3(Rawls \emph{et al.} \cite{Rawls})& 1.49 & 9.178 & 3  & 2.55 & 0.99 & 0.00663 & -0.98 & Case IIa.\\
\hline
4U 1538-52 ~(Rawls \emph{et al.} \cite{Rawls}) & 0.87 & 7.866 & 1.78& 2.915 & 0.4796 & 0.0056 &  0.23& Case IIb.\\
\hline
Her X-1 ~(Abubekerov \emph{et al.}) \cite{Abubekerov}& 0.85 & 8.1 & 3.1 & 0.8415 &  & 0.677 & 1 & Case IIc.\\
\hline
SAX J1808.4-3658 ~(Elebert \textit{et al.} \cite{Elebert}) & 0.9 & 7.951 & 2.1 & 1.759 &  & 0.048 & 0 & Case IId.\\
\hline
\end{tabular}
\end{center}
\end{table} 

\begin {table}
\caption{Energy densities, central pressure and Buchdahl limit for different compact star candidates for the above parameter values of Table \textbf{I}.}
\begin{center}
\begin{tabular}{|c|c|c|c|c|c|c|}
\hline
Compact star candidates& Central Density & Surface density & Central pressure& Buchdahl condition& Redshift&\\ 
&$(gm/cm^{3})$&$(gm/cm^{3})$&$(dyne/cm^{2})$&$(2M/R\leq8/9)$ & $(z_{S})$& Cases\\
\hline
EXO 1785-248 & $0.98\times 10^{15}  $& $ 8.33\times 10^{14}  $ &$ 0.99\times 10^{35}  $&0.21669& 0.328472 &Case Ia.\\
\hline
SMC X-1 & $1.109\times 10^{15}  $& $ 9.41\times 10^{14}  $ &$ 1.107\times 10^{35}  $&0.21546 & 0.325584 & Case Ia.\\
\hline
SAX J1808.4-3658 (SS2)& $4.06\times 10^{15}  $& $ 20.65\times 10^{14}  $ &$ 1.59\times 10^{35}  $&0.3071 & 0.648507 & Case Ib.\\
\hline
Her X-1& $0.91\times 10^{15}  $& $ 6.73\times 10^{14}  $ &$ 1.05\times 10^{35}  $&0.0618& 0.203489 & Case Ic.\\
\hline
4U 1538-52 & $1.03\times 10^{15}  $& $ 7.47\times 10^{14}  $ &$ 1.28\times 10^{35}  $&0.16314 & 0.218326 & Case Ic.\\
\hline
LMC X-4& $0.95\times 10^{15}  $& $ 6.59\times 10^{14}  $ &$ 1.45\times 10^{35}  $&0.18479 & 0.259444 & Case Ic.\\
\hline
SAX J1808.4-3658& $1.04\times 10^{15}  $& $ 7.46\times 10^{14}  $ &$ 1.34\times 10^{35}  $&0.16696 & 0.225259 & Case Ic.\\
\hline
PSR 1937+21& $2.97\times 10^{15}  $& $ 14.35\times 10^{14}  $ &$ 3.77\times 10^{35}  $&0.2692 & 1.049159 & Case Id.\\
\hline
Cen X-3& $3.24\times 10^{15}  $& $ 4.76\times 10^{14}  $ &$ 5.305\times 10^{35}  $&0.23945& 0.419898 & Case IIa.\\
\hline
4U 1538-52& $3.32\times 10^{15}  $& $ 4.27\times 10^{14}  $ &$ 2.63\times 10^{35}  $& 0.16314& 0.643074 & Case IIb.\\
\hline
Her X-1& $1.45\times 10^{15}  $& $ 5.27\times 10^{14}  $ &$ 1.59\times 10^{35}  $&0.15478& 0.203472 & Case IIc.\\
\hline
SAX J1808.4-3658& $2.34\times 10^{15}  $& $ 4.87\times 10^{14}  $ &$ 1.56\times 10^{35}  $&0.16696& 0.225316 & Case IId.\\
\hline
\end{tabular}
\label{T2}
\end{center}
\end{table}
\subsection{The Equation of state (EoS)} 
Here we derive the relation between most important features of neutron stars is an equation of state 
(EoS) i.e. a relation between pressure and density. The EoS of neutron star matter at the inner core where most
of the mass resides is not well constrained. It is worthwhile to mention that different EoS lead to different mass-radius (M-R) relations. To explain the structural properties of compact stars model at high densities, several authors
have proposed the EoS$P = P (\rho)$ should be well approximated by a linear function of the energy density $\rho$
\cite{Dey1998,harko2002,Gondek2000}. Some authors have also expressed more convincing
approximated forms of the EoS $P = P (\rho)$  as linear function of energy density $\rho$ about the ( for more details see 
\cite{Haensel1989,Frieman1989,Prakash1990}). In order to reach that aim, we start our calculation by writing the EoS in a
linear function form i.e. $P = P (\rho)$ 
\subsubsection*{\textbf{Case I. For $K < 0$  i.e  K is Negative}}
\begin{eqnarray}
 p_r&=& \frac{C}{\kappa\,(1+\tilde{\rho}_{1})K}\bigg[1-K+\frac{\,2(1+n^2)\big[\sinh(n\,\tilde{\rho})f_{1}(\tilde{\rho}_{s})+\cosh(n\,\tilde{\rho})f_{2}(\tilde{\rho}_{s})\big]}{\cosh(n\,\tilde{\rho})\big(n\,\cosh(n\,\tilde{\rho}_s)(-f_{3})+\sinh(n\,\tilde{\rho}_s)f_{4}\big)+\sinh(n\,\tilde{\rho})f_{5}} \bigg],~~~~~~~~\textbf{Case Ia.}\label{pd1a}\\
p_r&=& \frac{C}{\kappa\,\,(1+\tilde{\rho}_{1})K}\bigg[1-K+\frac{\,2(1-n^2)\big[-\sin(n\,\tilde{\rho})f_{6}(\tilde{\rho}_{s})+\cos(n\,\tilde{\rho})f_{7}(\tilde{\rho}_{s})\big]}{\cos(n\,\tilde{\rho})\big[n\,\cos(n\,\tilde{\rho}_s)f_{8}-\sin(n\,\tilde{\rho}_s)f_{9}+\sin(n\,\tilde{\rho})\big]f_{10}}\bigg],~~~~~~~~~~~~~~~~~~~\textbf{Case Ib.} \label{pd1b} \\
p_r&=&\frac{C}{\kappa\,(1+\tilde{\rho}_1)K}\,\bigg[1-K+\dfrac{8\sin(\tilde{\rho})\big[4\cos(\tilde{\rho})+(K-5)\,\cos(\tilde{\rho}-2\tilde{\rho}_s)+2\tilde{\rho}_s(K-1)\sin(\tilde{\rho})\big]}{8\tilde{\rho}-2\tilde{\rho}_s\,(K-1)\cos(2\tilde{\rho})+(K-5)\,f_{11}+4\sin(2\tilde{\rho})}\bigg], ~~~~~\textbf{Case Ic.} \label{pd1c}\\ 
p_r&=&\frac{C(3-K)\big[f_{12}+(1-K)\cos(\tilde{\rho})\,\sin(\tilde{\rho}_s)-f_{13}\,\sin(\tilde{\rho}_s)\big]}{\kappa\,(1+\tilde{\rho}_1)\,K\,\big[-f_{12}+((K-3)\cos(\tilde{\rho})+f_{13})\sin(\tilde{\rho_s}) \big]},~~~~~~~~~~~~~~~~~~~~~~~~~~~~~~~~~~~~~~~~~~~~~~~~\textbf{Case Id.} \label{pd1d}
\end{eqnarray} \\

where, for notational convenience, we use\\  
$ \tilde{\rho}=\sin^{-1}\sqrt{\frac{K+\tilde{\rho_1}}{K-1}}$,
~~$\tilde{\rho_1}=\frac{( K-1- 2\,\rho_1\, K)\pm \sqrt{1 - 2 K - 8 \rho_1\, K + K^2 + 8 \rho_1\, K^2}}{2 \rho_1\, K},~~~\rho_{1}=\frac{\kappa\,\rho}{C}$, \\ 
$\tilde{\rho_s}=\sin^{-1}\sqrt{\frac{K+\tilde{\rho}_{1s}}{K-1}}$,~~~~~$\tilde{\rho}_{1s}=\frac{( K-1- 2\rho_{1s} K)\pm \sqrt{1 - 2 K - 8 \rho_1\, K + K^2 + 8 \rho_{1s}\, K^2}}{2 \rho_{1s}\, K},~~~\rho_{1s}=\frac{\kappa\,\rho_s}{C}$, \\ 

$f_{1}(\tilde{\rho}_s)=(-2n^2+K-3)\,\cosh(n\,\tilde{\rho}_s)\,\sec(\tilde{\rho}_s)+n\,(K-1)\csc(\tilde{\rho}_s)\sinh(n\,\tilde{\rho}_s),$ \\

$f_{2}(\tilde{\rho}_s)=n(1-K)\cosh(n\,\tilde{\rho}_s)\csc(\tilde{\rho}_s)+(2n^2-K+3)\sec(\tilde{\rho}_s)\sinh(n\,\tilde{\rho}_s)$,\\

$f_{3}=(K-1)\csc(\tilde{\rho}_s)+(2n^2-K+3)\cot(\tilde{\rho})\sec(\tilde{\rho}_{s}),~~~~f_{4}=n^2(K-1)\cot(\tilde{\rho})\csc(\tilde{\rho}_s)+(2n^2-K+3)\sec(\tilde{\rho}_s),$ \\

$f_{5}=\big(\cosh(n\,\tilde{\rho}_s)\,(-f_{4})+n\,\sinh(n\,\tilde{\rho}_s)\,f_{3}\big),~~~ f_{6}(\tilde{\rho}_s)=(2n^2+K-3)\,\cos(n\,\tilde{\rho}_s)\,\sec(\tilde{\rho}_s)-n\,(K-1)\csc(\tilde{\rho}_s)\sin(n\,\tilde{\rho}_s), $\\

$ f_{7}(\tilde{\rho}_s)=n(K-1)\cos(n\,\tilde{\rho}_s)\csc(\tilde{\rho}_s)+(2n^2+K-3)\sec(\tilde{\rho}_s)\sin(n\,\tilde{\rho}_s),~~~f_{8}=(1-K)\csc(\tilde{\rho}_s)+(2n^2+K-3)\cot(\tilde{\rho})\sec(\tilde{\rho}_{s}),$ \\

$f_{9}=n^2(K-1)\cot(\tilde{\rho})\csc(\tilde{\rho}_s)+(2n^2+K-3)\sec(\tilde{\rho}_s) ,~~~~~f_{10}=\big(\cos(n\,\tilde{\rho}_s)\,f_{9}+n\,\sin(n\,\tilde{\rho}_s)\,f_{8}\big),$ \\

 $f_{11}=\big[2\,\tilde{\rho}\cos(2\tilde{\rho}_s)+\sin(2(\tilde{\rho}-\tilde{\rho}_s))\big],  ~~f_{12}=(K-1)\cos(\tilde{\rho}_s)\sin(\tilde{\rho}),~~~~~~~~f_{13}=(\tilde{\rho}-\tilde{\rho}_s)(K-3)\sin(\tilde{\rho}). $ \\

 %%%%%%%%%%%%%%%%%%%%%%%%%%%%%%%%%% 
 \subsubsection*{\textbf{Case II. For $K >1$  i.e  K is Positive}}
\begin{eqnarray}
  p_r&=&\dfrac{C\,}{\kappa\,(1+\tilde{\rho}_1)K}\bigg[ 1-K+\dfrac{2(1+n^2)\cosh(\bar{\rho})\big[n\,(K-1)\cos(n\,\bar{\rho}-n\,\bar{\rho}_s)\sec h(\bar{\rho}_s)+f_{14}\big]}{(3-K+2n^2)\csc h(\bar{\rho}_s)f_{15}+n\,(K-1)\sec h(\bar{\rho}_s)\,f_{16}}\bigg],~~~~~~~\textbf{Case IIa.}\label{pd2a}\\
    p_r&=&\dfrac{C\,}{\kappa\,(1+\tilde{\rho}_1)K}\bigg[ 1-K+\dfrac{2(1-n^2)\big[\cosh(n\,\bar{\rho})f_{17}(\bar{\rho}_s)-\sinh(n\,\bar{\rho})f_{18}(\bar{\rho}_s)\big]}{\sinh(n\,\bar{\rho})\big(n\,\sinh(n\,\bar{\rho}_s)f_{19}+\cosh(n\,\bar{\rho}_s)f_{20}\big)+\cosh(n\,\bar{\rho})f_{21}}\bigg],~~~~~~~~~\textbf{Case IIb.} \label{pd2b} \\
    p_r&=&\frac{C\,}{\kappa\,(1+\tilde{\rho}_1)K}\,\bigg[1-K+\dfrac{8\cosh(\bar{\rho})\big[2\bar{\rho}_s(K-1)\cosh(\bar{\rho})-4\sinh(\bar{\rho})+(K-5)\,\sin(\bar{\rho}-2\bar{\rho}_s)\big]}{8\bar{\rho}+2\bar{\rho}_s\,(K-1)\cosh(2\bar{\rho})-(K-5)\,f_{22}-4\sinh(2\bar{\rho})}\bigg], ~\textbf{Case IIc.} \label{pd2c}\\
  p_r&=&\dfrac{C\,(3-K)\big(f_{23}+\cosh(\bar{\rho})f_{24}\big)}{\kappa\,(1+\tilde{\rho}_1)\,K\,\big[(3-K)\cosh(\bar{\rho}_s)\sinh(\bar{\rho})+\cosh(\bar{\rho})f_{24} \big]}~~~~~~~~~~~~~  ~~~~~~~~~~~~~~~~~~~~~~~~~~~~~~~~~~~~\textbf{Case IId.}\label{pd2d}.  
\end{eqnarray} 

where, \\

$\bar{\rho}=\cosh^{-1}\sqrt{\frac{K+\tilde{\rho_1}}{K-1}}$,~~~~$\bar{\rho_s}=\cosh^{-1}\sqrt{\frac{K+\tilde{\rho}_{1s}}{K-1}}$,\\

$ f_{14}=(-3+K-2n^2)\,\textrm{csch}(\bar{\rho}_s)\,\sin[n(\bar{\rho}-\bar{\rho}_s)], ~~f_{15}=-\cosh(\bar{\rho})\,\sin(n(\bar{\rho}-\bar{\rho}_s))+n\,\cos(n(\bar{\rho}-\bar{\rho}_s))\sinh(\bar{\rho}) $,\\

$f_{16}=\cos[n(\bar{\rho}-\bar{\rho}_s)]\,\cosh(\bar{\rho})+n\,\sin[n(\bar{\rho}-\bar{\rho}_s)]\,\sinh(\bar{\rho})$,\\

$f_{17}(\bar{\rho}_s)=n\,(K-1)\cosh(n\,\bar{\rho}_s)\sec h(\bar{\rho}_s)+(-3+K+2n^2)\textrm{csch}(\bar{\rho}_s)\sinh(n\bar{\rho}_s)$,\\

$f_{18}(\bar{\rho}_s)=(-3+K+2n^2)\cosh(n\bar{\rho}_s)\textrm{csch}(\bar{\rho}_s)-n\,(K-1)\sec h(\bar{\rho}_s)\sinh(\bar{\rho}_s)$,\\

$f_{19}=(1-K)\sec h(\bar{\rho}_s)+(-3+K+2n^2)\,\textrm{csch}(\bar{\rho}_s)\tanh(\bar{\rho})$, ~~~$f_{20}=(-3+K+2n^2)\,\textrm{csch}(\bar{\rho}_s)+n^2\,(K-1)\,\textrm{sech}(\bar{\rho}_s)\tanh(\bar{\rho})$,\\

$f_{21}(\bar{\rho})=n\,\cosh(n\,\bar{\rho}_s)\,f_{19}-\sinh(n\,\bar{\rho}_s)\,f_{20}$, ~~$f_{22}=2\,\bar{\rho}\cosh(2\bar{\rho}_s)-\sinh(2(\bar{\rho}-\bar{\rho}_s))$,\\

$f_{23}=(K-1)\cosh(\bar{\rho}_s)\sinh(\bar{\rho}),~~~f_{24}=(\bar{\rho}-\bar{\rho}_s)(K-3)\cosh(\bar{\rho}_s)+(K-1)\sinh(\bar{\rho}_s)$.\\ \\
Form Eqs. (\ref{pd1a})-(\ref{pd2d}), one can observe that the radial pressures is purely density dependent,
which represents the simplest theoretical form of EoS for those stars. In an argument by
Dey \emph{et al} \cite{Dey1998} have proposed new types of EoSs for strange matter based on a model of
interquark potential. These EoSs have later been approximated to a linear
function of density by Gondek-Rosinska \emph{et al} \cite{Gondek2000}, as
%%%%%%%%%%%%%%%%%%%%%%%%%%%%%%%%%%%%%%%
\begin{eqnarray}
p = a\,(\rho-\rho_s),
\label{EOS}
\end{eqnarray}
 where $\rho_s$ denotes the energy density at zero pressure and $a$ is non-negative constant.
 Such an EoS has mainly been proposed to describe strange matter hypothesis
 built of $u$, $d$ and $s$ quarks. This was done by Harko and Cheng \cite{harko2002},
 who showed by using the Eq. (\ref{EOS}) when $\rho_s = 4B$ ( $B= 56 MeV fm^{3}$)
  the maximum mass of a strange star is $M_{max}$ = $1.83 M_{\odot}$.
  
\begin{figure} [h!]
 \begin{center}
 \includegraphics[width=4cm]{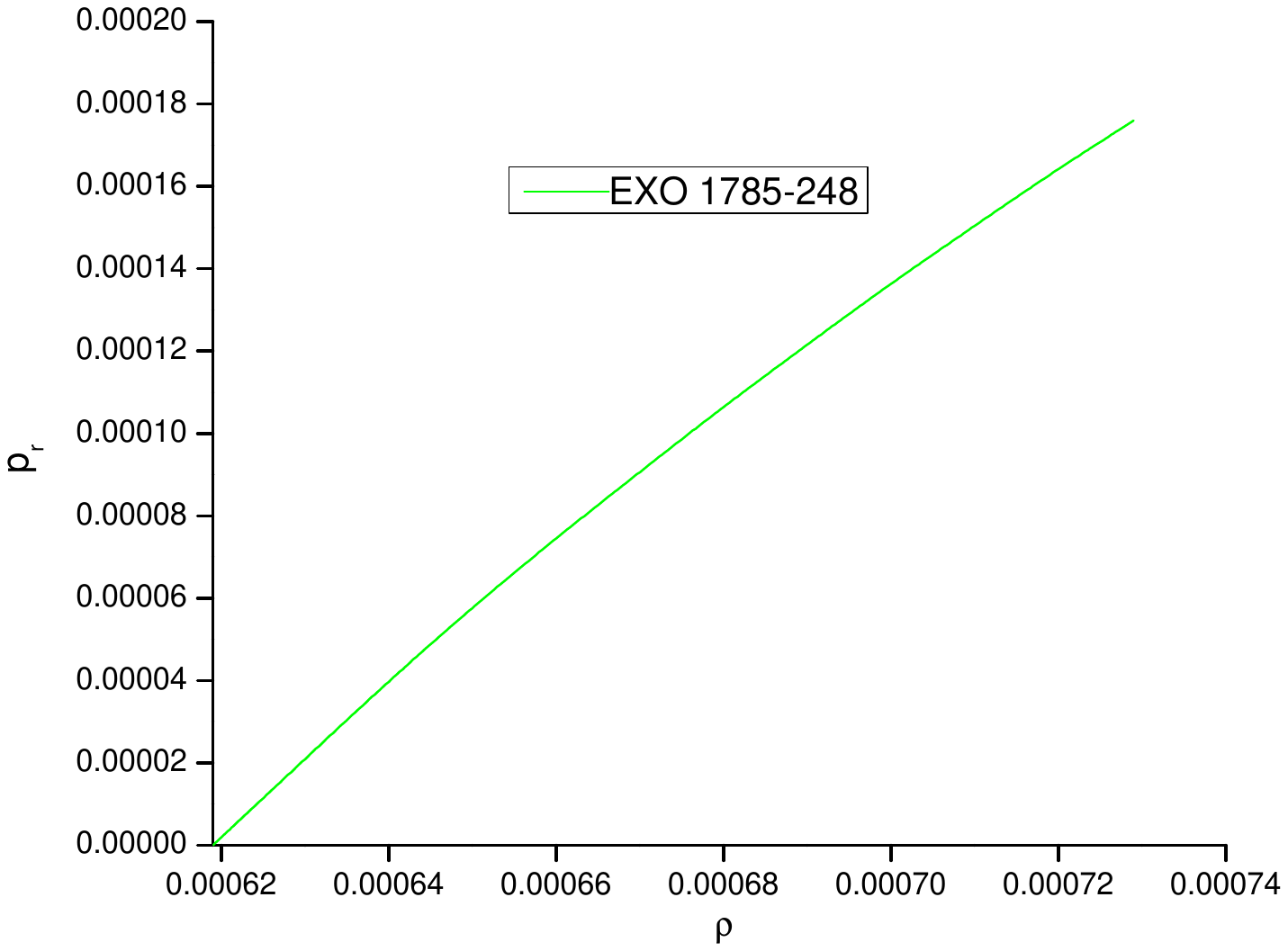}\includegraphics[width=4cm]{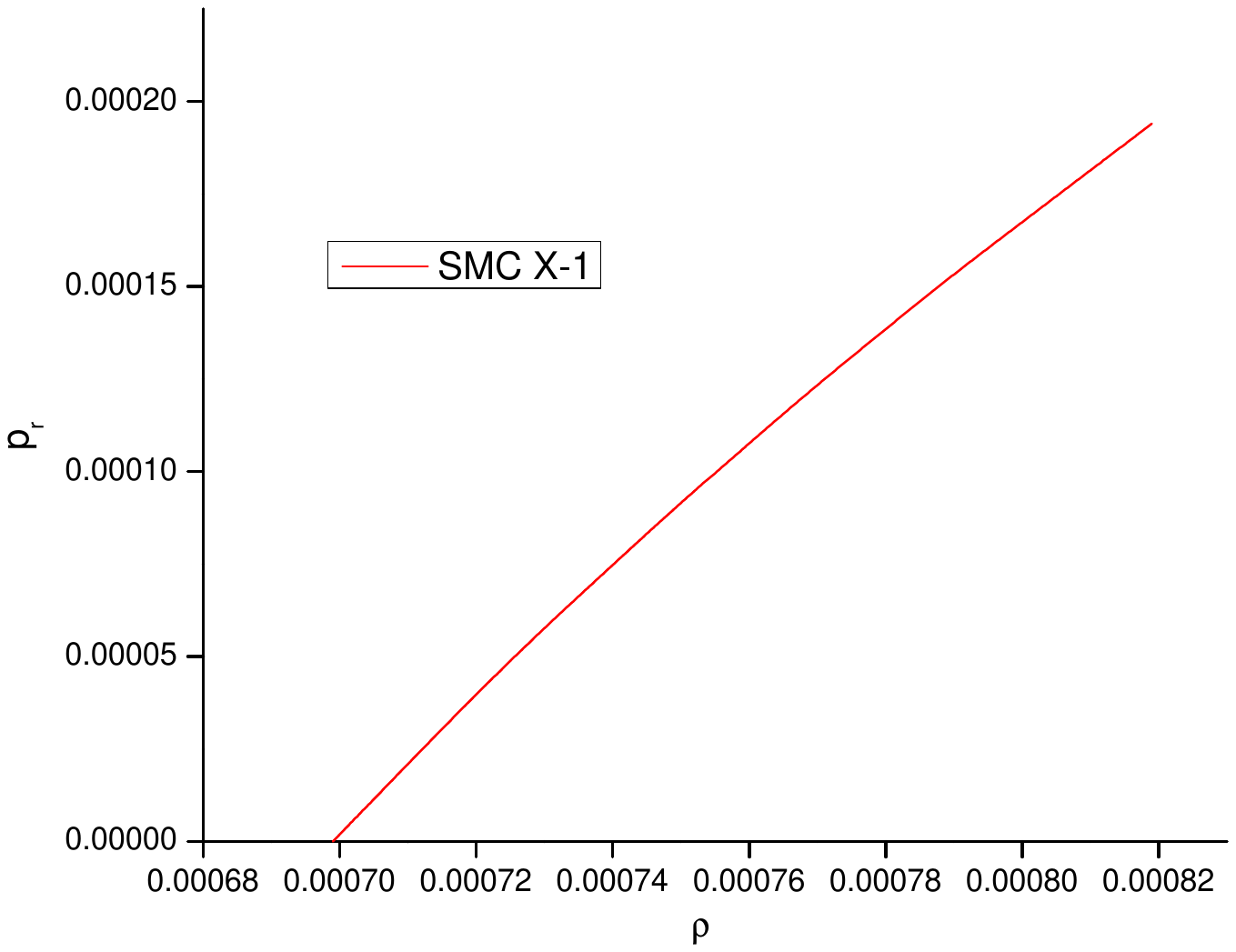}\includegraphics[width=4cm]{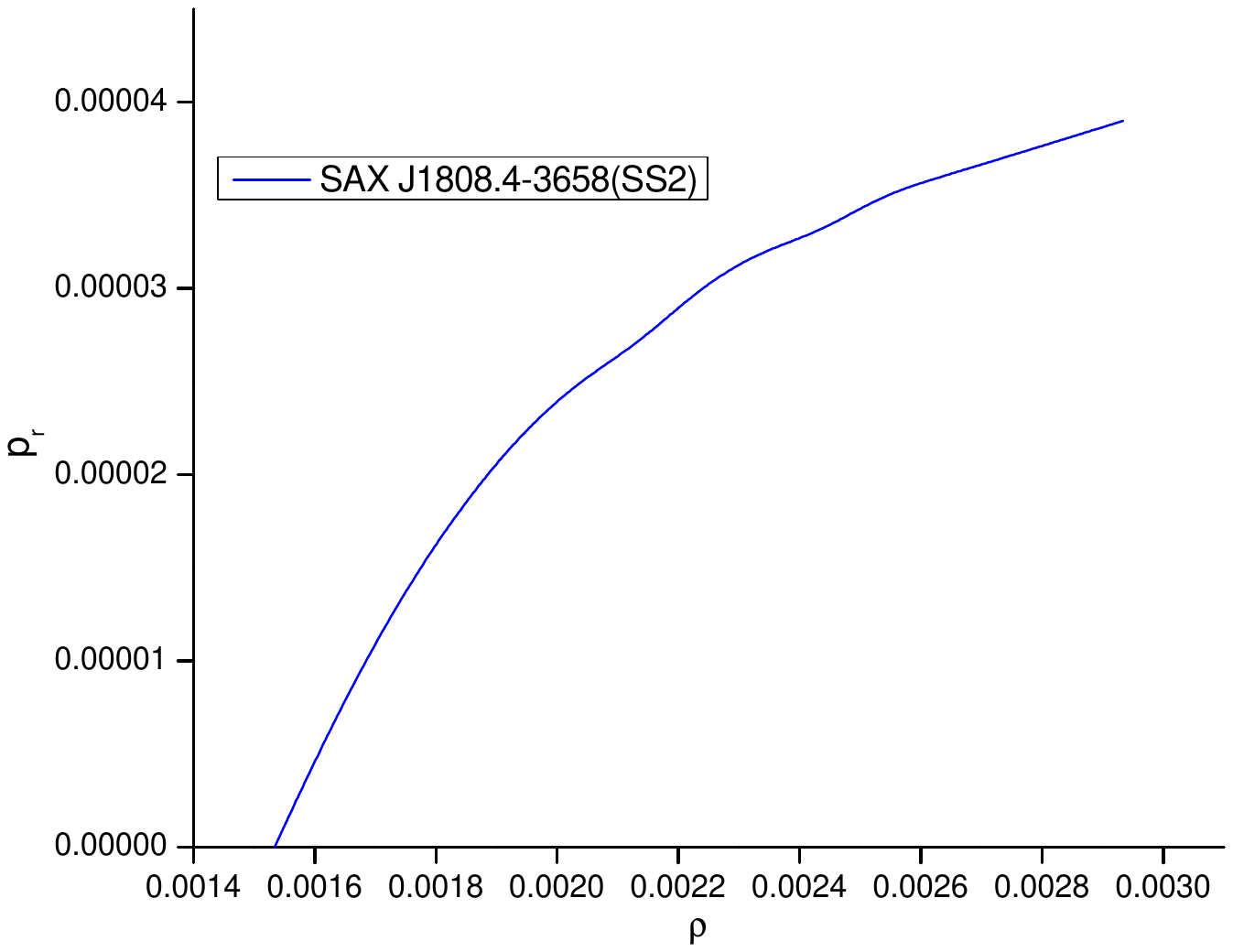}\includegraphics[width=4cm]{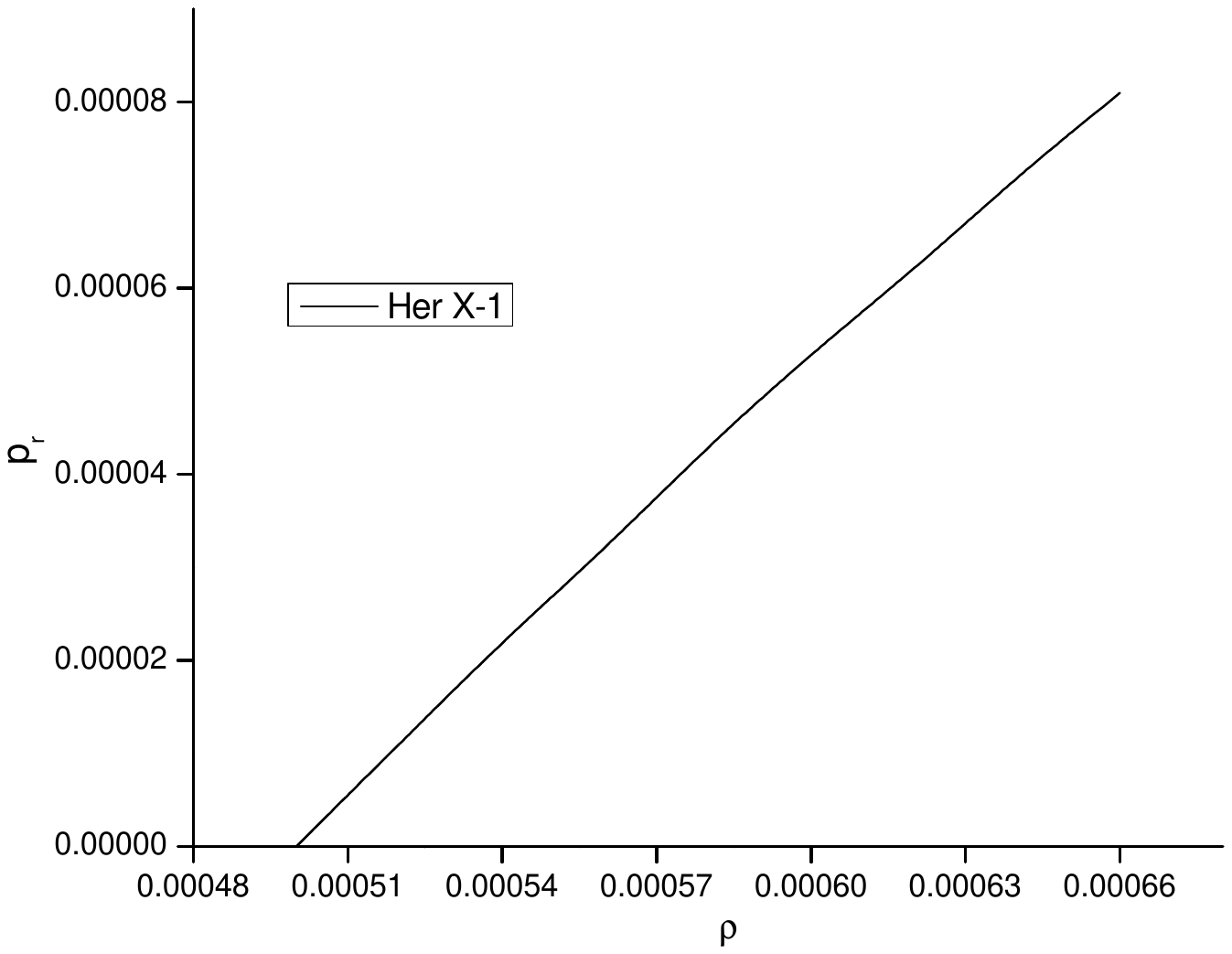}\\\includegraphics[width=4cm]{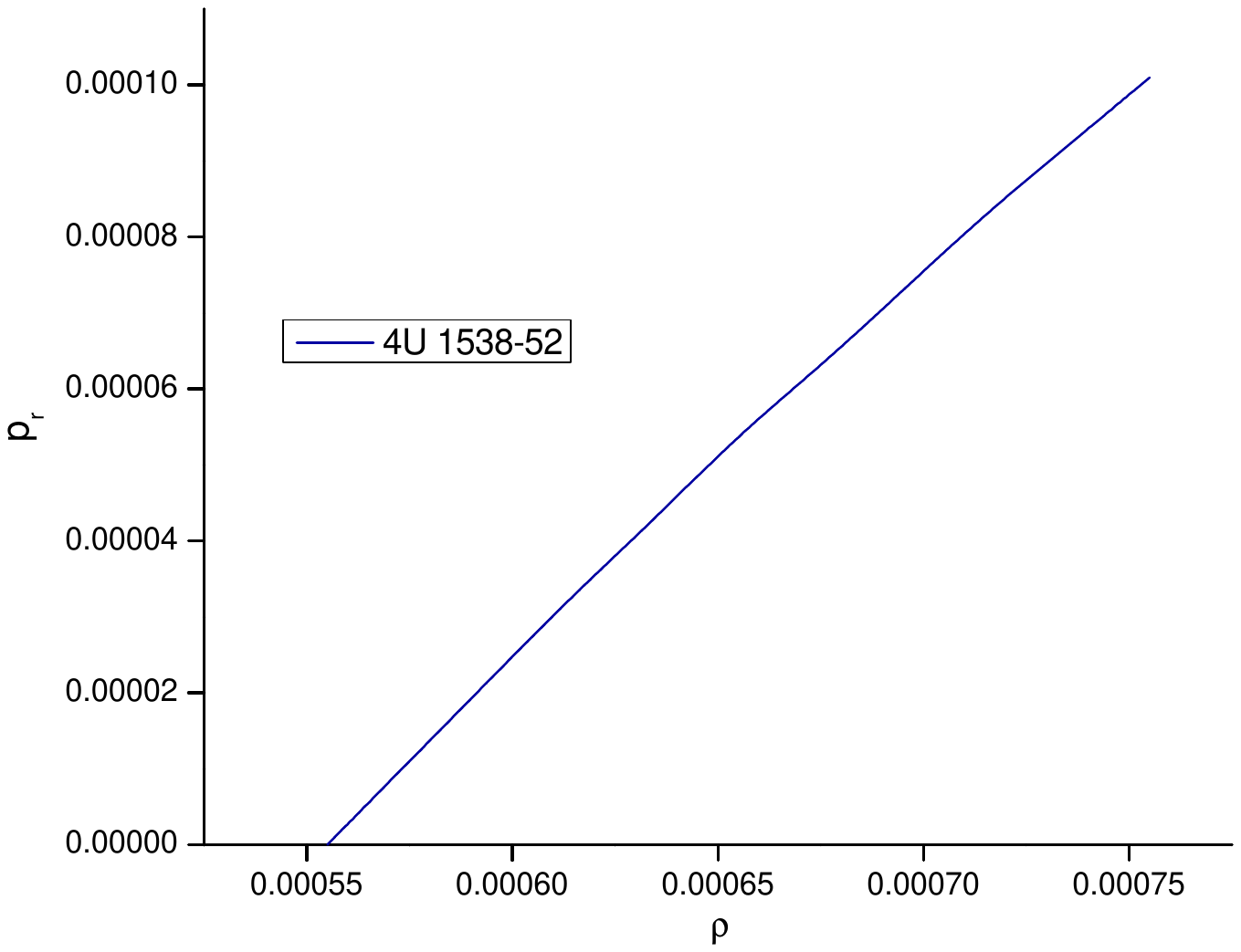}\includegraphics[width=4cm]{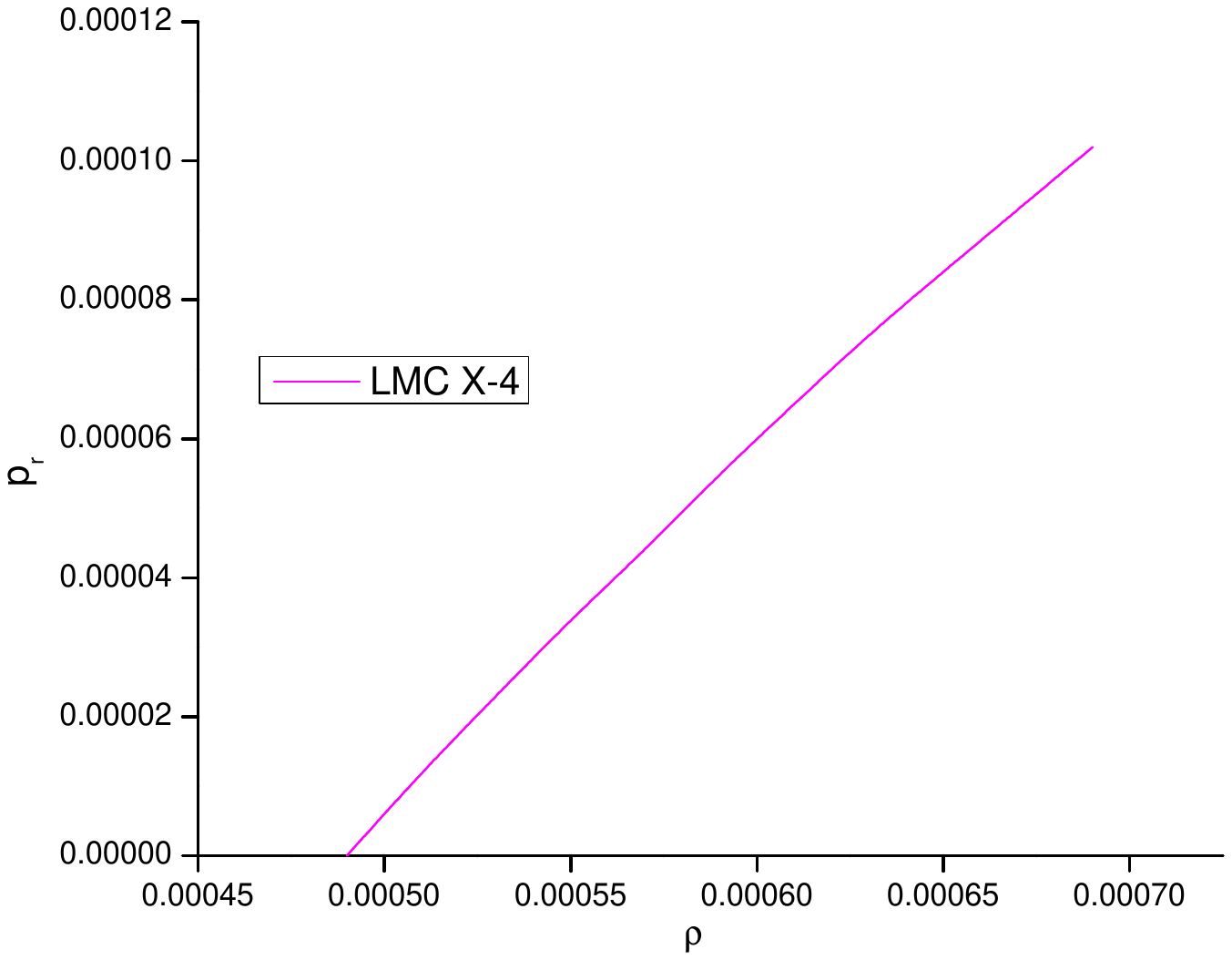}\includegraphics[width=4cm]{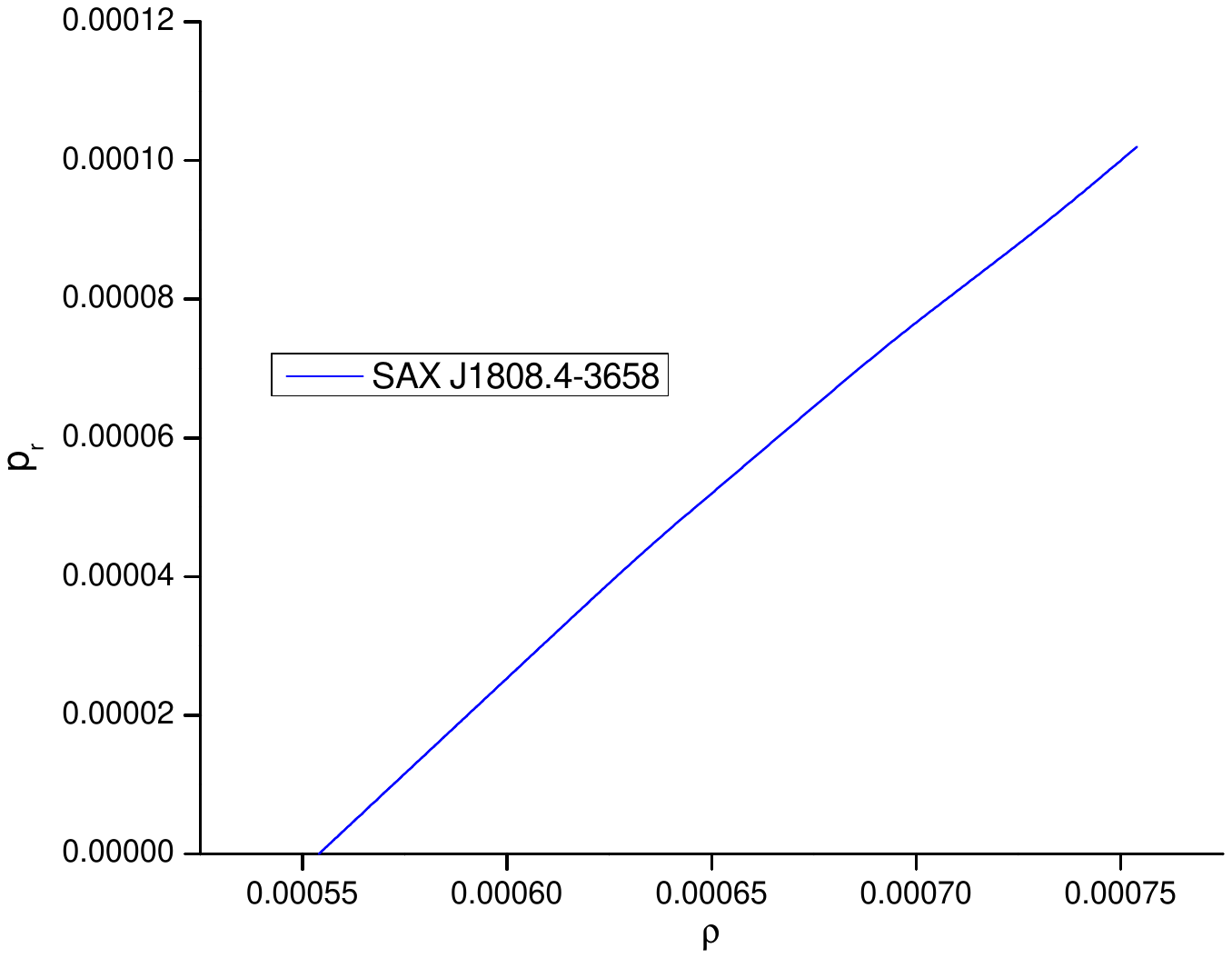}\includegraphics[width=4cm]{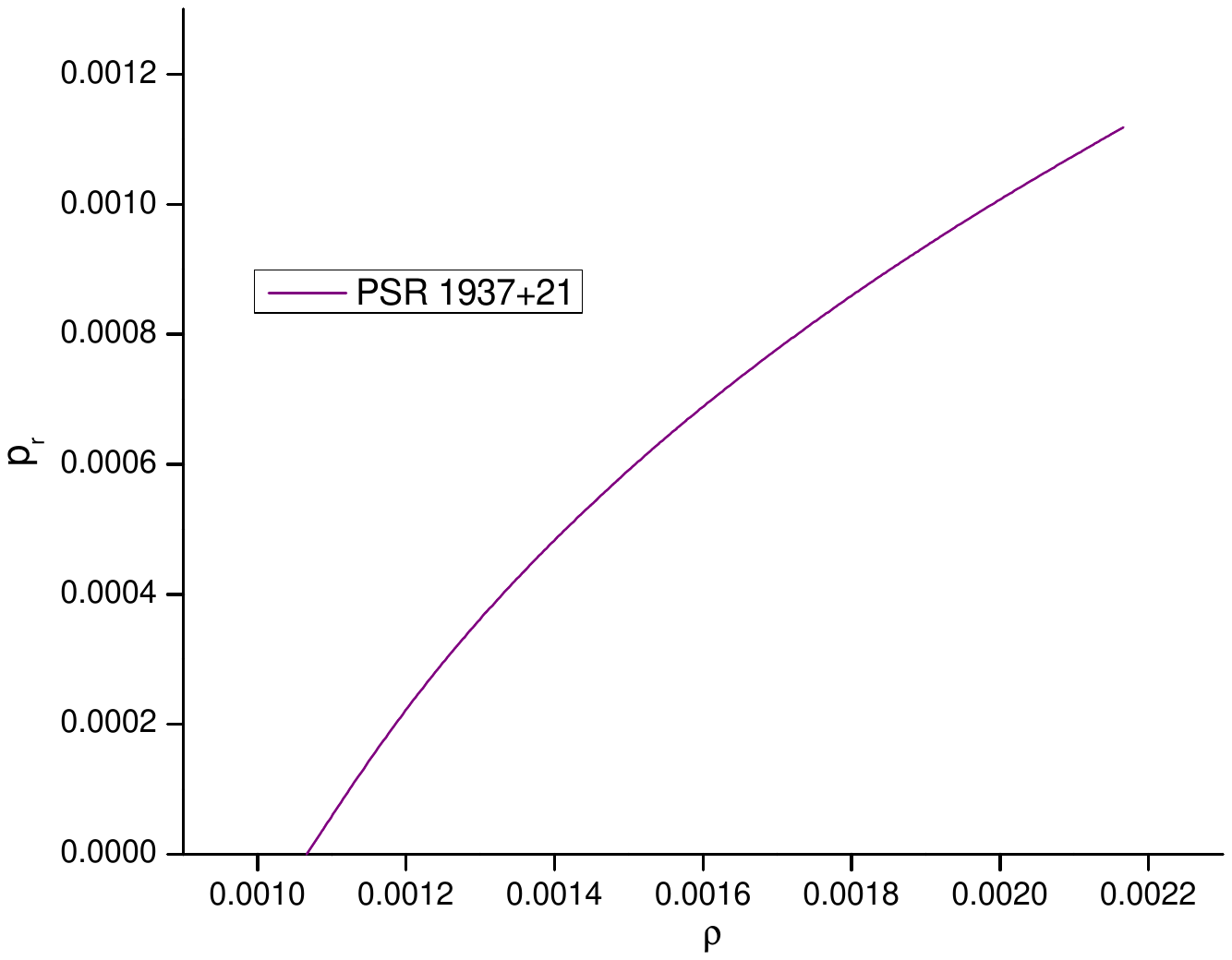}\\\includegraphics[width=4cm]{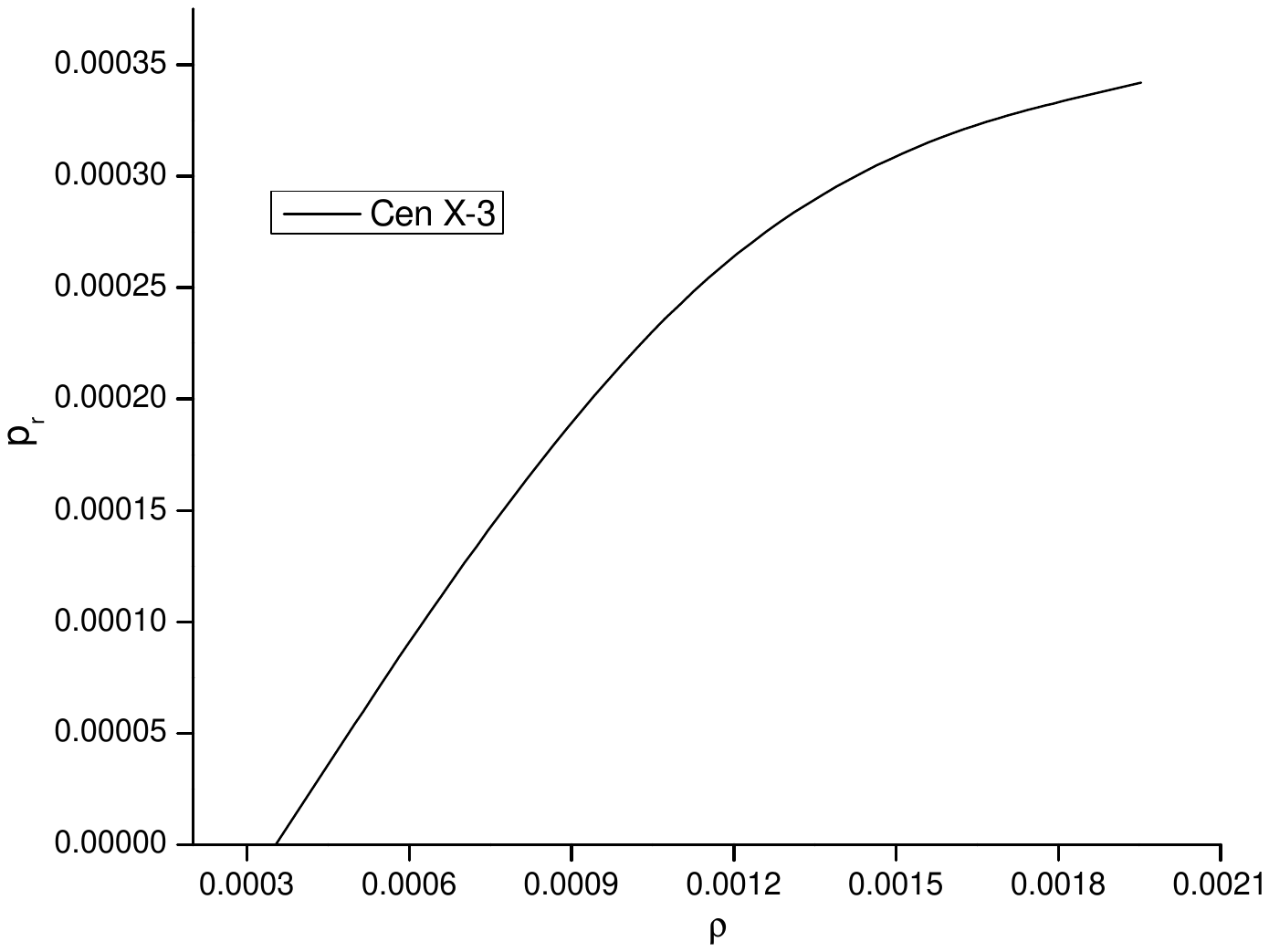}\includegraphics[width=4cm]{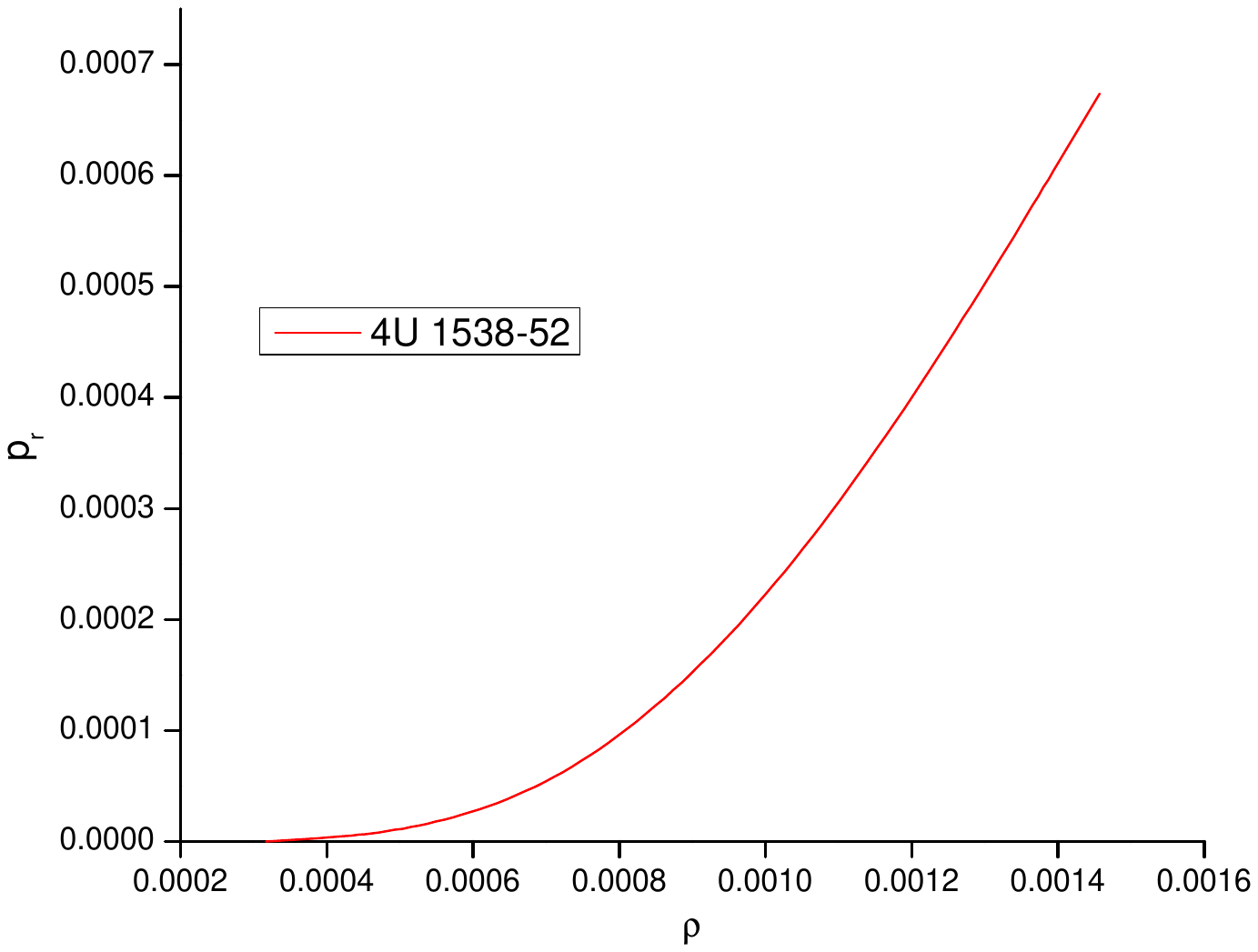}\includegraphics[width=4cm]{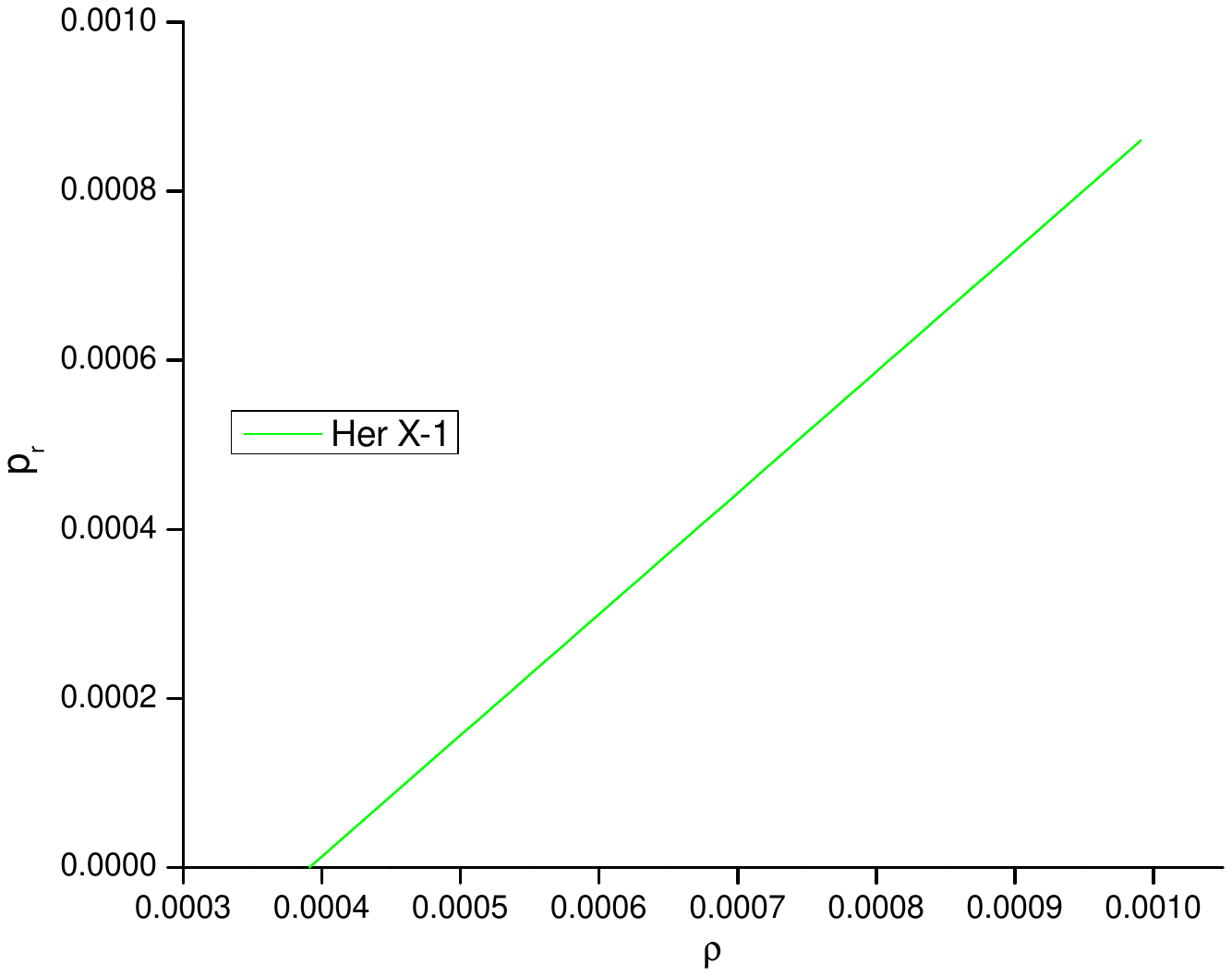}\includegraphics[width=4cm]{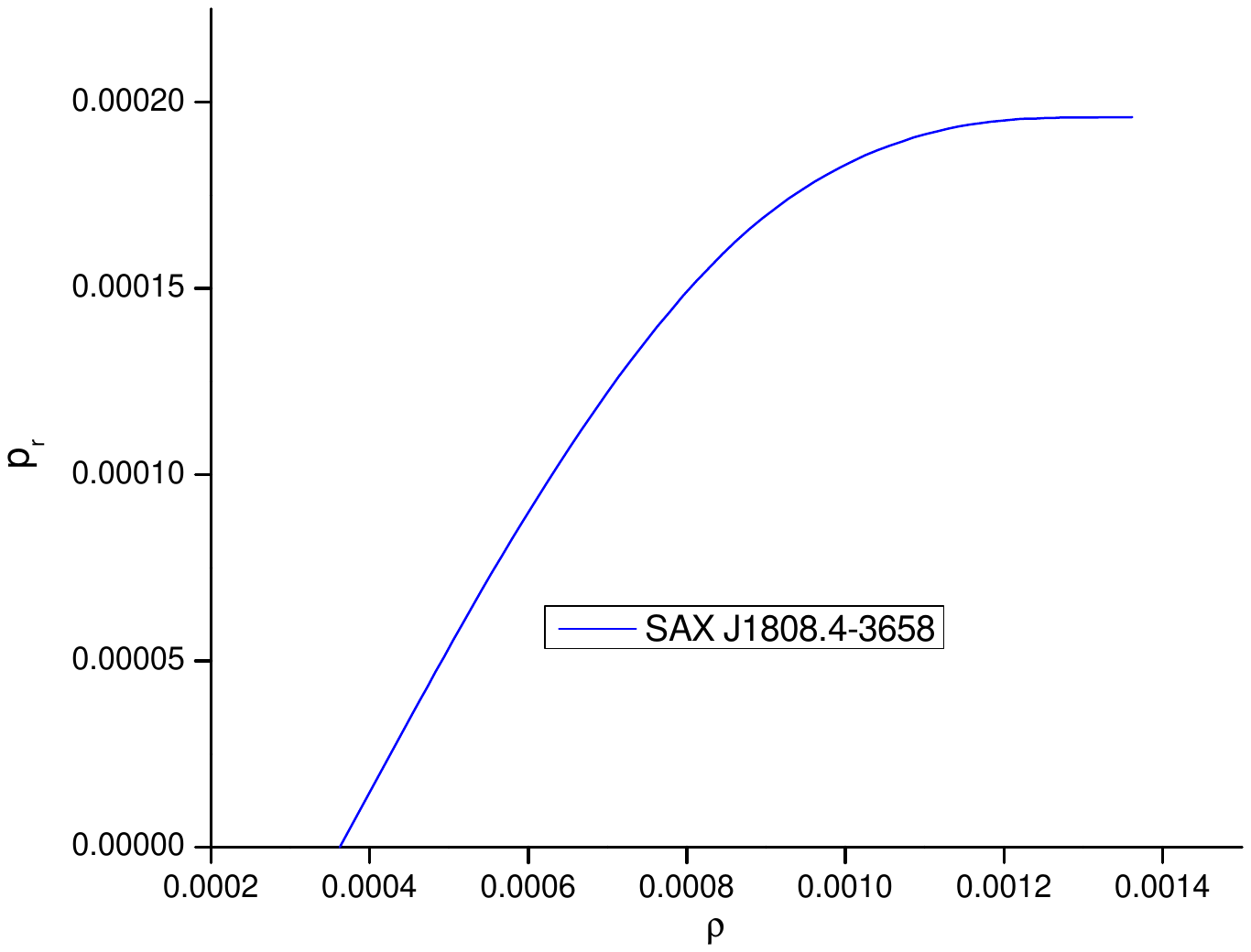} 
 \caption{Variation of radial pressure $p_r$ (in $km^{-2}$) with respect to energy density $\rho$ (in $km^{-2}$). For the Case \textbf{I}, we have plotted following stars in first and second row: (i) EXO 1785-248, (ii) SMC X-1, and (iii) SAX J1808.4-3658(SS2)-1, (iv) Her X-1, (v) 4U 1538-52, (vi) LMC X-4, (vii) SAX J1808.4-3658, (viii) PSR 1937+21. For the Case \textbf{II}, we have plotted the following stars in third row: (ix) Cen X-3, (x) 4U 1538-52, (xi) Her X-1, (xii) SAX J1808.4-3658}.\label{prho} 
 \end{center}
 \end{figure}

 Here, we are developing our consideration same as in \cite{Gondek2000}.
 In that work, authors showed that Eq. (\ref{EOS}) corresponds to self-bound matter at the density 
 $\rho_s$ at zero pressure, and with a fixed sound velocity. We start from 
 a certain value of $\rho_s$ where the pressure is zero i.e. at the boundary for our model.
  The dependence of pressure on the density diagram for neutron stars with realistic EoS are represented in Fig. \ref{prho} (see Table \textbf{II} for considering values). For example, By approximation of Eq. (\ref{pd2d}) in linear power of $\rho-\rho_s$, we obtain $p_r \approx a\,(\rho-\rho_s)$, where $a=\frac{(3-K)}{2+\tilde{\rho_1}\,(1-K)^{3/2}}\,\Big(1+\frac{\tilde{\rho_1}}{2K}\Big)\,\Big(1+\frac{\tilde{\rho}_{1s}}{2K}\Big)$. In the Fig. \ref{prho}, we observe that the radial pressure $p_r$ vanishes at surface density $\rho_s$. This implies that $p_r$ can be expressed by interpolation in power of $\rho-\rho_s$. Such parametrization is very convenient for stellar modelling, 
 which also relevant to the interior of stable stellar configurations \cite{Gondek2000}.  
 
\section{Final Remarks}
\label{FR}
In this article we have considered Buchdahl ansatz \cite{Buchdahl} for representing a
class of neutron stars in the standard framework of General Relativity. The main catch point is an extendable analytic solution for positive and negative values of spheroidal parameter $K$. We have focused on characterizing several exotic astrophysical objects with similar mass and radii, like LMC X-4, SMC X-1, EXO 1785-248 etc, which are confirmed by observations of gamma-ray repeaters and anomalous X-ray pulsars. To make the set of equations more tractable, we use Gupta-Jasim \cite{Gupta2004} two step method to solve the system of hypergeometric equation. Furthermore, pressure inside relativistic compact objects is most likely isotropic, but here we investigate anisotropic fluid model
that plays a significant role in the strong-field regime \cite{Raposo}. Considering the motivation, we choose anisotropy parameter $\Delta$, which is increasing for small $r$, and decreasing after
reaching at maximum in the interior of the star \cite{Mak}. Overall, we find eight different solutions depending on the choice for the metric potential which leads to
solutions of the condition of pressure anisotropy.

Based on physical requirements, we match the interior solution to an exterior vacuum Schwarzschild spacetime on the boundary surface at $r = R$, and from the comparison of both side metrics, all constants are determined which are enlisted in Table \textbf{I}. By using these constants in our investigation for several analogue objects with similar mass and radii, namely, EXO 1785-248, SMC X-1, SAX J1808.4-3658 (SS2), Her X-1, 4U 1538-52, LMC X-4, SAX J1808.4-3658, PSR 1937+21, Cen X-3, 4U 1538-52, and SAX J1808.4-3658, we have studied the anisotropy affects physical properties, such as: energy density, radial and tangential pressure. To illustrate these behaviour, we have generated a plot in Fig. \ref{f1} against the radial
coordinate $r/R$ in Km. The density and pressures are positive and
remain finite at the interior of stars. Interestingly, energy density attends its maximum value at the centre of stars and central densities close to the order of $\sim 10^{15}$. Since this bound is consistent with the argument by Ruderman \cite{Ruderman} for anisotropic matter in certainly very high density ranges. Mainly, this situation admits the theory
that the core of exotic astrophysical objects, is intensely compact, particularly in case of millisecond pulsar SAX J1808.4-3658 (SS2). We have succeeded to determine the central density of the  massive pulsars $4.06\times 10^{15}  $ with masses 1.3237 $M_{\odot}$.

To refine the model further, we have analyzed mass-radius ($M-R$) relationship, generalised
TOV equations, the surface redshift, energy conditions and the EoS in linear approximation form, respectively
The obtained mass-radius ratio for anisotropic stars are consistent with the Buchdahl's \cite{Buchdahl} bound, though
he proposed for isotropic object for which the energy density is non-increasing outwards the
boundary.  On the otherhand, bases on the work by Gondek-Rosinska \emph{et al} \cite{Gondek2000}
in which EoSs have been approximated to a linear function of density, we have plotted dependence of pressure on the density diagram in Fig. \ref{prho}, and utilizing values are summarized in Table \textbf{II}. Such EoS is  very convenient for stable stellar modelling. At the same time, we have checked the velocity of sound ($v^2_i$) 
which is less than the light's velocity as evident in Fig. \ref{f2}.

 Aside from the influence on the $M-R$ ratio for anisotropic stars, we present the variation
of total mass $M$ (normalized in solar mass $M_{\odot}$) with the
total radius $R$ for different chosen parametric values (see Fig. \ref{f7}). We have also studied the stability of the configurations with respect to generalized-TOV equation and found the equilibrium configuration where the gravitational force ($F_g$) is dominating over the hydrostatic ($F_h$) and anisotropic ($F_a$) forces, as seen from Fig. \ref{f4}. As a concluding remark is that our proposed model satisfies all physical requirements as well as horizon-free and stable configuration that helps us further understanding about anisotropic compact objects.

%%%%%%%%%%%%%%%%%%%%%%%%%%%%%%%%%%%%%%%%%%%%%%%%%%%%%%%%%%%%%%%%%%%%%%%%%%%%%%%%%%%%%%%%%%%%%%%%
\section*{Appendix~(A): Gupta-Jasim two step Method}
This appendix is devoted in solving hypergeometric differential equation (HDE).
Note that HDE can be solved
directly in terms of hypergeometric series. However, some hypergeometric equations can be solved in closed form.
In our preset article, we use Gupta-Jasim \cite{Gupta2004} two step method for solving
the system of equations.

\textbf{Step-I:}
In this section, we provide \emph{Gupta-Jasim} Method
in detail to supplement the results presented in the main text. Starting with Eq. (\ref{eq10}),
which is 
\begin{equation}
(1-Z^2)\,\frac{d^2Y}{dZ^2}+Z\,\frac{dY}{dZ}+(1-K+\Delta_0\,K)\,Y=0, \label{Diff12}
\end{equation}
Now, differentiate the equation with respect to $Z$, we get
\begin{equation}
(1-Z^2)\,\frac{d^3Y}{dZ^3}-Z\,\frac{d^2Y}{dZ^2}+(2-K+\Delta_0\,K)\,\frac{dY}{dZ}=0. \label{Diff13}
\end{equation}
Here, substitute a new variable $G=dY/dZ$, yields
\begin{equation}
(1-Z^2)\,\frac{d^2G}{dZ^2}-Z\,\frac{dG}{dZ}+(2-K+\Delta_0\,K)\,G=0. \label{Diff14}
\end{equation}
In the $Z< 1$ case, we use the transformation $Z=\sin x$ (which corresponds the case $K < 0$, as  $0 < K < 1$ is not a valid solution) into the Eq. (\ref{Diff14}), and the above turns out to be (note that the first derivative term vanishes) 
\begin{equation}
\frac{d^2G}{dx^2}+(2-K+\Delta_0\,K)\,G=0. \label{Diff15}
\end{equation}
Thus we have the solution for Eq. (\ref{Diff15}) takes the following form 
\begin{equation}
\frac{dY}{dZ}=G=A_1\,\cosh(nx)+B_1\,\sinh(nx),~~~~~~~~~\text{where}~~~ 2-K+\Delta_0\,K =-n^2  
\label{Diff16}
\end{equation}

\textbf{Step-2:} In this step we find $\frac{d^2Y}{dZ^2}$ from Eq. (\ref{Diff16}) this
yields
\begin{eqnarray}
\frac{d^2Y}{dZ^2}
&=&\frac{dG}{dZ}=\frac{dG}{dx}.\,\frac{dx}{dZ}=[A_1\,n\,\sinh(nx)+B_1\,n\,\cosh(nx)]\,\sec x. \label{Diff17}
\end{eqnarray}
Now, inserting the expression (\ref{Diff16}) and (\ref{Diff17}) into the hypergeometric Eq. (\ref{Diff12}),
 and using $2-K+\Delta_0\,K =-n^2$, we finally arrive at
\begin{equation}
 Y(x)=\frac{1}{(n^2+1)} \left[\cosh (n x)\,(A_{1} \sin x +B_{1}\,n\,\cos x ) + \sinh(n x)\, (A_{1}\,n\,\cos x + B_{1} \sin x \,) \right].
\end{equation}
which determines the $e^{\nu}=Y^2$. Similarly, one can obtain the other solutions of hypergeometric equation. \\ 
\\
\section*{Appendix (B:) The expressions for  used coefficients in Eqs. (\ref{62})-(\ref{77})}

We listed here all the expression that have been used to find the velocity of sound in Eqs. (\ref{62})-(\ref{77}) as follows\\
\\
$\, N_1=\,\frac{4\,(n^{2}+1)\,\tan x}{\,K\,(1-K)\,\cos^{2} x}\,\left[\,\frac{A_{1}\,\cosh(n\,x)+B_{1}\,\sinh(n\,x)}{\cosh(n x)\,(\,A_{1} +B_{1}\,n\,\cot x \,) + \sinh(n\,x)\,(\,A_{1}\,n\,\cot x + B_{1} \,) }\,\right]+\frac{2\,\tan x}{K\,\cos^2 x}+\frac{2\,(n^{2}+1)}{\,K\,(1-K)\,\cos^{2} x}\,\frac{L_1}{M_1}, $ \\ \\
$ L_1=\big[\cosh(n\,x)\,(A_{1}+B_{1}\,n\,\cot x) + \sinh(n\,x)\,(A_{1}\,n\,\cot x+B_{1})\big] \big[A_1\,n\,\sinh(n\,x)+B_1\,n\,\cosh(n\,x)\big]-\big[A_{1}\,\cosh(n\,x)+B_{1}\,\sinh(n\,x)\big]\big[n\,\sinh(n\,x)(\,A_{1} +B_{1}\,n\,\cot x \,)-B_1\,\cosh(n\,x)\,n\,\csc^2 x+n\,\cosh(n\,x)(\,A_{1}\,n\,\cot x + B_{1} \,)-n\,\sinh(n\,x)\,\csc^2 x\big], $\\ \\
$ M_1=\big[\cosh(n\,x)\,(A_{1}+B_{1}\,n\,\cot x) + \sinh(n\,x)\,(A_{1}\,n\,\cot x+B_{1})\big]^2 $,\\ \\
$\, N_2=\,\frac{4\,(1-n^{2})\,\tan x}{\,K\,(1-K)\,\cos^{2} x}\,\left[\,\frac{C_{1}\,\cos(n\,x)+D_{1}\,\sin(n\,x)}{C_1\,\cos(n x)+D_{1}\,\sin(n\,x)-\,n\,\cot x\,(\,C_{1}\,\sin(n\,x) - D_{1}\,\cos(n\,x)) }\,\right]+\frac{2\,\tan x}{K\,\cos^2 x}+\frac{2\,(1-n^{2})}{\,K\,(1-K)\,\cos^{2} x}\,\frac{L_2}{M_2}, $ \\ \\
$ L_2=\big[C_1\,\cos(n x)+D_{1}\,\sin(n\,x)-\,n\,\cot x\,\big(\,C_{1}\,\sin(n\,x) - D_{1}\,\cos(n\,x)\big)\big] \big[-n\,C_1\,\sin(n\,x)+D_1\,n\,\cos(n\,x)\big]-\big[C_{1}\,\cos(n\,x)+D_{1}\,\sin(n\,x)\big]\big[-n\,C_1\,\sin(n\,x)+D_1\,n\,\cos(n\,x)+n\,\csc^2 x\big(\,C_{1}\sin(n\,x) -D_{1}\,\cos(n\,x )\big)-n\,\cot x\big(\,C_{1}\,n\,\cos(n\, x) + D_{1}\,n\,\sin(n\,x) \big)\big], $\\ \\
$ M_2=\big[C_1\,\cos(n x)+D_{1}\,\sin(n\,x)-\,n\,\cot x\,\big(\,C_{1}\,\sin(n\,x) - D_{1}\,\cos(n\,x)\big)\big]^2 .$\\\\
$ N_3=\frac{8\,(2-\cos^2 x)}{K(1-K)\,\cos^3 x}\left[\frac{E_{1}\,\cos (x)+F_{1}\,\sin (x)}{ \,E_{1}\,(2x+\sin 2x)-F_{1}\,\cos 2x}\right]+\frac{2\tan x}{K\,\cos^2 x}+\frac{8\,\sin x}{K(1-K)\,\cos^2 x},\\\\
M_3=\Bigg[\frac{\big(E_{1}\,(2x+\sin 2x)-F_{1}\,\cos 2x \big)(-E_1\,\sin x+F_1\,\cos x)-(E_1\,\cos x+F_1\,sin x)\big( 4\,E_1\,\cos^2 x+2\,F_1\,\sin\, 2x\big)}{\big(E_{1}\,(2x+\sin 2x)-F_{1}\,\cos 2x \big)^2}\Bigg]. $ \\\\
$ N_4=\frac{2\tan x}{K\,\cos^2 x}+\frac{2\,\sin x}{K(1-K)\,\cos^2 x}
\,\Bigg[\frac{\big(G_{1}\,(\cos x + x\,\sin x)+H_{1}\,\sin x \big)\,G_1-(G_1\, x+H_1)\big( G_1\,x\,\cos x+H_1\,\cos x\big)}{\big(G_{1}\,(\cos x+x\,\sin x)+H_{1}\,\sin x \big)^2}\Bigg]\\+\frac{2\,(2-\cos^2 x)}{K(1-K)\,\cos^3 x}\left[\frac{G_{1}\,(x)+H_{1}}{G_{1}\,(\cos x + x\,\sin x)+H_{1}\,\sin x}\right].$\\\\
$\, N_5=\,\frac{-4\,(n^{2}+1)\,\cosh x}{\,K\,(K-1)\,\sinh^{3} x}\,\left[\,\frac{A_{2}\,\cos(n\,x)+B_{2}\,\sin(n\,x)}{A_{2}\,\cos (n\,x) +B_{2}\,\sin(n\,x) + n\,\tanh x\,\big(\,A_{2}\,\sin(n\,x) - B_{2}\,\cos(n\,x)\big) } \,\right]+\frac{2\,\cosh x}{K\,\sinh^3 x}+\frac{2\,(n^{2}+1)}{\,K\,(1-K)\,\sinh^{2} x}\,\frac{L_5}{M_5}, $ \\ \\
$ L_5=\big[A_{2}\,\cos (n\,x) +B_{2}\,\sin(n\,x) + n\,\tanh x\,\big(\,A_{2}\,\sin(n\,x) - B_{2}\,\cos(n\,x)\big) \big] \big[-A_2\,n\,\sin(n\,x)+B_2\,n\,\cos(n\,x)\big]-\big[A_{2}\,\cos(n\,x)+B_{2}\,\sin(n\,x)\big]\big[-n\,A_{2}\,\sin(n\,x) +B_{2}\,n\,\cos (n\,x) +n\,\sec h^2 x\big(A_{2}\,\sin(n\,x) - B_{2}\,\cos(n\,x)\big)+n\,\tanh x\big(A_2\,n\,\cos(n\,x) +B_2\,n\,\sin(n\,x)\big)\big], $\\ \\
$ M_5=\big[A_{2}\,\cos (n\,x) +B_{2}\,\sin(n\,x) + n\,\tanh x\,\big(\,A_{2}\,\sin(n\,x) - B_{2}\,\cos(n\,x)\big) ]^2 $,\\\\
$ S_2=\,\frac{2\,\cosh x\,\sinh^2 x\,(K-1)-4\,\cosh x\,\big(3-K+(K-1)\cosh^2 x\big)}{K(K-1)\,\sinh^5 x}. $\\\\
$\, N_6=\,\frac{-4\,(1-n^{2})\,\coth x}{\,K\,(K-1)\,\sinh^{2} x}\,\left[\,\frac{C_{2}\,\cosh (n\,x)+D_{2}\,\sinh (n\,x)}{~ [C_{2}\,\cosh(n x) + D_{2}\, \sinh(n x)~ ] - 
 n\,\tanh x~[~ C_{2}\, \sinh(n x)+D_{2}\, \cosh(n x) ~] }\,\right]+\frac{2\,\cosh x}{K\,\sinh^3 x}+\frac{2\,(1-n^{2})}{\,K\,(K-1)\,\sinh^{2} x}\,\frac{L_6}{M_6}, $ \\\\ 
$ L_6=\big[ C_{2}\,\cosh(n x) + D_{2}\,\sinh(n x) - 
 n\,\tanh x\,\big( C_{2}\,\sinh(n x)+D_{2}\,\cosh(n x)\big)\big] \big[C_2\,n\,\sinh(n\,x)+D_2\,n\,\cosh(n\,x)\big]-\big[C_{2}\,\cosh(n\,x)+D_{2}\,\sinh(n\,x)\big]\big[n\,C_2\,\sinh(n\,x)+D_{2}\,n\,\cosh(n\,x) -\,n\sec h^2 x\,\big(C_2\sinh(n\,x)+D_2\cosh(n\,x)\big)-n\,\tanh x \big(\,C_{2}\,n\,\cosh(n\,x)+D_{2}\,n\,\sinh(n\,x)\,\big)\big] $\\ \\
$ M_5=\big[ C_{2}\,\cosh(n x) + D_{2}\,\sinh(n x) - 
 n\,\tanh x\,\big( C_{2}\,\sinh(n x)+D_{2}\,\cosh(n x)\big)\big]^2 $,\\\\
$ N_7=\frac{-8\,(1+\cosh^2 x)}{K(K-1)\,\sinh^3 x}\left[\frac{E_{2}\,\cosh x + F_{2}\,\sinh x}{ \,E_{2}\,\cosh 2x+F_{2}\,(\sinh 2x-2x)}\right]+\frac{2\cosh x}{K\,\sinh^3 x}+\frac{8\,\cosh x}{K(K-1)\,\sinh^2 x}\,M_7,\\\\
M_7=\Bigg[\frac{\big(\,E_{2}\,\cosh 2x+F_{2}\,(\sinh 2x-2x) \big)(E_2\,\sinh x+F_2\,\cosh x)-(E_2\,\cosh x+F_2\,\sinh x)\big( 2\,E_2\,\sinh 2x+2\,F_2(\cosh 2x\,-1)\big)}{\big(E_{2}\,\cosh 2x+F_{2}\,(\sinh 2x-2x) \big)^2}\Bigg]. $\\\\
$ N_8=\frac{2\cosh x}{K\,\sinh^3 x}+\frac{2\,\cosh x}{K(K-1)\sinh^2 x}\,\Bigg[\frac{\big(G_{2} (x\,\cosh x -\sinh x)+H_{2}\,\cosh x \big)\,G_2-(G_2\, x+H_2)\big( G_2\,x\,\sinh x+H_2\,\sinh x\big)}{\big(G_{2}\,(x\,\cosh x-\sinh x)+H_{2}\,\cosh x \big)^2}\Bigg]\\+\frac{-2\,(1+\cosh^2 x)}{K(K-1)\,\sinh^3 x}\left[\frac{G_{2}\,(x)+H_{2}}{G_{2}\,(x\,\cosh x - \sinh x)+H_{2} \cosh x}\right]. $ \\\\
$ S_1=\,\frac{2\,\cos^2 x\,\sin x\,(K-1)+4\,\sin x\,\big(3-K+(K-1)\sin^2 x\big)}{K(K-1)\,\cos^5 x}, $ \\\\
$ S_2=\,\frac{2\,\cosh x\,\sinh^2 x\,(K-1)-4\,\cosh x\,\big(3-K+(K-1)\cosh^2 x\big)}{K(K-1)\,\sinh^5 x}. $\\\\

\textbf{Acknowledgments}: SKM acknowledges support from the Authority of University of Nizwa, Nizwa, Sultanate of Oman. The authors would like give special thanks to anonymous referee for suggesting several pertinent issues which have enabled us to improve the manuscript substantially. 

\end{document}